\documentclass{aa}
\usepackage[varg]{txfonts}
\usepackage{natbib}
\usepackage{color}
\usepackage{amsmath, amssymb}
\usepackage{pifont}
\usepackage{graphicx}
\usepackage{rotating}
\usepackage{anysize}
\usepackage{longtable}
\usepackage{times}
\usepackage{mathptmx}
\usepackage{multirow}
\usepackage[breaklinks=true]{hyperref}
\usepackage{float} 
\usepackage{verbatim}
\usepackage{footnote}

     \def\aap{A\&A}
     \def\apjl{ApJL}
     \def\apjs{ApJS}

\hypersetup{
	pdftoolbar=false,
	pdfmenubar=true,
	pdfstartview={FitH},
	breaklinks=true,
	frenchlinks=false,
        colorlinks=true,
        linkcolor=red,
        citecolor=blue,
        filecolor=cyan,
        urlcolor=magenta,
        pdftitle={Modelling the local and global cloud formation on HD 189733b},
	pdfsubject={planets and satellites: individual: HD 189733b - planets and satellites: atmospheres - methods: numerical.},
	pdfauthor={Mr. Graham Lee}
        }

\begin{document}

\title{Modelling the local and global cloud formation on HD 189733b}
\author{G. Lee\inst{1}\thanks{E-mail:
gl239@st-andrews.ac.uk} \and Ch. Helling\inst{1} \and I. Dobbs-Dixon\inst{2} \and D. Juncher\inst{3}}
\institute{SUPA, School of Physics and Astronomy, University of St Andrews, North Haugh, St Andrews, Fife KY16 9SS, UK \and
NYU Abu Dhabi, PO Box 129188, Abu Dhabi, UAE \and
Niels Bohr Institute \& Centre for Star and Planet Formation, University of Copenhagen, DK-2100 Copenhagen, Denmark}

\date{Received: 27/02/2015  / Accepted: 21/05/2015}

\abstract{Observations suggest that exoplanets such as HD 189733b form clouds in their atmospheres which have a strong feedback onto their thermodynamical and chemical structure, and overall appearance.}
{
Inspired by mineral cloud modelling efforts for Brown Dwarf atmospheres, we present the first spatially varying kinetic cloud model structures for HD 189733b.}
{
We apply a 2-model approach using results from a 3D global radiation-hydrodynamic simulation of the atmosphere as input for a detailed, kinetic cloud formation model. 
Sampling the 3D global atmosphere structure with 1D trajectories allows us to model the spatially varying cloud structure on HD 189733b.
The resulting cloud properties enable the calculation of the scattering and absorption properties of the clouds.}
{
We present local and global cloud structure and property maps for HD 189733b. 
The calculated cloud properties show variations in composition, size and number density of cloud particles which are strongest between the dayside and nightside. 
Cloud particles are mainly composed of a mix of materials with silicates being the main component.
Cloud properties, and hence the local gas composition, change dramatically where temperature inversions occur locally. 
The cloud opacity is dominated by absorption in the upper atmosphere and scattering at higher pressures in the model.
The calculated 8$\mu$m single scattering Albedo of the cloud particles are consistent with \textit{Spitzer} bright regions.   
The cloud particles scattering properties suggest that they would sparkle/reflect a midnight blue colour at optical wavelengths.}
{}


\maketitle

\section{Introduction}
\label{sec:Introduction}
The atmospheres of exoplanets are beginning to be characterised in great detail. 
Spectroscopic measurements from primary and secondary transits of hot Jupiters have opened up the first examinations into the composition of their atmospheres. 
HD 189733b is one of the most studied hot Jupiters due to its bright parent star and large planet-to-star radius ratio. 
Observational data of this object has been collected by various groups from X-ray to radio wavelengths \citep{Knutson2007,Tinetti2007,Charbonneau2008,Grillmair2008,Swain2008,Swain2009,Danielski2014}.
Transit spectra from the \textit{Hubble} space telescope \citep{Lecavelier2008, Sing2011, Gibson2012} between 0.3$\mu$m and 1.6$\mu$m show a featureless spectrum consistent with Rayleigh scattering from the atmosphere.
These observations sample the outer limbs (dayside-nightside terminator regions) of the planet's atmosphere during primary eclipse.
\citet{Lecavelier2008} suggests haze/clouds to be responsible for the Rayleigh slope and estimate cloud particle radii of 10$^{-2}$$\ldots$10$^{-1}$ $\mu$m at $\sim$10$^{-6}$$\ldots$10$^{-3}$ bar local gas pressures. 
They suggest MgSiO$_{3}$[s] as the most likely cloud particle material due to its strong scattering properties. 
\citet{Pont2013} who reanalysed and combined the \textit{Hubble} and \textit{Spitzer} \citep{Agol2010,Desert2011,Knutson2012} observations found that the spectrum was `Dominated by Rayleigh scattering over the whole visible and near-infrared range$\ldots$' and also suggest cloud/haze layers as a likely scenario.
Subsequent analysis by \citet{Wakeford2015} showed that single composition cloud condensates could fit the Rayleigh scattering slope with grain sizes of 0.025$\mu$m.
\citet{Evans2013} measured the geometric albedo of HD 189733b over 0.29$\ldots$0.57 $\mu$m and infer the planet is likely to be a deep blue colour at visible wavelengths, which they attribute to scattering mid-altitude clouds.
These albedo observations sample the dayside face of the planet by comparing measurements before, during and after the secondary transit (occulation).
Star spots \citep{McCullough2014}, low atmospheric metallicity \citep{Huitson2012} and photo-chemical, upper atmosphere haze layers \citep{Pont2013} have been proposed as alternative explanations to these observations.

Clouds present in optically thin atmospheric regions are thought to significantly impact the observed spectra of exoplanets by flattening the ultra-violet and visible spectrum through scattering by small cloud particles, by depletion of elements and by providing an additional opacity source (e.g GJ 1214b; \citet{Kreidberg2014}).
The strong radiative cooling (or heating) resulting from the high opacity of the cloud particles also affect the local pressure which influences the local velocity field \citep{Helling2004}.
These effects have been observed and modelled in Brown Dwarf atmospheres which are the inspiration for the current study.
Clouds are also important for ionisation of the atmosphere by dust-dust collisions \citep{Helling2011} and cosmic ray ionisation \citep{Rimmer2013}.
This leaves open the possibility of lightning discharge events in exo-clouds \citep{Bailey2014}. 
Furthermore, \citet{Stark2014} suggest that charged dust grains could aid the synthesis of prebiotic molecules by enabling the required energies of formation on their surfaces. 
Multi-dimensional atmosphere simulations have been used to gain a first insight into the atmospheric dynamics of HD 189733b (e.g. \citet{Showman2009, Fortney2010, Knutson2012, Dobbs-Dixon2013}). 
Output from \citet{Showman2009} has been used for kinetic, non-equilibrium photo-chemical models of HD 189733b \citep{Moses2011,Venot2012, Agundez2014}. 

In this work, we present cloud structure calculations for HD 189733b based on 3D radiation-hydrodynamic [3D RHD] atmosphere results of \citet{Dobbs-Dixon2013}. 
With this 2-model approach, we investigate how the specific cloud properties vary locally and globally throughout HD 189733b's atmosphere.
In Sect. \ref{sec:Method} we outline our 2-model approach.
We summarise our cloud formation model and discuss our approach to atmospheric mixing, cloud particle opacities and model limitations.
In Sect. \ref{sec:Application} the spatially varying cloud properties (particle sizes, number density, material composition, nucleation rate, growth rate and opacity) are presented.
In Sect. \ref{sec:Opacity} we use the results of the  locally and spatially varying cloud properties to calculate the wavelength dependent opacity and contribution of the clouds to the reflected light and transit spectra for HD 189733b.
Section \ref{sec:Discussion} discusses our results with respect to observational findings for HD 189733b.
Our conclusions are outlined in Sect. \ref{sec:Conclusion}.

\section{2-model approach}
\label{sec:Method}

\begin{figure*}
\centering
 \includegraphics[width=0.49\textwidth]{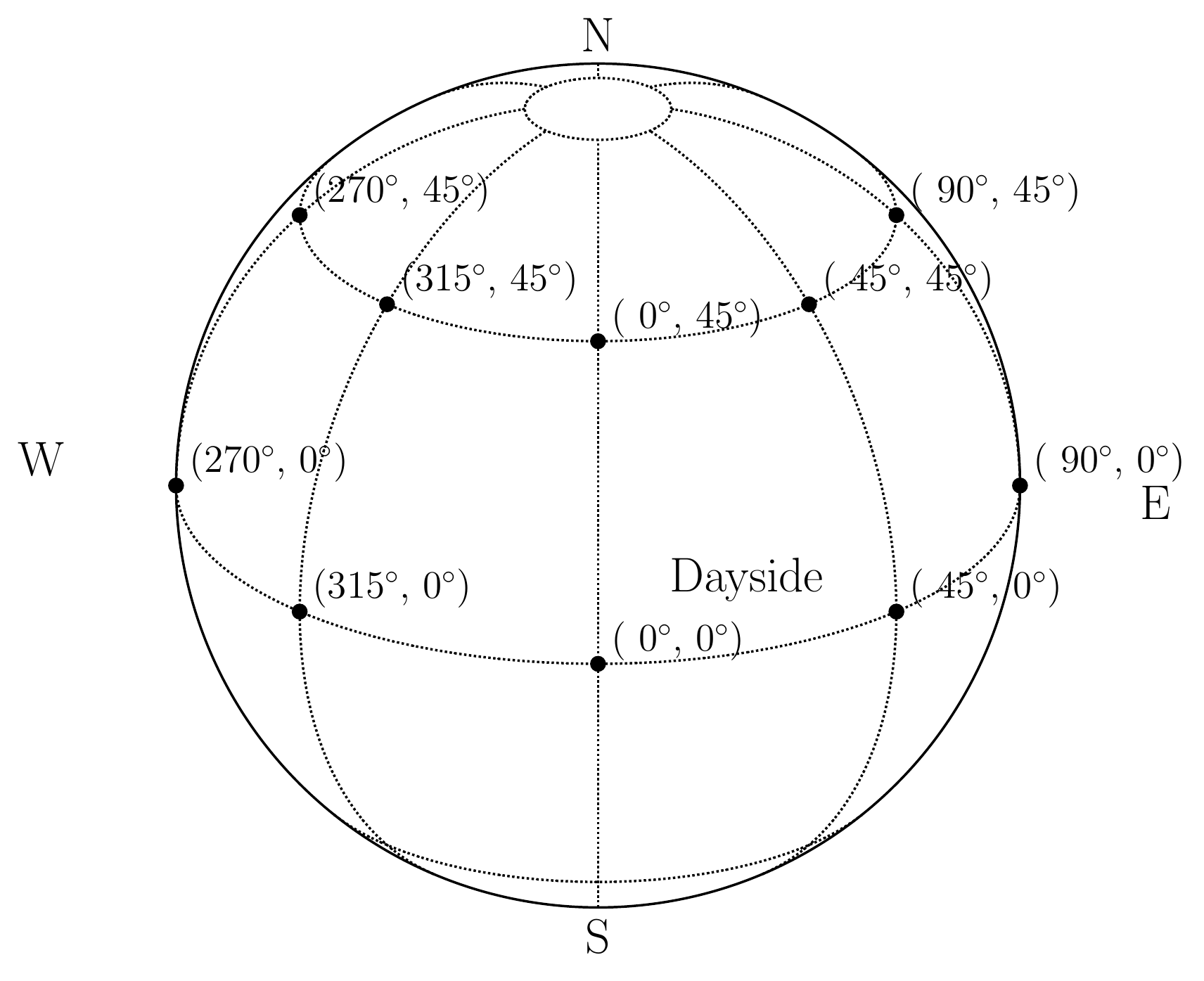}
 \includegraphics[width=0.49\textwidth]{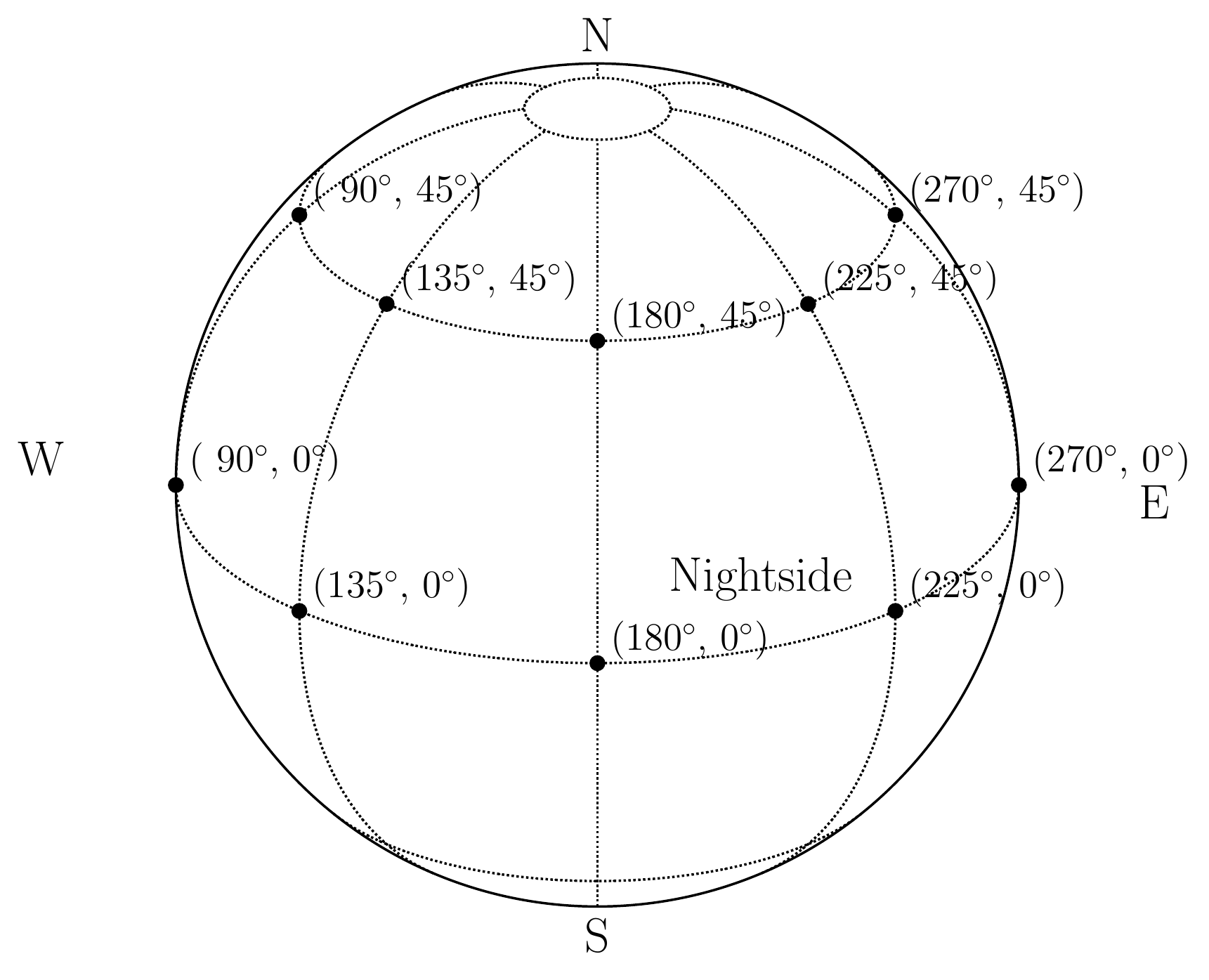}
 \caption{Illustration of the sample trajectories (black points) taken from the 3D radiative-hydrodynamic model atmosphere of HD 189733b \citep{Dobbs-Dixon2013}; longitudes $\phi$ = 0\degr$\ldots$315\degr in steps of $\Delta$$\phi$ = +45\degr, latitudes $\theta$ = 0\degr, +45\degr. 
  The sub-stellar point is at $\phi$ = 0\degr, $\theta$ = 0\degr.
 We assume that the sample trajectories are latitudinal (north-south) symmetric. 
 }
 \label{fig:globe}
\end{figure*}

We apply our kinetic dust cloud formation model to results from 3D radiation-hydrodynamic [3D RHD] simulations of HD 189733b's atmosphere \citep{Dobbs-Dixon2013} and present a first study of spatially varying cloud formation on a hot Jupiter.
We briefly summarise the main features of the kinetic cloud formation model (see: \citet{Woitke2003,Woitke2004,Helling2008,Helling2013}). 
This cloud model has been successfully combined with 1D atmosphere models (\textsc{Drift-Phoenix}; \citet{Dehn2007,Helling2008b,Witte2009,Witte2011}) to produce synthetic spectra of M Dwarfs, Brown Dwarfs and \emph{non-irradiated} hot Jupiter exoplanets.
We then summarise the multi-dimensional radiative-hydrodynamical model from \citet{Dobbs-Dixon2013} used as input for the kinetic cloud formation model.
Finally, we outline our approach in calculating the absorption and scattering properties of multi-material, multi-disperse, mixed cloud particles.

\subsection{Cloud formation modelling}
\label{sec:Cloud}

\begin{table}
\caption{Input quantities for the cloud formation model. Local T$_{\rm gas}$, p$_{\rm gas}$, $\rho_{\rm gas}$, v$_{\rm z}$ and z are taken from the 3D radiation-hydrodynamic model.
}
 \begin{tabular}{l l l} \hline
 \textbf{Input} & \textbf{Definition} & \textbf{Units} \\ \hline
 T$_{\rm gas}$ ($\vec{r}$) & local gas temperature & K \\
 p$_{\rm gas}$ ($\vec{r}$) & local gas pressure & dyn cm$^{-2}$ \\
 $\rho_{\rm gas}$ ($\vec{r}$) & local gas density & g cm$^{-3}$ \\
 v$_{\rm z}$ ($\vec{r}$) & local vertical gas velocity& cm s$^{-1}$ \\
 z & vertical atmospheric height & cm \\
  $\epsilon^{0}_{\rm x}$/$\epsilon_{\textrm{H} (\vec{r})}$  & element abundance & - \\
 g ($\vec{r}$) & surface gravity & cm s$^{-2}$ \\
 
 \hline
\end{tabular} 
\label{tab:Drift}
\end{table}

Cloud formation proceeds via a sequence of processes that are described kinetically in our cloud formation model:

\begin{enumerate}
 \item Formation of seed particles by homomolecular homogeneous nucleation \citep{Jeong2003,Lee2015}. 
 \item Growth of various solid materials by gas-grain chemical surface reactions on the existing seeds or grains \citep{Gail1986,Helling2006,Helling2008}.
 \item Evaporation of grains when the materials that they are composed of become thermally unstable \citep{Helling2006,Helling2008}.
 \item Gravitational settling allowing a continuation of grain growth through transport of grains out of under-saturated regions \citep{Woitke2003,Woitke2004}.
 \item Element depletion in regions of cloud formation which can stop new grains from forming \citep{Helling2006}.
 \item Convective/turbulent mixing from deeper to higher atmospheric regions to provide element replenishment \citep{Woitke2003,Woitke2004}.
\end{enumerate}

Without the nucleation step 1., grains do not form and a stationary atmosphere remains dust-free. 
We refer the reader to \citet{Helling2013} for a summary of the applied theoretical approach to model cloud formation in an oxygen-rich gas.  
The nucleation rate (formation of seed particles) J$_{*}$ [cm$^{-3}$ s$^{-1}$] for homomolecular homogeneous nucleation \citep{Helling2006} is 

\begin{multline}
 J_{*}(t, \vec{r}) = \frac{f(1,t)}{\tau_{\rm gr}(1,N_{*},t)}Z(N_{*}) \\ \exp\left((N_{*} - 1) \ln S(T) - \frac{\Delta G(N_{*})}{RT}\right),
\end{multline}

where f(1,t) is the number density of the seed forming gas species, $\tau_{\rm gr}$ the growth timescale of the critical cluster size $N_{*}$, $Z(N_{*})$ the Zeldovich factor, S(T) the supersaturation ratio and $\Delta G(N_{*})$ the Gibbs energy of the critical cluster size.
We consider the homogeneous nucleation of TiO$_{2}$ seed particles.
A new value for the surface tension $\sigma_{\infty}$ = 480 [erg cm$^{-2}$] is used based on updated TiO$_{2}$ cluster calculations from \citet{Lee2015}. 

The net growth/evaporation velocity $\chi^{\rm net}$ [cm s$^{-1}$] of a grain (after nucleation) due to chemical surface reactions \citep{Helling2006} is

\begin{equation}
\label{eq:chinet}
 \chi^{\rm net}(\vec{r}) = \sqrt[3]{36\pi}\sum_{s}\sum_{r = 1}^{R} \frac{\Delta V_{r}n_{r}v_{r}^{\rm rel}\alpha_{r}}{\nu_{r}^{\rm key}}\left(1 - \frac{1}{S_{r}}\frac{1}{b^{s}_{\rm surf}}\right),
\end{equation}

where `r' is the index for the chemical surface reaction, $\Delta$V$_{r}$ the volume increment of the solid `s' by reaction r, n$_{r}$ the particle density of the reactant in the gas phase, v$_{r}^{\rm rel}$ the relative thermal velocity of the gas species taking part in reaction r, $\alpha_{r}$ the sticking coefficient of reaction r and $\nu_{r}^{\rm key}$ the stoichiometric factor of the key reactant in reaction r \citep{Helling2006}. 
S$_{r}$ is the reaction supersaturation ratio and 1/b$^{s}_{\rm surf}$ = V$_{\rm s}$/V$_{\rm tot}$ the volume ratio of solid s. 
Our cloud formation method allows the formation of mixed heterogeneous grain mantles.
We simultaneously consider 12 solid growth species (TiO$_{2}$[s], Al$_{2}$O$_{3}$[s], CaTiO$_{3}$[s], Fe$_{2}$O$_{3}$[s], FeS[s], FeO[s], Fe[s], SiO[s], SiO$_{2}$[s], MgO[s], MgSiO$_{3}$[s], Mg$_{2}$SiO$_{4}$[s]) with 60 surface chemical reactions \citep{Helling2008}. 
The local dust number density n$_{\rm d}$ [cm$^{-3}$] and mean grain radius $\langle$a$\rangle$ [cm] are given by

\begin{equation}
\label{eq:nd}
 n_{\rm d}(\vec{r}) = \rho_{\rm gas}(\vec{r}) L_{0}(\vec{r}),
\end{equation}

\begin{equation}
\label{eq:meana}
 \langle{a}\rangle(\vec{r}) = \sqrt[3]{\frac{3}{4\pi}}\frac{L_{1}(\vec{r})}{L_{0}(\vec{r})},
\end{equation}

respectively; where $\rho_{\rm gas}(\vec{r})$ [g cm$^{-3}$] is the local gas volume density and L$_{0}(\vec{r})$, L$_{1}(\vec{r})$ are derived by solving for L$_{j}(\vec{r})$ [cm$^{j}$g$^{-1}$] in the dust moment equations \citep{Woitke2003}. 

\subsection{3D radiative-hydrodynamical model}
\label{sec:3Dmodel}

The 3D radiative-hydrodynamical [3D RHD] model solves the fully compressible Navier-Stokes equations coupled to a wavelength dependent two-stream radiative transfer scheme.
The model assumes a tidally-locked planet with $\phi$ = 0\degr, $\theta$ = 0\degr denoting the sub-stellar point (the closest point to the host star).
There are three components to the wavelength dependent opacities: molecular opacities consistent with solar composition gas, a gray component representing a strongly absorbing cloud deck, and a strong Rayleigh scattering component.
Equations are solved on a spherical grid with pressures ranging from $\sim$10$^{-4.5}$ to $\sim$10$^{3}$ bar.
Input parameters were chosen to represent the bulk observed properties of HD 189733b, but (with the exception of the opacity) were not tuned to match spectroscopic observations.
The dominate dynamical feature is the formation of a super-rotating, circumplanetary jet \citep{Tsai2014} that efficiently advects energy from day to night near the equatorial regions.
This jet forms from the planetary rotation (Rossby waves) coupled with eddies which pump positive angular momentum toward the equator \citep{Showman2011a}. 
A counter-rotating jet is present at higher latitudes. 
Significant vertical mixing, discussed later, is seen throughout the atmosphere. 
Calculated transit spectra, emission spectra, and light curves agree quite well with current observations from $0.3 \mu\mathrm{m}$ to $8.0 \mu\mathrm{m}$. 
Further details can be found in \citet{Dobbs-Dixon2013}.

\subsection{Model set-up and input quantities}
\label{sec:setup}

\begin{figure*}
\centering
\includegraphics[width=0.49\textwidth]{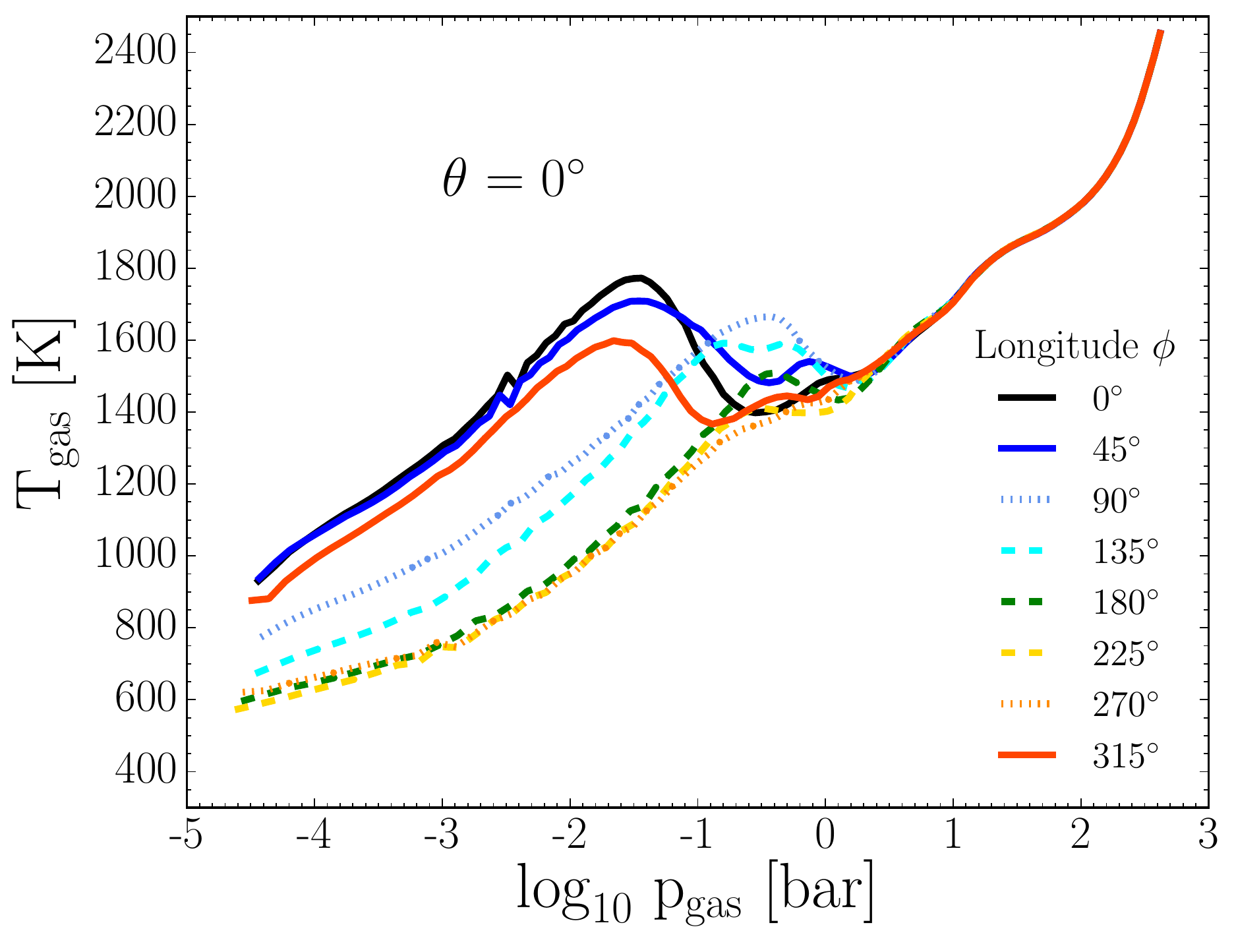}
\includegraphics[width=0.49\textwidth]{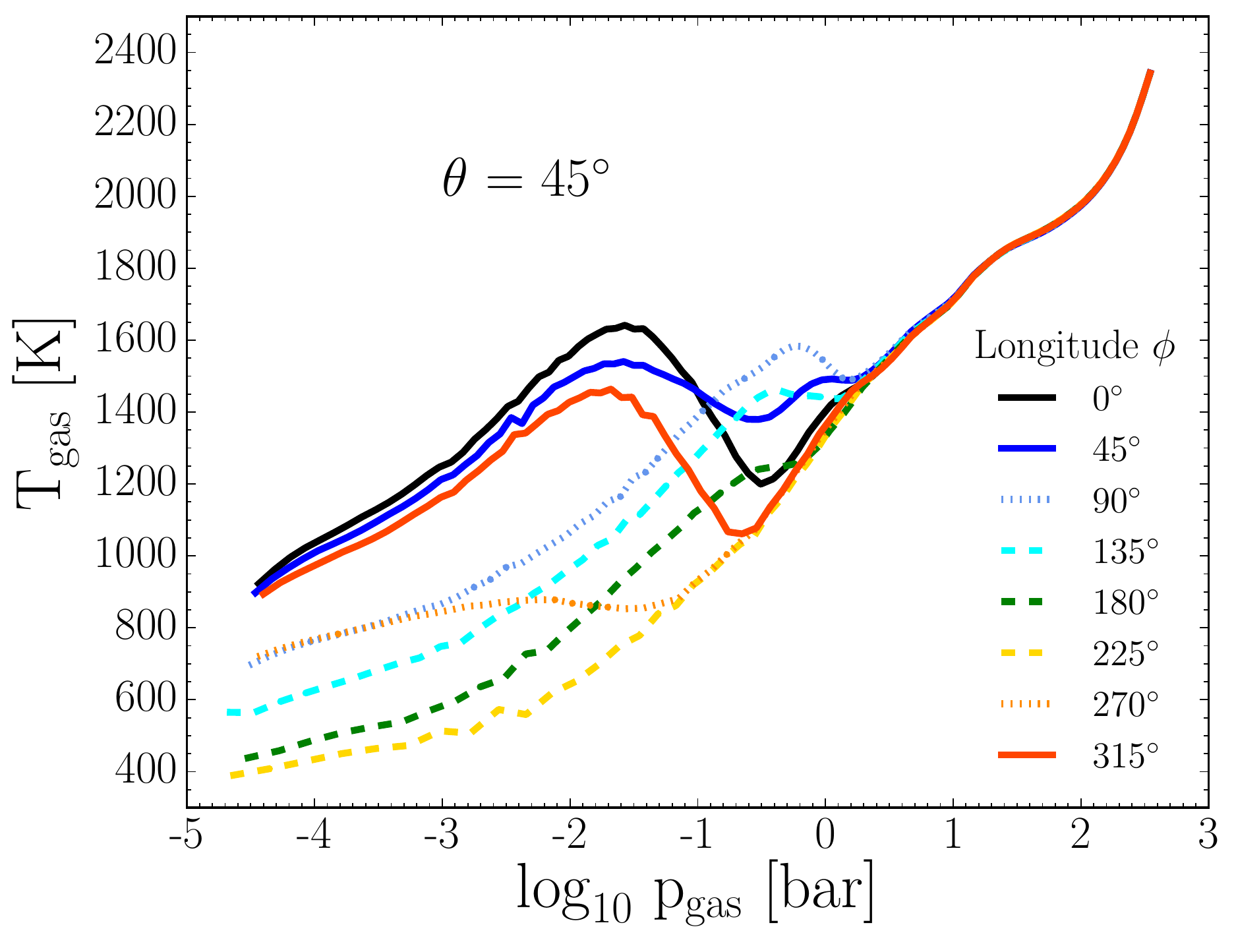}
\includegraphics[width=0.49\textwidth]{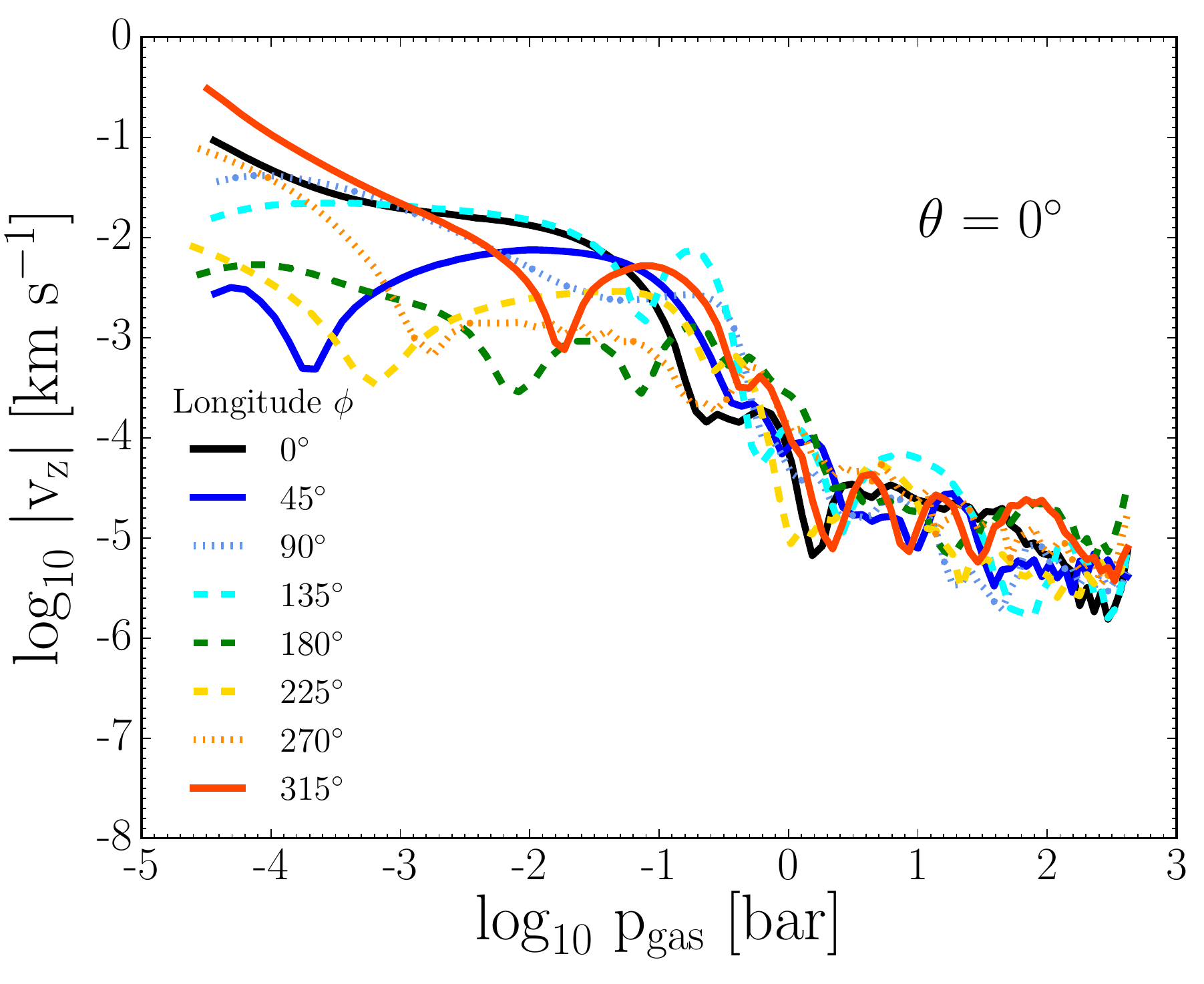}
\includegraphics[width=0.49\textwidth]{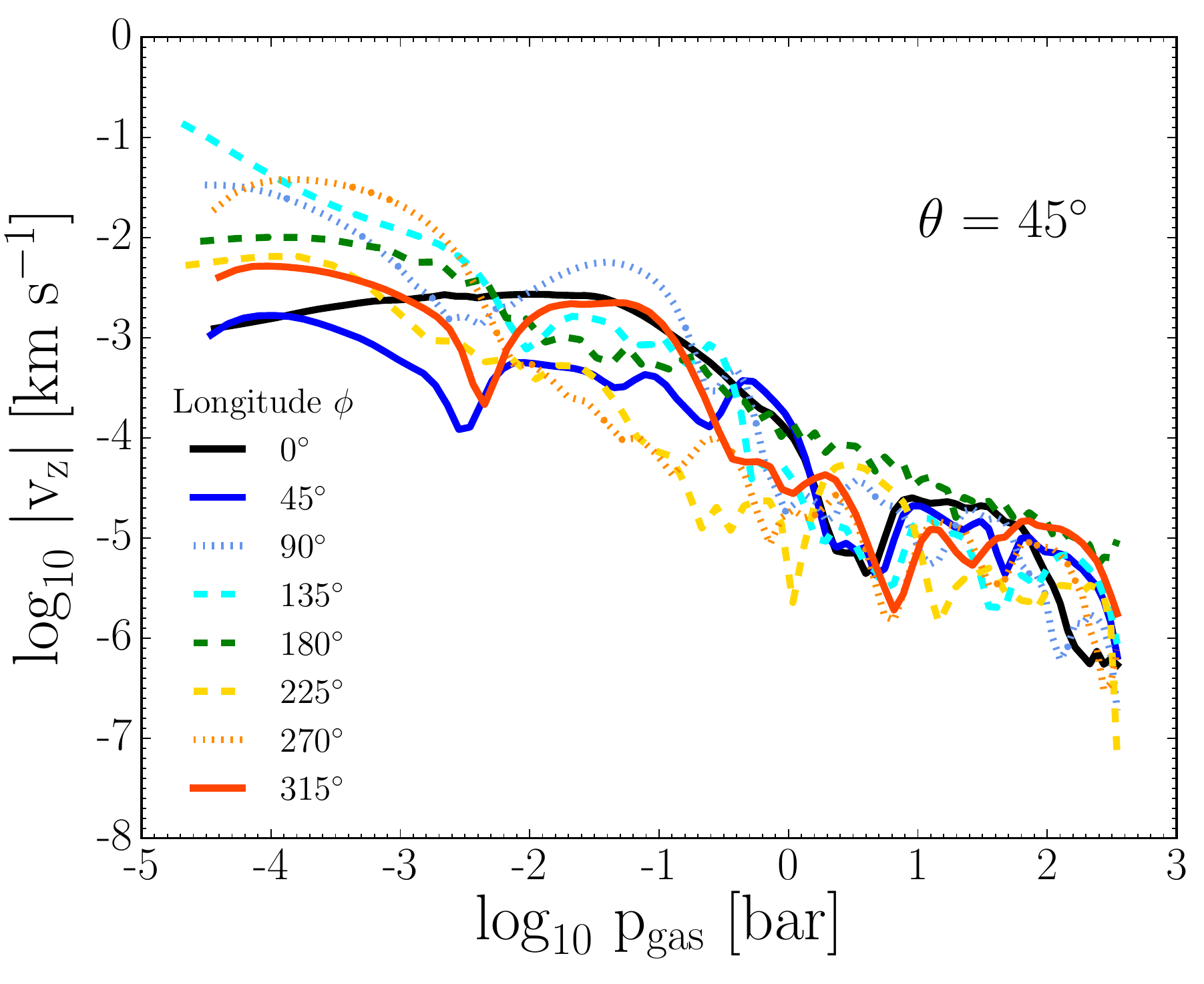}
\caption{1D trajectory results from the 3D RHD HD 189733b atmosphere simulation used as input for the cloud formation model.
\textbf{Top Row:} (T$_{\rm gas}$, p$_{\rm gas}$) profiles in steps of longitude $\Delta\phi$ = +45\degr at equatorial, $\theta$ = 0\degr and latitude, $\theta$ = 45\degr profiles.
\textbf{Bottom Row:} Smoothed vertical gas velocities $\vec{v_{\rm z}}$ [km s$^{-1}$] in steps of longitude $\Delta\phi$ = +45\degr at equatorial, $\theta$ = 0\degr and latitude, $\theta$ = 45\degr profiles.
Both latitudes show a temperature inversion on the dayside.
The sub-stellar point is at $\phi$ = 0\degr, $\theta$ = 0\degr.
Solid, dotted and dashed lines indicate dayside, day-night terminator and nightside profiles respectively.}
\label{fig:inputq}
\end{figure*}

We sampled vertical trajectories of the 3D RHD model at longitudes of $\phi$ = 0\degr$\ldots$315\degr in steps of $\Delta\phi$ = +45\degr and latitudes of $\theta$ = 0\degr, +45\degr (Fig. \ref{fig:globe}).   
The horizontal wind velocity moves in the positive $\phi$ direction. 
Figure \ref{fig:inputq} shows the (T$_{\rm gas}$, p$_{\rm gas}$) and vertical velocity profiles used to derive the cloud structure at the equator $\theta$ = 0\degr and latitude $\theta$ = 45\degr. 
Local temperature inversions are present in the dayside (T$_{\rm gas}$, p$_{\rm gas}$) profiles which are exposed to the irradiation of the host star without the need for an additional opacity source. 
The local temperature maximum migrates eastward with increasing atmospheric depth. 
This is due to hydrodynamical flows funnelling gas towards the equator causing viscous, compressive and shock heating.
The (T$_{\rm gas}$, p$_{\rm gas}$) profile at latitude $\theta$ = 45\degr has steeper temperature inversions on the dayside.
Sample trajectories converge to equal temperatures at p$_{\rm gas}$ $>$ 1 bar for all longitudes and latitudes.
We apply a 3-point boxcar smoothing to the vertical velocity profiles in order to reduce the effects of unresolved turbulence.
Further required input quantities are a constant surface gravity of log g = 3.32 and initial solar element abundances $\epsilon^{0}_{\rm x}$/$\epsilon_{\rm H}$ (element `x' to Hydrogen ratio) with C/O = 0.427 \citep{Anders1989} for all atmospheric layers. 
However, we note the element abundance of the gas phase will increase or decrease due to cloud formation or evaporation (Fig. \ref{fig:thetaelement}).
We assume local thermal equilibrium (LTE) for all gas-gas and dust-gas chemical reactions.
The required input quantities for the kinetic cloud formation model are summarised in Table \ref{tab:Drift}.

\subsection{Atmospheric vertical mixing}
\label{sec:Mixing} 
Vertical mixing is important to resupply the upper atmosphere with elements which have been depleted by cloud particle formation and their subsequent gravitational settling into deeper atmospheric layers \citep{Woitke2003}. 
Without mixing, the cloud particles in the atmosphere would rain out and remove heavy elements from the upper atmosphere. 
This leaves a metal poor gas phase where no future cloud particles could form (\citealt{Woitke2004}; Appendix A). 

\subsubsection{Previous Brown Dwarf approach}
\label{sec:previousmixing}

The main energy transport in the core of a brown dwarf is convection. The atmosphere is convectively unstable in the inner, hotter parts. This atmospheric convection causes substantial overshooting into even higher, and radiation dominated parts (e.g. \citet{Ludwig2002}).
\citet{Woitke2003,Woitke2004} define the mixing timescale in low mass stellar atmospheres as the time for convective motions $v_{\rm conv}$ to travel a fraction of the pressure scale height, H$_{\rm p}(\vec{r})$

\begin{equation}
\label{eq:taumix}
 \tau_{\rm mix, conv}(\vec{r}) = \textrm{const} \cdot \frac{H_{\rm p}(\vec{r})}{{\rm v}_{\rm conv}(\vec{r})}.
\end{equation}

\citet{Helling2008,Witte2009,Witte2011} represent values for v$_{\rm conv}$($\vec{r}$), in their 1D Drift-Phoenix model atmospheres, above the convective zone (defined by the Schwarzschild criterion) from inertially driven \textit{overshooting} of ascending gas bubbles. 

\subsubsection{Previous Hot Jupiter approach}

On hot Jupiters, in contrast to Earth, Jupiter and Brown Dwarf atmospheres, the intensity of stellar irradiation can suppress convective motions down to very large pressures \citep{Barman2001}. 
Therefore turbulent diffusion is likely the dominant vertical transport mechanism.
The use of the 3D RHD model results allows us to apply the local vertical velocity component, v$_{\rm z}$($\vec{r}$), in our cloud formation model. 
No additional assumptions are required.
Hence, v$_{\rm z}$($\vec{r}$) is known for each atmospheric trajectory chosen, as visualised in Fig. \ref{fig:inputq}.

Some chemical models of hot Jupiters (e.g. \citet{Moses2011}) use an approximation of the vertical diffusion coefficient K$_{\rm zz}$($\vec{r}$) [cm$^{2}$ s$^{-1}$] of the gaseous state.

\begin{equation}
\label{eq:Kzz1}
 K_{\rm zz}(\vec{r}) = H_{\rm p}(\vec{r}) \cdot {\rm v_{\rm z}}(\vec{r}).
\end{equation}

The diffusion timescale is then
\begin{equation}
\label{eq:taudiff}
 \tau_{\rm mix, diff}(\vec{r}) = {\rm const} \cdot \frac{H_{\rm p}^{2}(\vec{r})}{K_{\rm zz}(\vec{r})}.
\end{equation}

\citet{Moses2011} apply the root-mean-square (rms) vertical velocities derived from global horizontal averages at each atmospheric layer to their chemical models. This results in a horizontally homogenous  K$_{\rm zz}(\vec{r})$ mixing parameter.
\citet{Parmentier2013} derive a global mean value for K$_{\rm zz}$($\vec{r}$) as a function of local pressure using a passive tracer in their global circulation model [GCM] of HD 189733b resulting in K$_{\rm zz}$($\vec{r}$) = 10$^{7}$ p$^{-0.65}$ (p [bar]). Using a global mean smooths away all horizontally dependent vertical velocity variations.
Any parameterisation of the vertical mixing will depend on the details of the underlying hydrodynamical structure. 
\citet{Parmentier2013} use a 3D primitive equation model where vertical hydrostatic equilibrium is assumed.
The radiation hydrodynamic simulations performed by \citet{Dobbs-Dixon2013}, applied in this paper, solve the full Navier-Stokes equations and will produce larger vertical velocities compared to the primitive equations.
In summary, vertical velocity may be substantially damped in models using the primitive equations.
Both Eq. \eqref{eq:Kzz1} and \citet{Parmentier2013} definitions for K$_{\rm zz}$($\vec{r}$) assume that the dominant mixing occurs in the vertical direction, possible mixing from horizontal flows are neglected.
The time-scale comparison in Sect. \ref{sec:cloudtimescale} supports this assumption for cloud formation processes.
It is worthwhile noting that the idea of diffusive mixing originates from shallow water approximations as applied in solar system research where a 2D velocity field produces shear which creates a turbulent velocity component. \citet{Hartogh2005} outline how local wind shear and the hydrostatic gas pressure are used to represent a vertical mixing.

Some arguments reinforce why vertical mixing is important:
\begin{itemize}
 \item Regions with low vertical velocity can be replenished of elements by the horizontal winds from high vertical velocity regions in a 3D situation. 
 \item Large Hadley cell circulation is present and here vertical velocities can be significant and element replenishment to the upper atmosphere may be more efficient. 
 \item Vertical transport is a key mixing process on Earth which has been successfully applied to hot Jupiter chemical models \citep{Moses2011, Venot2012, Agundez2014}.
\end{itemize}

\begin{figure*}
\centering
\includegraphics[width=0.48\textwidth]{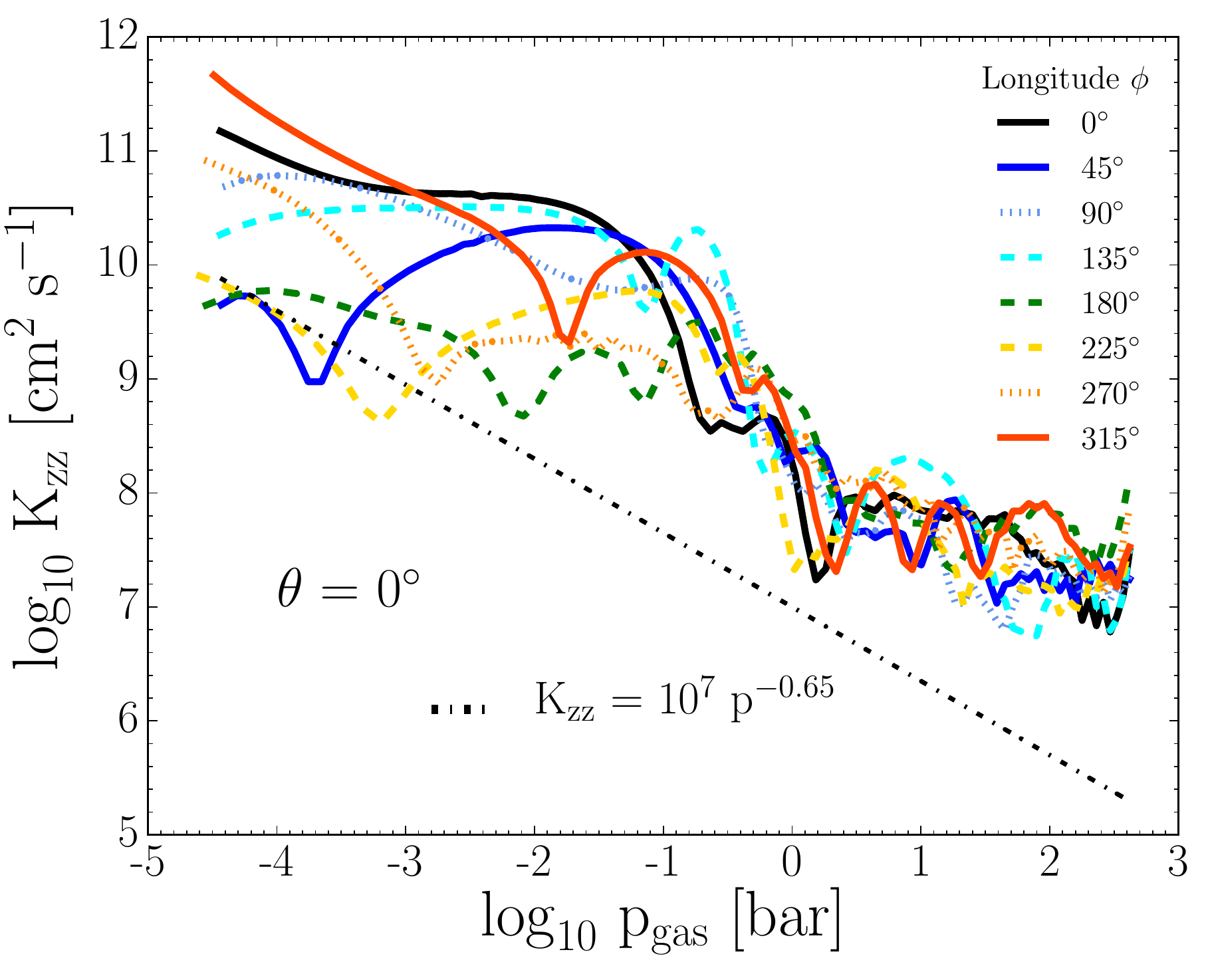} 
\includegraphics[width=0.48\textwidth]{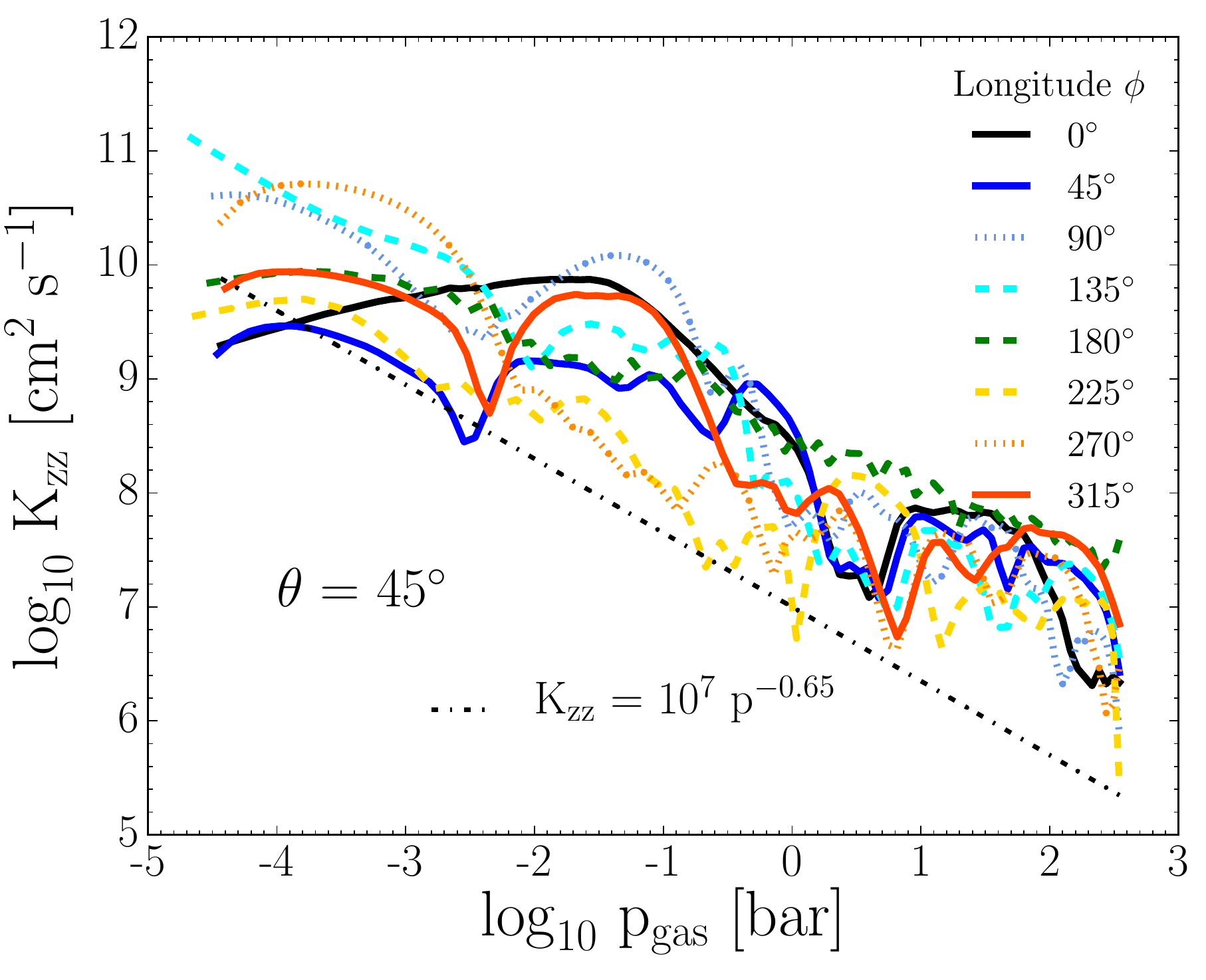}
\includegraphics[width=0.48\textwidth]{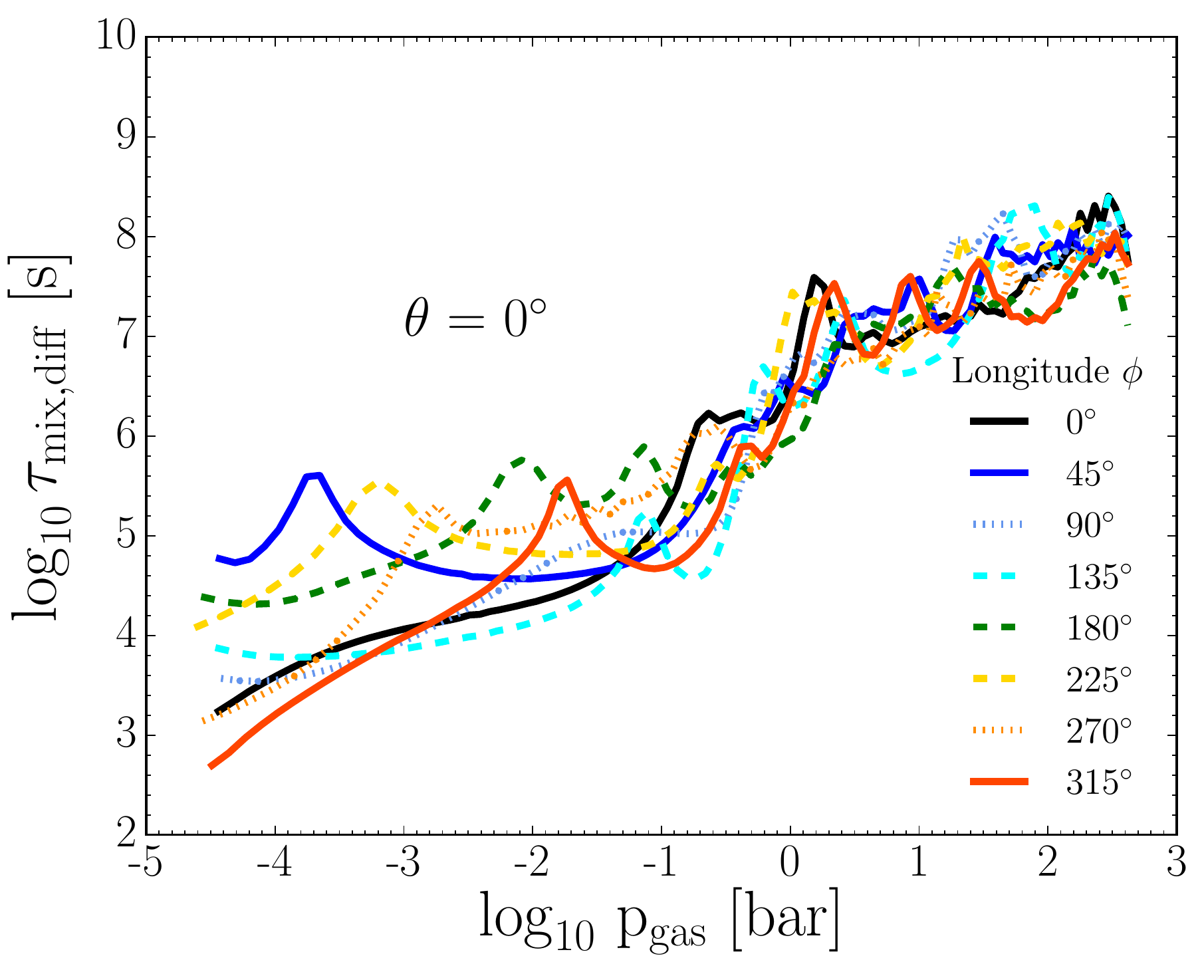}  
\includegraphics[width=0.48\textwidth]{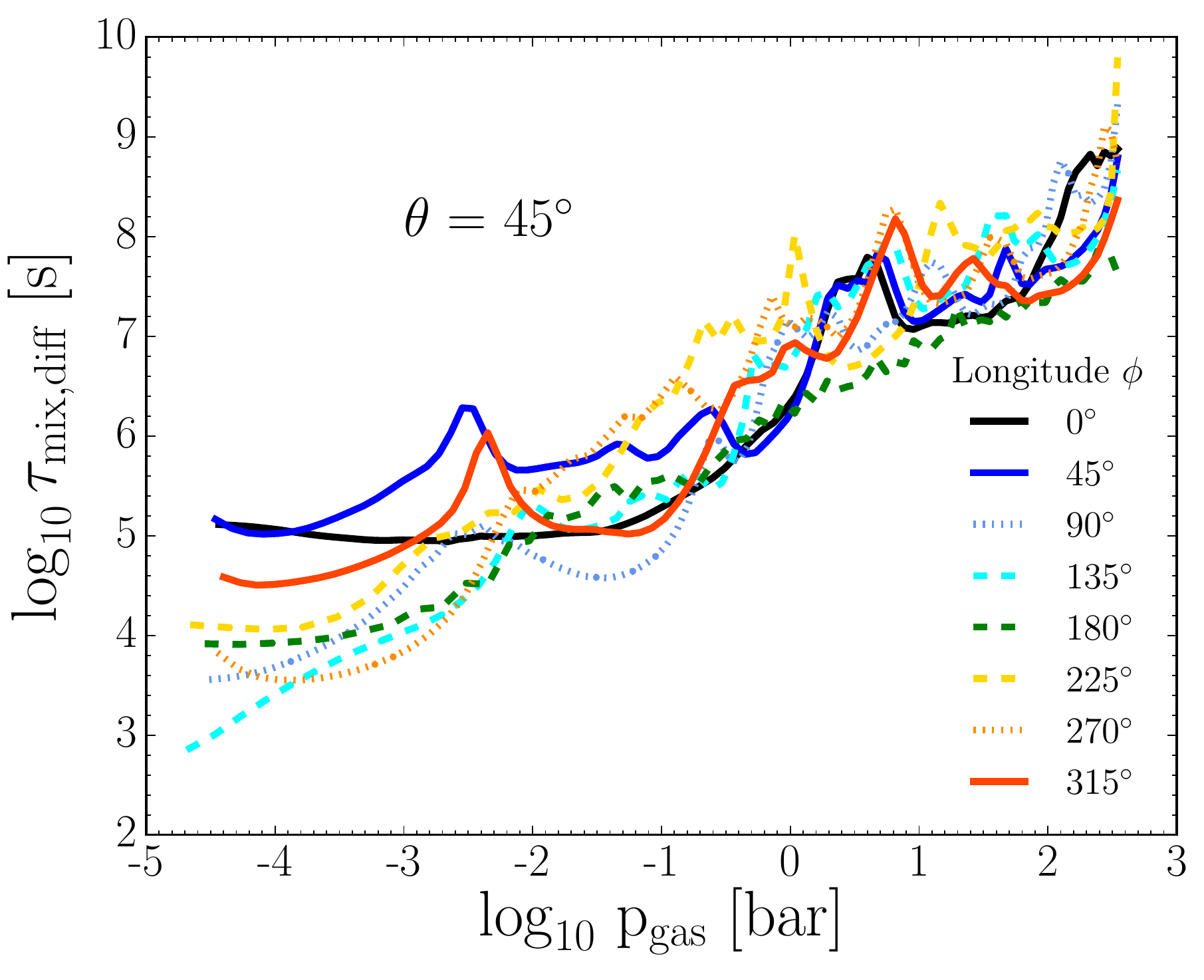} 
\includegraphics[width=0.48\textwidth]{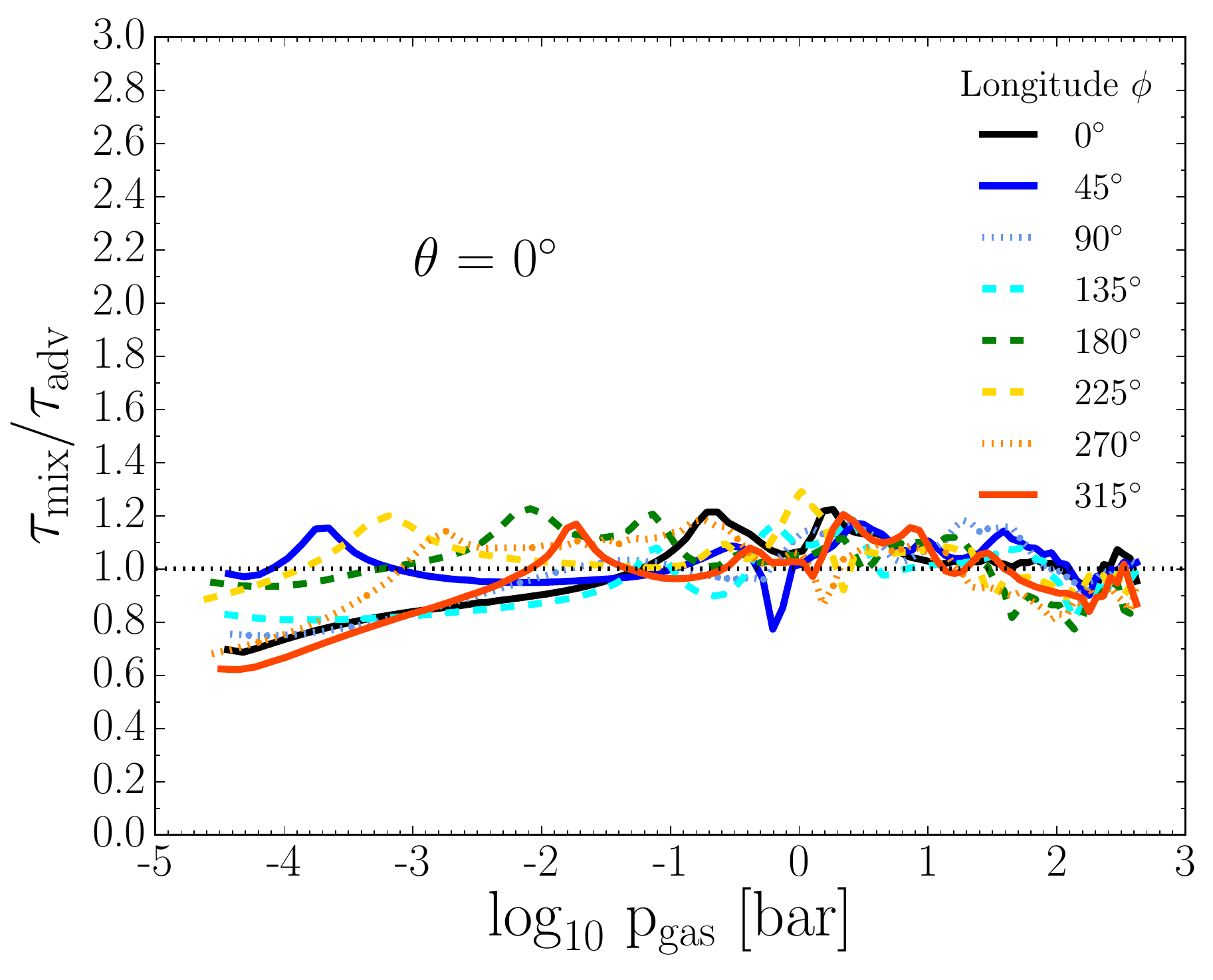} 
\includegraphics[width=0.48\textwidth]{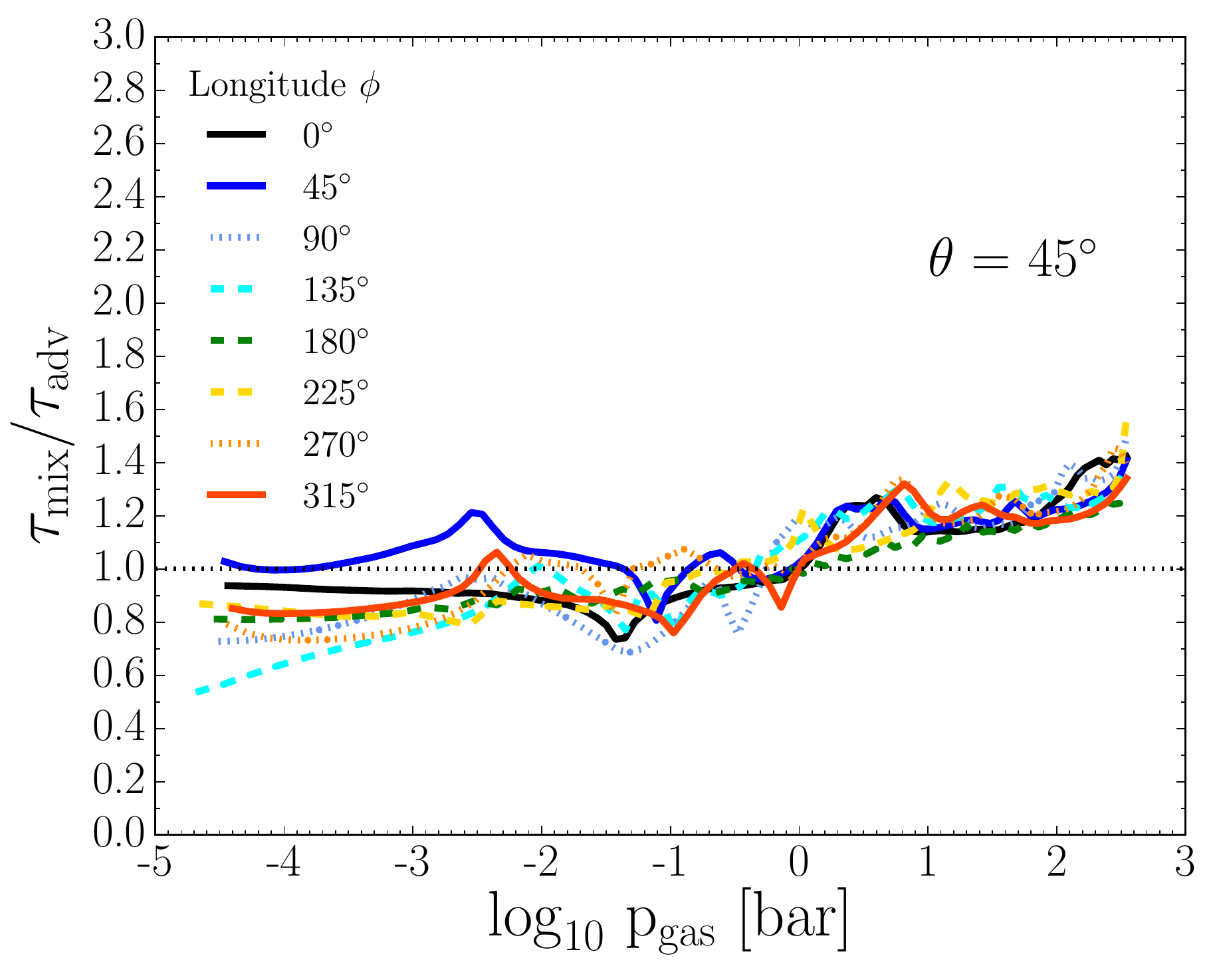} 
\caption{
\textbf{Top Row:}
Vertical diffusion coefficient K$_{\rm zz}$($\vec{r}$) = H$_{\rm p}$ $\cdot$ v$_{\rm z}$($\vec{r}$) [cm$^{2}$ s$^{-1}$] applying the smoothed vertical velocities of the radiative-hydrodynamic HD 189733b model at latitudes $\theta$ = 0\degr and $\theta$ = 45\degr, as a function of pressure at $\Delta\phi$ = +45\degr longitude intervals. 
The K$_{\rm zz}$($\vec{r}$) = 10$^{7}$p$^{-0.65}$  (p [bar]) expression from \citet{Parmentier2013} is shown as dash-dotted lines.
\textbf{Middle Row:}
The vertical mixing timescales $\tau_{\rm {mix, diff}}$ [s] derived from the HD 189733b radiative-hydrodynamic model results at latitudes $\theta$ = 0\degr, $\theta$ = 45\degr, as a function of pressure at $\Delta\phi$ = +45\degr longitude intervals. 
\textbf{Bottom Row:}
Ratio of the mixing and advective timescales at $\theta$ = 0\degr, 45\degr respectively. 
The ratio of the timescales stays approximately equal at all pressures.
Solid, dotted and dashed lines indicate dayside, day-night terminator and nightside profiles respectively.}
\label{fig:timescales}
\label{fig:Kzzfig}
\end{figure*}

We use Eq. \eqref{eq:taudiff} (const = 1) as the 1D mixing timescale input for our kinetic dust formation model, with Eq. \eqref{eq:Kzz1} as the definition of the diffusion coefficient. 
We adopt the local vertical velocities that result from the \citet{Dobbs-Dixon2013} 3D RHD atmosphere simulations for HD 189733b as the values for v$_{\rm z}$($\vec{r}$) (Fig. \ref{fig:inputq}). 
A 3-point boxcar smoothing was applied to these velocities to reduce the effects of unresolved turbulence.
The longitude dependent K$_{\rm zz}$ ($\phi$, r) for latitudes $\theta$ = 0\degr, 45\degr is shown in Fig. \ref{fig:Kzzfig} (first row) and the resulting vertical mixing timescales (second row).
We include the K$_{\rm zz}$ relation from \citet{Parmentier2013} in Fig. \ref{fig:Kzzfig} (dash-dotted line) for comparison. 
Their linear fit is approximately one order of magnitude lower which is similar to the difference found in HD 189733b and HD 209458b chemical models \citep{Agundez2014} who compared the two K$_{\rm zz}$($\vec{r}$) expressions for \citet{Showman2009} GCM simulations. 
The current approach aims to capture the unique vertical mixing and thermodynamic conditions at each trajectory, while also accounting for practical modelling of atmospheric mixing.

\subsubsection{Advective timescales}

An important timescale to consider is the charcteristic advective timescale which is a representative timescale for heat to redistribute over the circumference of the globe.
The advective timescale is given by

\begin{equation}
 \tau_{\rm adv}(\vec{r}) = \frac{\pi r}{{\rm v}_{\rm horiz}(\vec{r})},
\end{equation}

Where $r(z)$ is the radial height of the planet and v$_{\rm horiz}$($\vec{r}$) the local gas velocity in the horizontal direction ($\phi$).
This timescale gives an idea of how fast thermodynamic conditions can change in the longitudinal direction at a particular height z in the atmosphere.
Figure. \ref{fig:timescales} compares the advective timescale to the vertical diffusion timescale (Eq. \eqref{eq:taudiff}) and find that both time scales are of the same order of magnitude. 
This suggests that the convective/turbulent diffusion of upward gaseous material as described in Sect. \ref{sec:previousmixing} occurs approximately at the same timescales as gas advected across the globe.
This suggests that, to a first approximation, the 1D (T$_{\rm gas}$, p$_{\rm gas}$) trajectories used as input for the cloud formation model do not fluctuate rapidly enough to substantially influence the cloud properties.

\subsubsection{Cloud formation timescales}
\label{sec:cloudtimescale}

\begin{figure*}
\centering
 \includegraphics[width=0.49\textwidth]{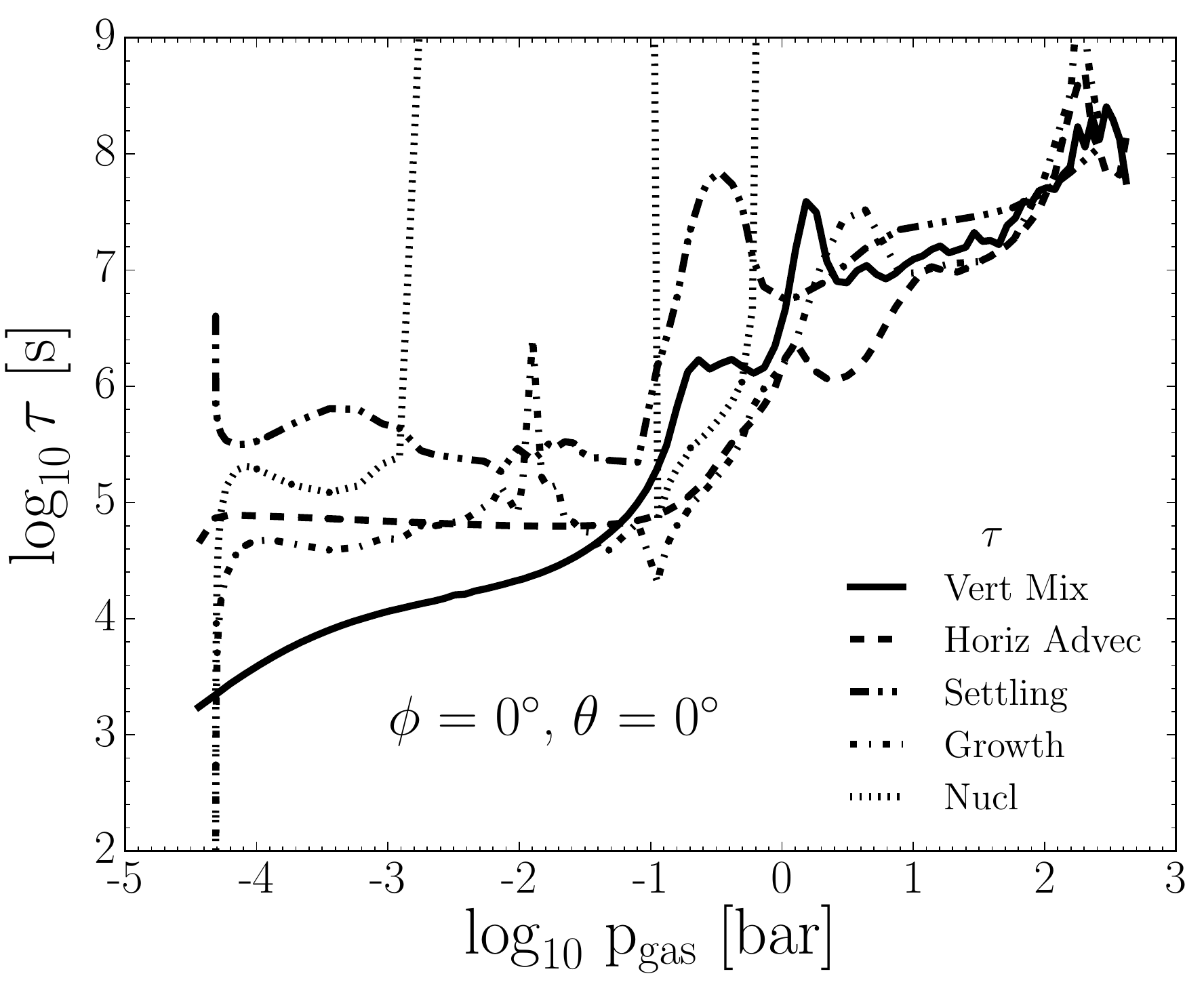} 
\includegraphics[width=0.49\textwidth]{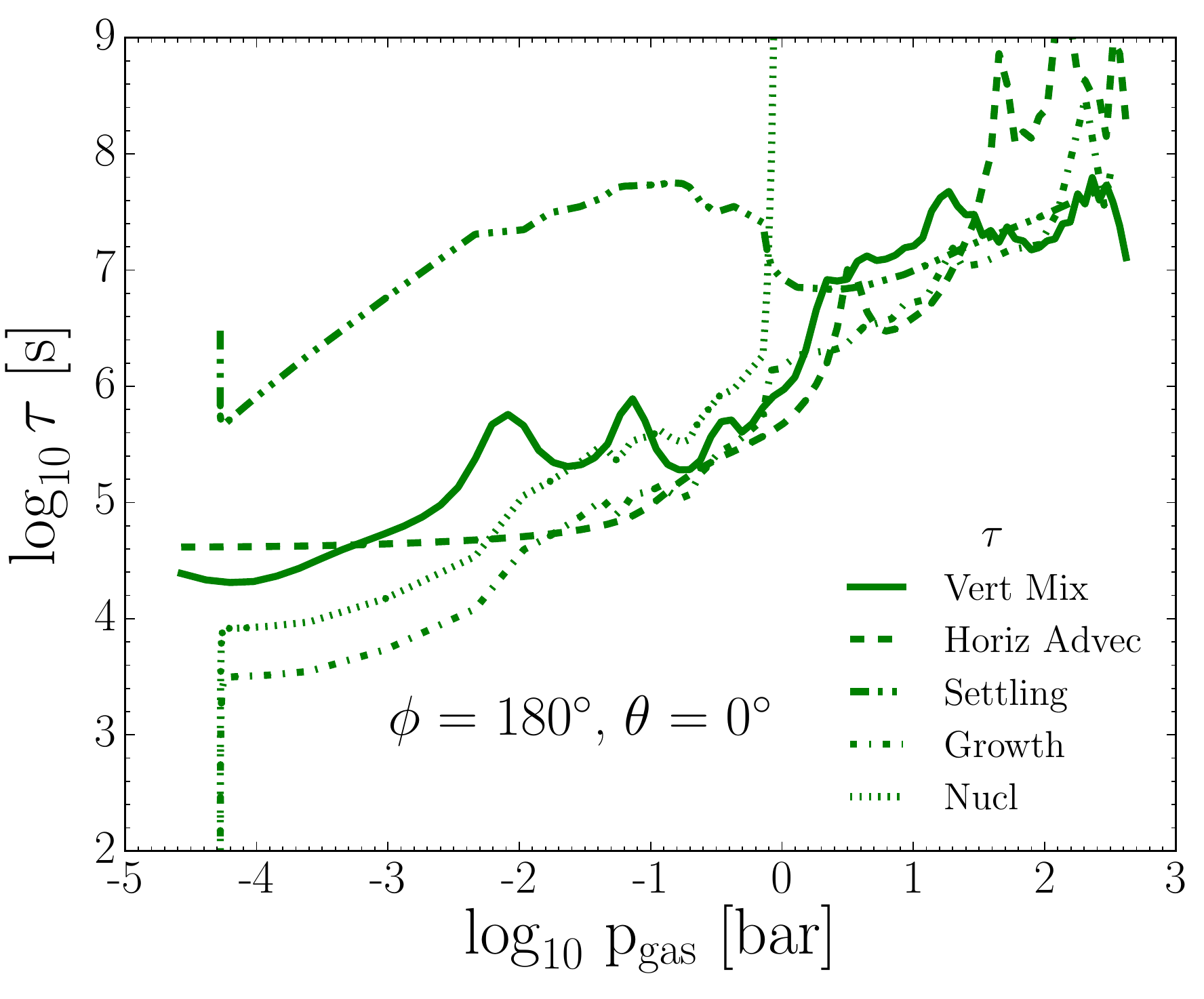}
\caption{The vertical mixing, horizontal advection, cloud settling, cloud growth and cloud nucleation timescales at the sub-stellar point (Left) and $\phi$ = 180\degr, $\theta$ = 0\degr (Right).}
\label{fig:timescalecomp}
\end{figure*}

We compare the cloud particle settling, growth and nucleation timescales that result from our cloud model (Sect. \ref{sec:Application}) to the mixing and advection timescales that are derived from the hydrodynamic fluid field.
The nucleation timescale $\tau_{\rm nuc}$ is defined as
\begin{equation}
\tau_{\rm nuc} = \frac{n_{\rm d}}{J_{\star}} ,
\end{equation}
the growth timescale $\tau_{\rm gr}$ by
\begin{equation}
\tau_{\rm gr} = \frac{\sqrt[3]{36\pi}\langle a\rangle}{3 |\chi^{\rm net}|} ,
\end{equation}
and the cloud particle settling timescale $\tau_{\rm setl}$ by 
\begin{equation}
\tau_{\rm setl}  = \frac{H_{\rm p}}{|\langle \textrm{v}_{\rm dr}\rangle|} ,
\end{equation}
where $\langle$v$_{\rm dr}$$\rangle$ is the large Knudsen number frictional regime (Kn $>>$ 1) mean drift velocity \citep[Eq. 63]{Woitke2003}.
Figure \ref{fig:timescalecomp} shows the timescales at the sub-stellar and anti-stellar points.
Our results agree with earlier timescale analysis in \citet{Woitke2003} who point out an hierarchical dominance of the cloud formation processes through the atmosphere.
In the upper atmosphere nucleation is the most efficient process:

\begin{equation*}
 \tau_{\rm nuc} \lesssim \tau_{\rm gr} << \tau_{\rm mix} \sim \tau_{\rm adv}  << \tau_{\rm setl} .
\end{equation*}

Deeper in the atmosphere nucleation eventually becomes inefficient and all timescales become comparable

\begin{equation*}
  \tau_{\rm gr} \lesssim \tau_{\rm mix} \sim \tau_{\rm adv} \lesssim \tau_{\rm setl} .
\end{equation*}

The chemical processes that determine the cloud particle formation occur on shorter timescales than the large scale hydrodynamical timescales.
This emphasises that the cloud particle formation is a local process.

\subsection{Dust opacity}

\begin{savenotes}
\begin{table}
\caption{References for the (n, k) optical constants of the 12 different solids.}
\label{tab:opticalconst}
\begin{tabular}{c c } \hline
\textbf{Solid species} & \textbf{Source} \\ \hline
TiO$_{2}$[s] & Ribarsky in \citet{Palik1985} \\
Al$_{2}$O$_{3}$[s] & \citet{Zeidler2013} \\
CaTiO$_{3}$[s] & \citet{Posch2003} \\
Fe$_{2}$O$_{3}$[s] & Unpublished\footnote{\url{http://www.astro.uni-jena.de/Laboratory/OCDB/oxsul.html}} \\
FeS[s] & \citet{Henning1995} \\
FeO[s] & \citet{Henning1995} \\
Fe[s] & \citet{Posch2003} \\
SiO[s] & Philipp in \citet{Palik1985} \\
SiO$_{2}$[s] & \citet{Posch2003} \\
MgO[s] & \citet{Palik1985} \\
MgSiO$_{3}$[s] & \citet{Dorschner1995} \\
Mg$_{2}$SiO$_{4}$[s] & \citet{Jager2003} \\
\hline 
\end{tabular}

\end{table}
\end{savenotes}

Based on the results of the spatially varying cloud properties we calculate the opacity of the cloud particles.
We determine the cloud particle absorption and scattering coefficients in each atmospheric layer for wavelengths 0.3$\mu$m, 0.6$\mu$m, 1.1$\mu$m, 1.6$\mu$m, 3.6$\mu$m, 4.5$\mu$m, 5.8$\mu$m, 8.0$\mu$m, 24.0$\mu$m, corresponding to the \textit{Hubble} and \textit{Spitzer} average band passes. 
Since the cloud particles are made of mixed solids, the effective (n, k) optical constants is calculated for each particle using effective medium theory. 
We follow the approach of \citet{Bruggeman1935} where

\begin{equation}
 \sum_{s}\left(\frac{V_{\rm s}}{V_{\rm tot}}\right)\frac{\epsilon_{s} - \epsilon_{\rm av}}{\epsilon_{s} + 2\epsilon_{\rm av}} = 0,
 \label{eq:bruggeman}
\end{equation}

where V$_{\rm s}$/V$_{\rm tot}$ is the volume fraction of solid species s, $\epsilon_{s}$ the dielectric function of solid species s and $\epsilon_{\rm av}$ the average dielectric function over the total cloud particle volume.
The average dielectric function is then found by Newton-Raphson minimisation of Eq. \eqref{eq:bruggeman}.
The scattering and extinction cross sections are then calculated using Mie theory for spherical particles \citep{Mie1908}.
We follow the approach of \citet{Bohren1983} where the scattering and extinction cross sections are defined as

\begin{equation}
 C_{\rm sca}(\lambda, a) = \frac{2 \pi a^{2}}{x^{2}}\sum_{n = 1}^{\infty} (2n + 1)(|a_{\rm n}|^{2} + |b_{\rm n}|^{2}),
\end{equation}

\begin{equation}
 C_{\rm ext}(\lambda, a) = \frac{2 \pi a^{2}}{x^{2}}\sum_{n = 1}^{\infty} (2n + 1)Re(a_{\rm n} + b_{\rm n}),
\end{equation}

respectively; where x = $2\pi a/\lambda$ is the wavelength dependent size parameter. 
The scattering coefficients a$_{\rm n}$ and b$_{\rm n}$ are calculated from the material optical k constant \citep{Bohren1983}.
The wavelength-dependent absorption and scattering efficiency of a cloud particle is then

\begin{equation}
 Q_{\rm sca}(\lambda, a) =  \frac{C_{\rm sca} (\lambda, a)}{\pi a^{2}},
\end{equation}
\begin{equation}
 Q_{\rm abs}(\lambda, a) =  \frac{C_{\rm ext} (\lambda, a)}{\pi a^{2}} - Q_{\rm sca},
\end{equation}

respectively. 
The total absorption and scattering efficiency $\kappa$ [cm$^{2}$ g$^{-1}$] can then be derived by multiplying the corresponding efficiencies with the area and occurrence rate of each cloud particle.

\begin{equation}
 \kappa_{\rm sca}(\lambda, a) = Q_{\rm sca}(\lambda, a) \pi a^{2} n_{\rm d} /\rho_{\rm gas},
\end{equation}
\begin{equation}
 \kappa_{\rm abs}(\lambda, a) = Q_{\rm abs}(\lambda, a) \pi a^{2} n_{\rm d} /\rho_{\rm gas}.
\end{equation}

We include all 12 solid dust species in the opacity calculations.
References for their optical constants are presented in Table \ref{tab:opticalconst}.
We assume that for all Mie calculations the grain size is given by the mean grain radius $\langle$a$\rangle$ (Eq. \ref{eq:meana}) with number density n$_{\rm d}$ (Eq. \ref{eq:nd}) for each atmospheric layer. 

\section{Mineral clouds in the atmosphere of HD 189733b}
\label{sec:Application}

We apply the dust formation theory developed by \citet{Woitke2003,Woitke2004,Helling2006} and \citet{Helling2008} in its 1D stationary form to 1D output trajectories of a 3D HD 189733b radiation-hydrodynamical [3D RHD] model, as described in \citet{Dobbs-Dixon2013}. 
In the following, we present local and global cloud structures for HD 189733b and discuss detailed results on cloud properties such as grain sizes, material composition, element abundances, dust-to-gas ratio and C/O ratio. 
We investigate general trends of the cloud structure as it varies throughout the atmosphere and make first attempts to study the cloud properties across the planetary globe.

\subsection{The cloud structure of HD 189733b at the sub-stellar point}
\label{sec:sub-stellar}
\begin{figure*}
\centering
\includegraphics[width=0.7\textwidth]{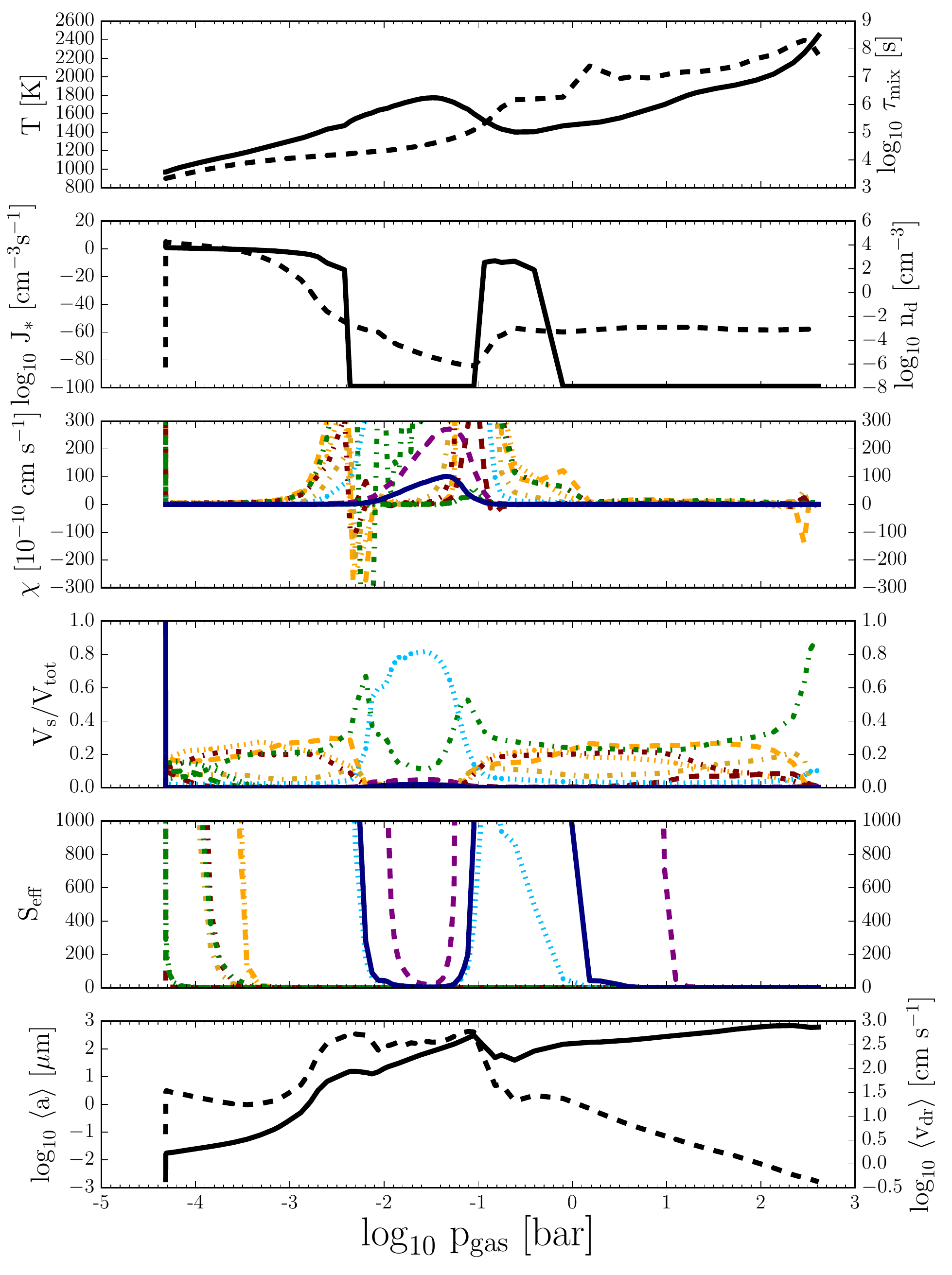} 
\includegraphics[width=0.29\textwidth]{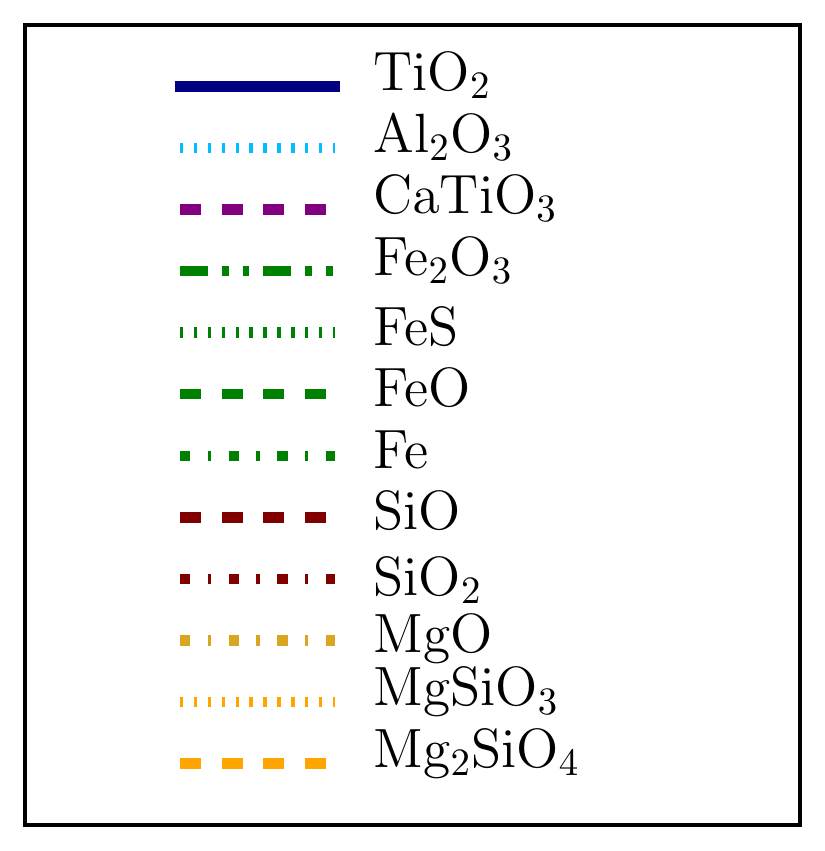} 
 \caption{HD 189733b's calculated cloud structure at the sub-stellar point ($\theta$ = 0\degr, $\phi$ = 0\degr).
 \textbf{Left}: 
 \textit{1st panel}: Local gas temperature T$_{\rm gas}$ [K] (solid, left) and mixing timescale $\tau_{\rm mix}$ [s] (dashed, right).
 \textit{2nd panel}: Nucleation rate J$_{*}$ [cm$^{-3}$s$^{-1}$] (solid, left) and dust number density n$_{\rm d}$ [cm$^{-3}$] (dashed, right). 
 \textit{3rd panel}: Growth velocity of material s $\chi$ [cm s$^{-1}$].
 \textit{4th panel}: Volume fraction V$_{\rm s}$/V$_{\rm tot}$ of solid s. 
 \textit{5th panel}: Effective supersaturation ratio S$_{\rm eff}$ of individual solids. 
 \textit{6th panel}: Cloud particle mean radius $\langle$a$\rangle$ [$\mu$m] (solid, left) and mean drift velocity $\langle$v$_{\rm dr}$$\rangle$ [cm s$^{-1}$] (dashed, right). 
 \textbf{Right}: Key showing line-style and colour of our considered dust species.}
 \label{fig:sub-stellar}
\end{figure*}

The substellar point ($\phi$ = 0\degr, $\theta$ = 0\degr) is the most directly irradiated point in atmospheres of hot Jupiters such as HD 189733b which is measured by observing before, during and after secondary transit (occulatation).
We use this well defined location to provide a detailed description of the vertical cloud structure.
We compare this atmospheric trajectory to other longitudes in Sec. \ref{sec:longitude} and to other latitudes in Sec. \ref{sec:latitude}. 
Figure \ref{fig:sub-stellar} shows that the cloud structure starts with the formation of seed particles (J$_{*}$ [cm$^{-3}$ s$^{-1}$] - nucleation rate) occurring at the upper pressure boundary of $\sim$10$^{-4}$ bar.
After the first surface growth processes occur on the seed particles, the cloud particles then gravitationally settle into the atmospheric regions below (toward higher density/pressure).
Primary nucleation is efficient down to $\sim$10$^{-2.5}$ bar where it drops off significantly, indicating that the local temperature is too high for further nucleation and that the seed forming species has been substantially depleted.
The gas-grain surface chemical reactions that form the grain mantle (Eq. \eqref{eq:chinet}) increase in rate as the cloud particles fall inward.
This is due to the cloud particles encountering increasing local gas density, and therefore more condensible material is available to react with cloud particles.
This surface growth becomes more efficient from $\sim$10$^{-3}$$\ldots$$\sim$10$^{-2}$ bar until the local temperature is so hot that the materials become thermally unstable and evaporate.
The evaporation region results in a half magnitude decrease of the cloud particle sizes (negative $\chi$) in the center region of the cloud.
The relative volume fractions of the solid `s', V$_{\rm s}$/V$_{\rm tot}$, represents the material composition of the cloud particles.
The cloud particle composition is dominated by silicates and oxides such as MgSiO$_{3}$[s]($\sim$27\%), Mg$_{2}$SiO$_{4}$[s]($\sim$20\%), SiO$_{2}$[s]($\sim$21\%) at the upper regions $\lesssim$10$^{-2.5}$ bar.
Fe[s] contributes $\lesssim$20\% to the volume of the cloud particle in this region.
The other 8 growth species (TiO$_{2}$[s]($\sim$0.03\%), Al$_{2}$O$_{3}$[s]($\sim$2\%), CaTiO$_{3}$[s]($\sim$0.15\%), Fe$_{2}$O$_{3}$[s]($\sim$0.001\%), FeS[s]($\sim$1.6\%), FeO[s]($\sim$0.35\%), SiO[s]($\sim$0.05\%), MgO[s]($\sim$7\%)) constitute the remaining grain volume. 
An evaporation region at $\sim$10$^{-2.5}$ bar before the temperature maximum at $\sim$10$^{-1.5}$ bar alters the grain composition dramatically.
Al$_{2}$O$_{3}$[s] and Fe[s] dominate the grain composition in this region as the less stable silicates and oxides have evaporated from the grain surface.
At the temperature maximum the grain is composed of Al$_{2}$O$_{3}$[s]($\sim$80\%) and Fe[s]($\sim$15\%).
This suggests the presence of a cloud section more transparent than surrounding layers (Fig. \ref{fig:opacity}).
We address the issue of transparency in Sect. \ref{sec:Discussion}.
Deeper in the atmosphere, a temperature inversion occurs (Fig. \ref{fig:inputq}) starting from $\sim$10$^{-1.5}$ bar.
The temperature drops by $\sim$400 K from $\sim$10$^{-1.5}$$\ldots$$\sim$10$^{-0.5}$ bar.  
This temperature decrease allows a secondary nucleation region from $\sim$10$^{-1}$$\ldots$1 bar.
The number density of grains jumps by $\sim$2 orders of magnitude at the secondary nucleation region, as a result of formation of many new cloud particles.
This causes a dip in the mean grain size at $\sim$10$^{-1}$ bar.
Such secondary nucleation has not been seen in our cloud model results in Brown Dwarf nor non-irradiated giant gas planet atmospheres.
Silicates and oxides are now thermally stable and once again form the grain mantle. 
The grain composition is approximately a 70-20-10 \% mix of silicates and oxides, iron and other material respectively in this region, similar to the composition before the temperature maximum.
At the cloud base, Fe[s] dominates the composition ($\sim$35\%)  with MgO[s] and Mg$_{2}$SiO$_{4}$[s] making up $\sim$16\% and $\sim$23\% respectively.
Table. \ref{tab:volfrac0} shows the percentage volume fraction V$_{\rm s}$/V$_{\rm tot}$ of the 12 dust species at the sub-stellar point and the $\phi$ = 180\degr, $\theta$ = 0\degr \ trajectory at 10$^{-4}$, 10$^{-3}$, 10$^{-2}$, 10$^{-1}$, 1 and 10 bar.

Our results show that the entire vertical atmospheric range considered here ($\sim$10$^{-4.5}$$\ldots$10$^{3}$ bar) is filled with dust.
The exact properties of this dust such as size, composition and number density change depending on the local thermodynamical state.
The 3D RHD model does not expand to such low pressures and densities that the cloud formation processes becomes inefficient (Sect. \ref{sec:Discussion}).
This suggests that clouds can be present at a higher and lower pressure than the present 3D RHD model boundary conditions allow.

\subsection{Cloud structure changes with longitude (East-West)}
\label{sec:longitude}
\begin{figure*}
 \centering
\includegraphics[width=0.43\textwidth]{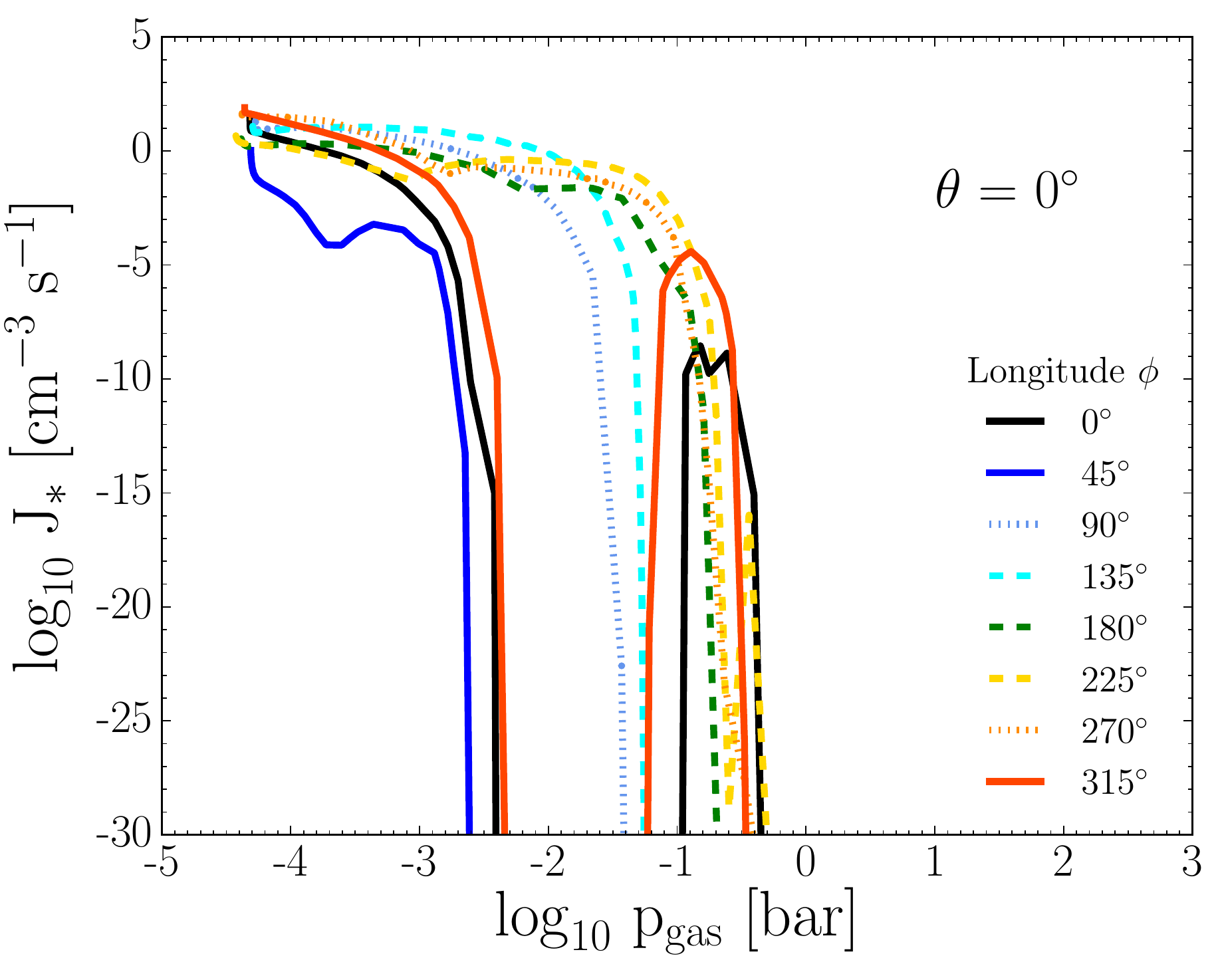}
\includegraphics[width=0.43\textwidth]{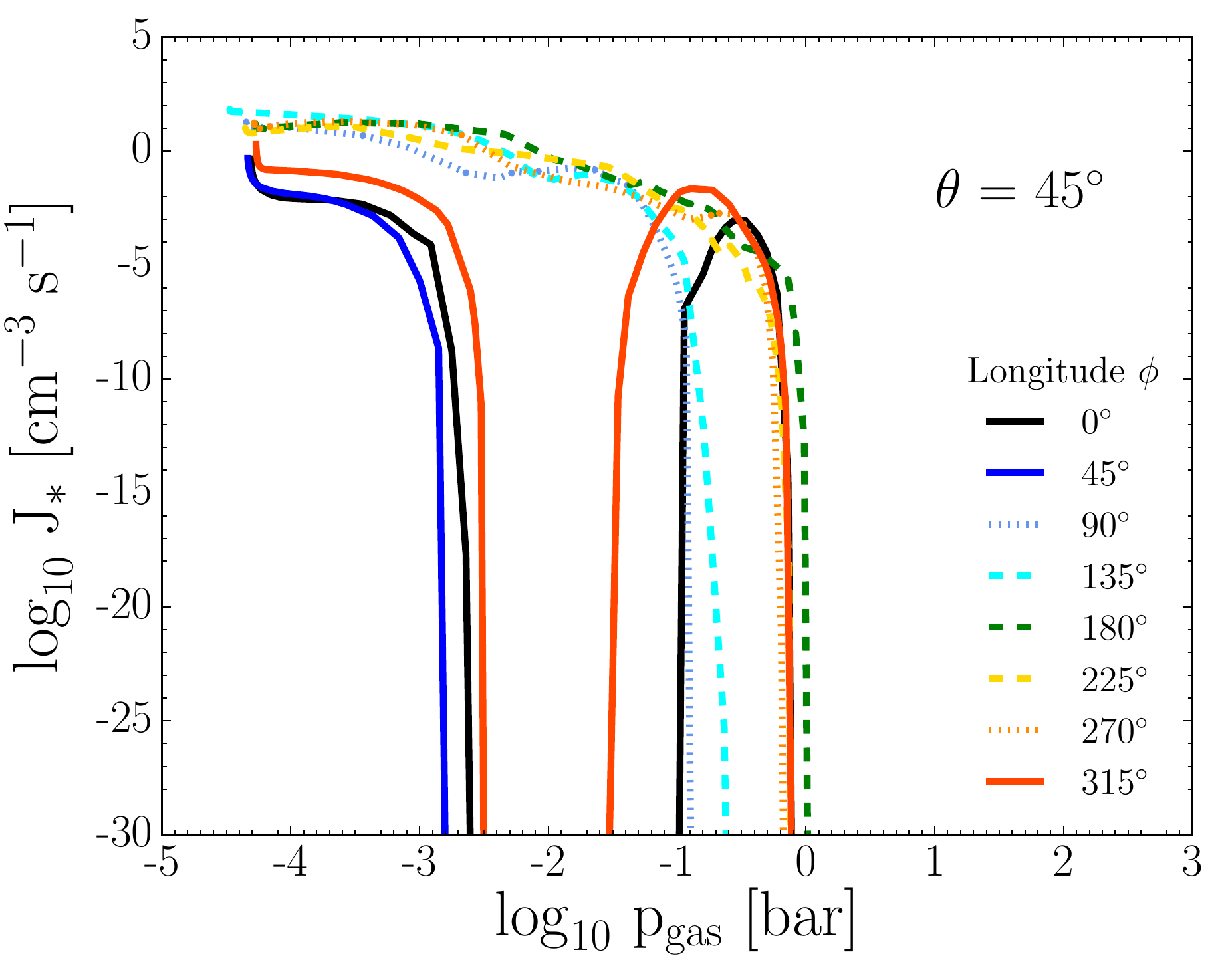}
\includegraphics[width=0.43\textwidth]{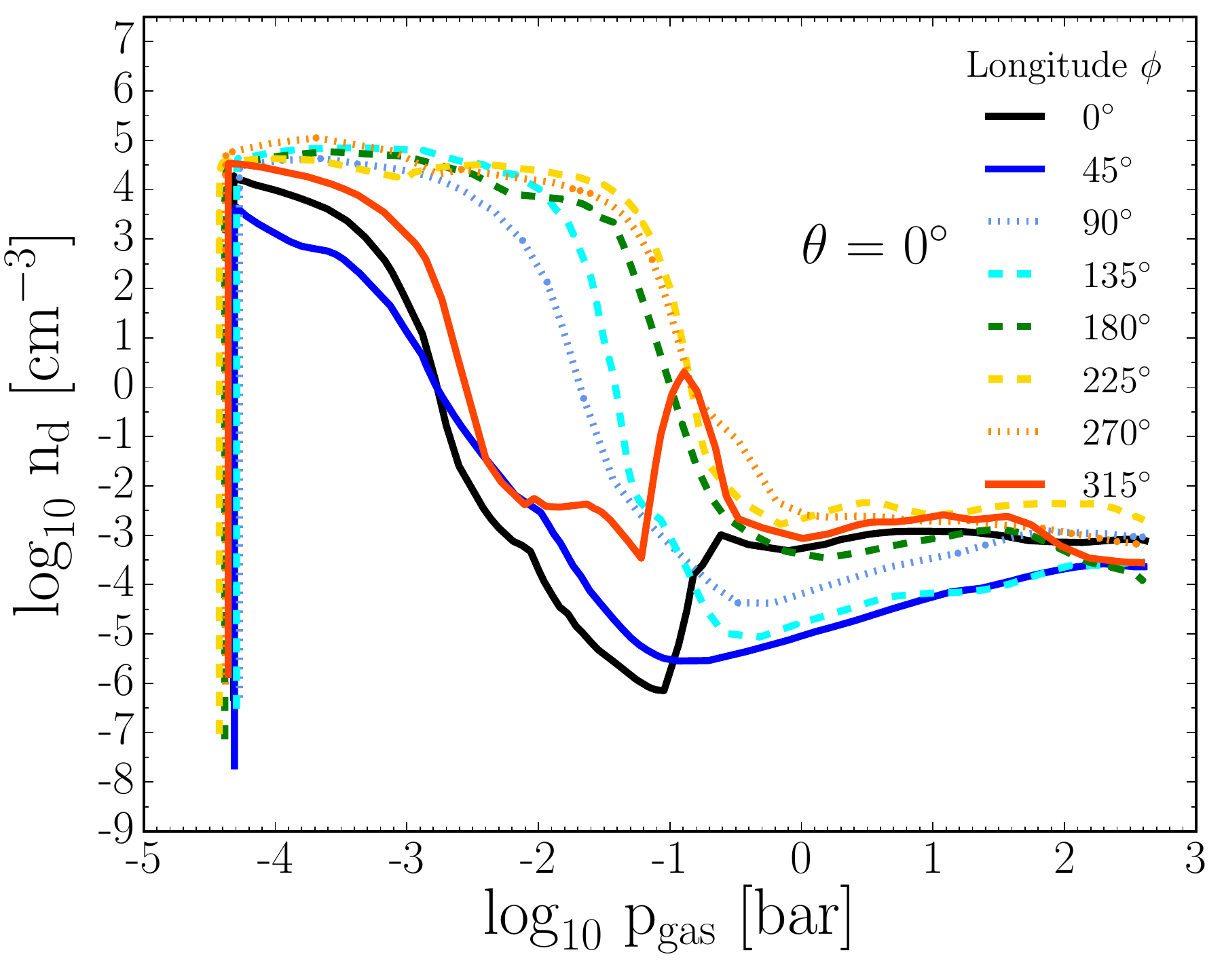}
\includegraphics[width=0.43\textwidth]{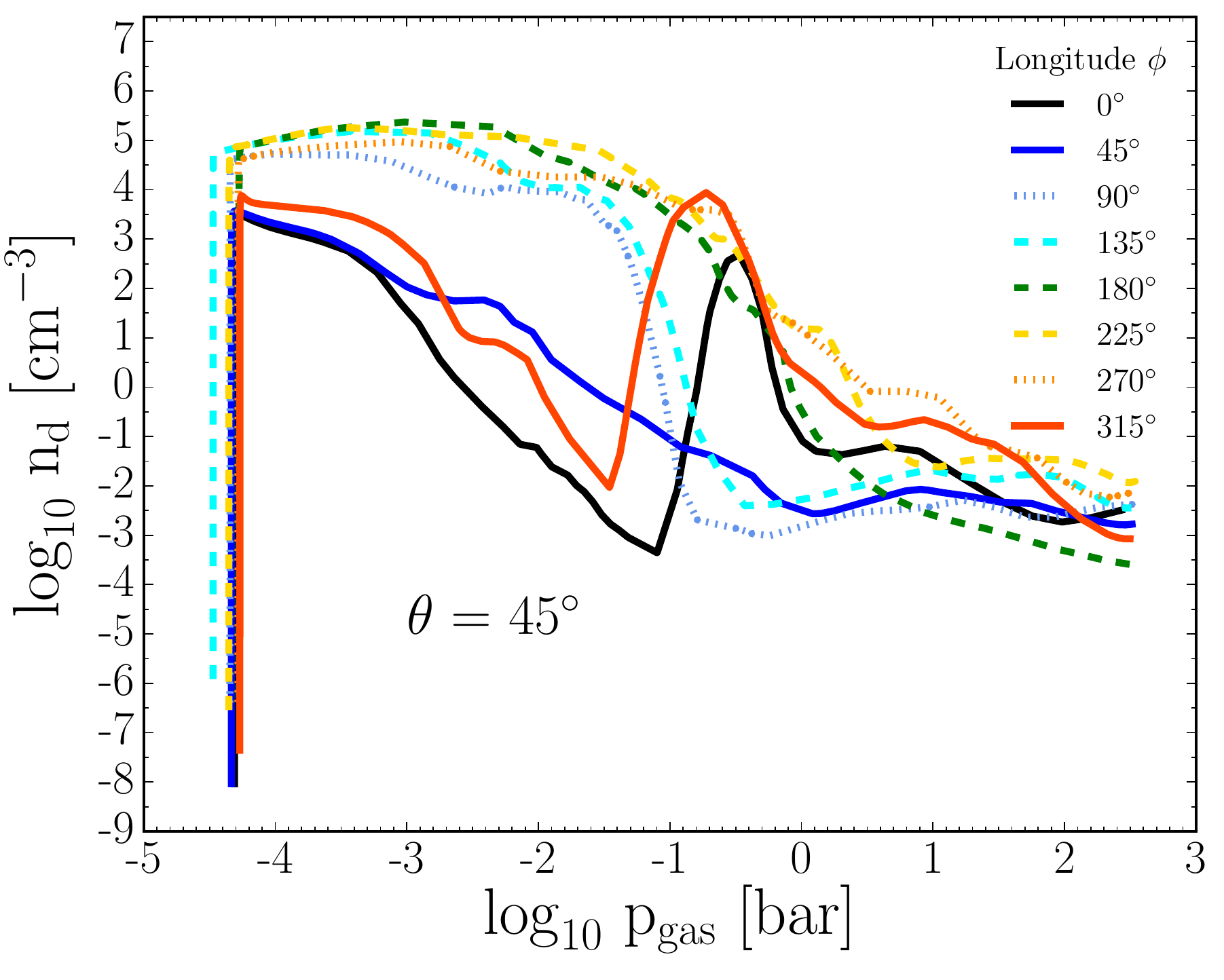}
\includegraphics[width=0.43\textwidth]{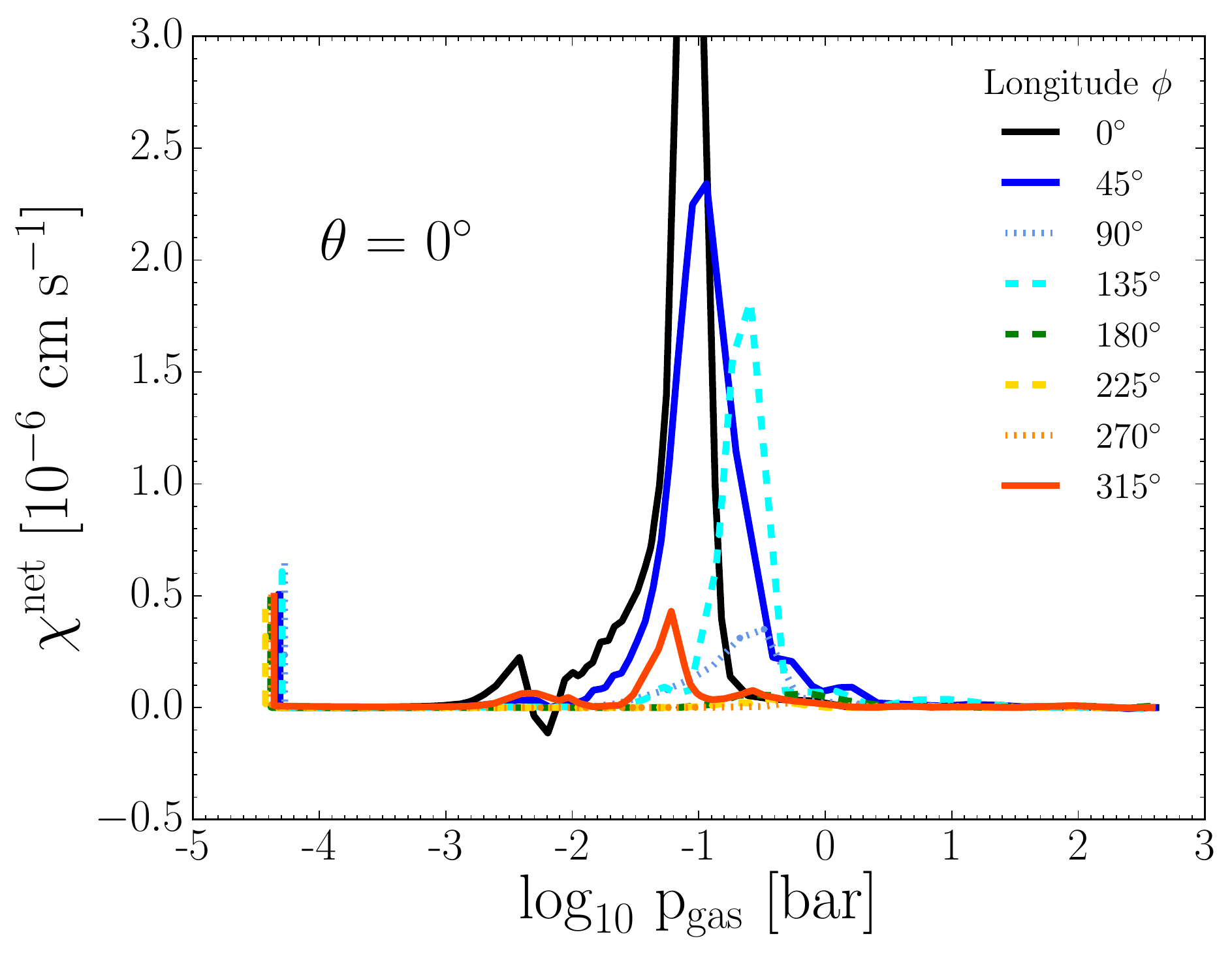}
\includegraphics[width=0.43\textwidth]{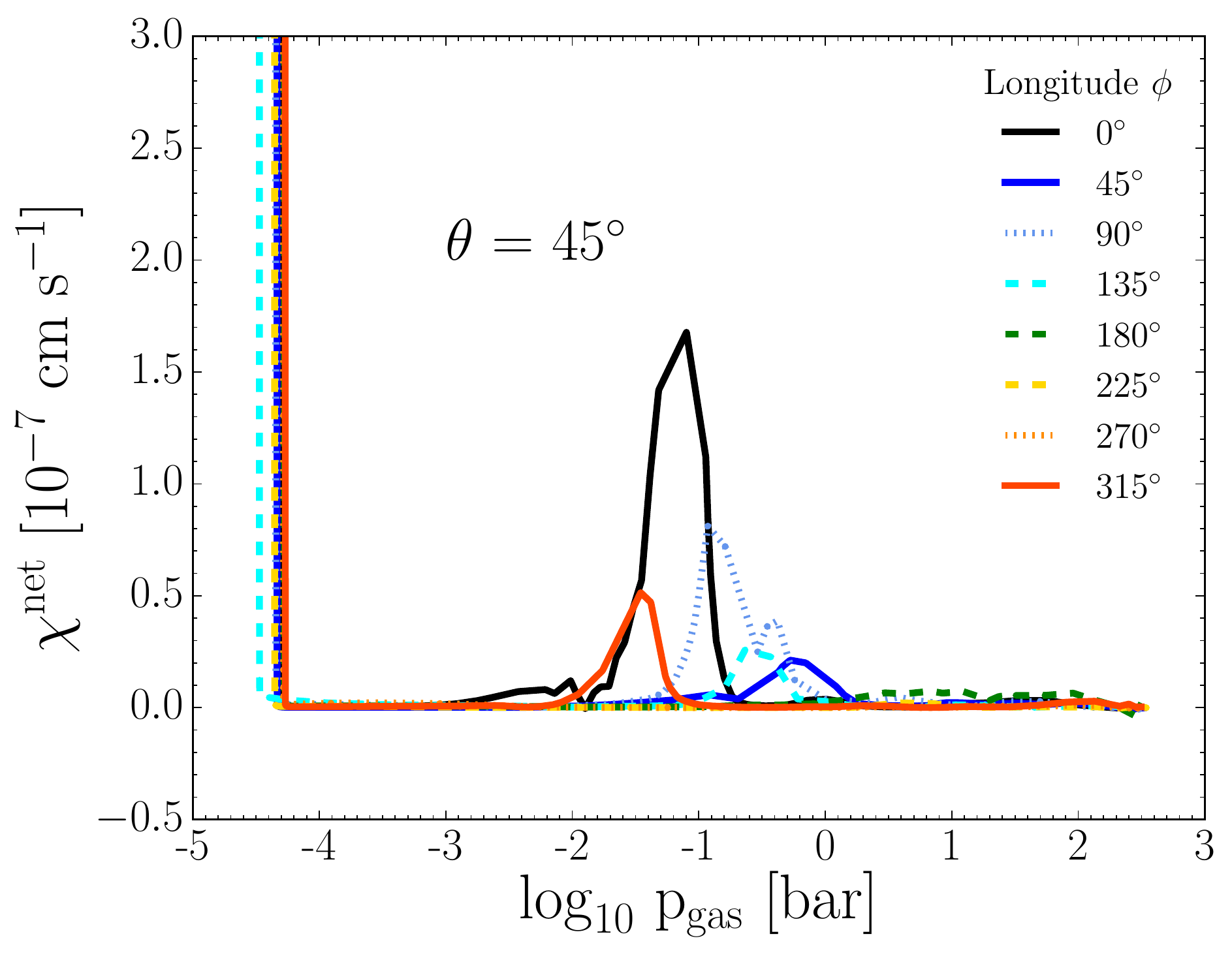}
\includegraphics[width=0.43\textwidth]{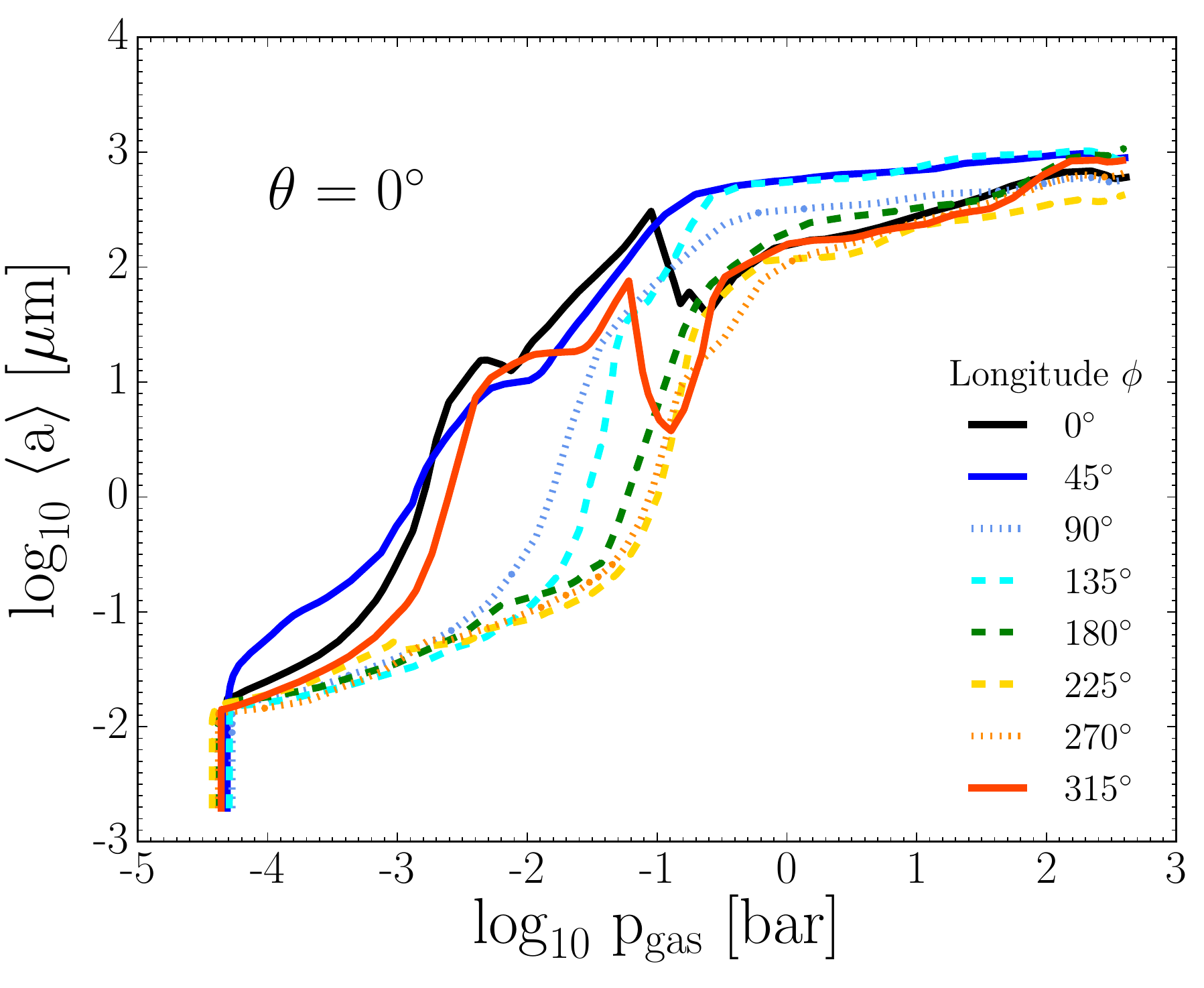}
\includegraphics[width=0.43 \textwidth]{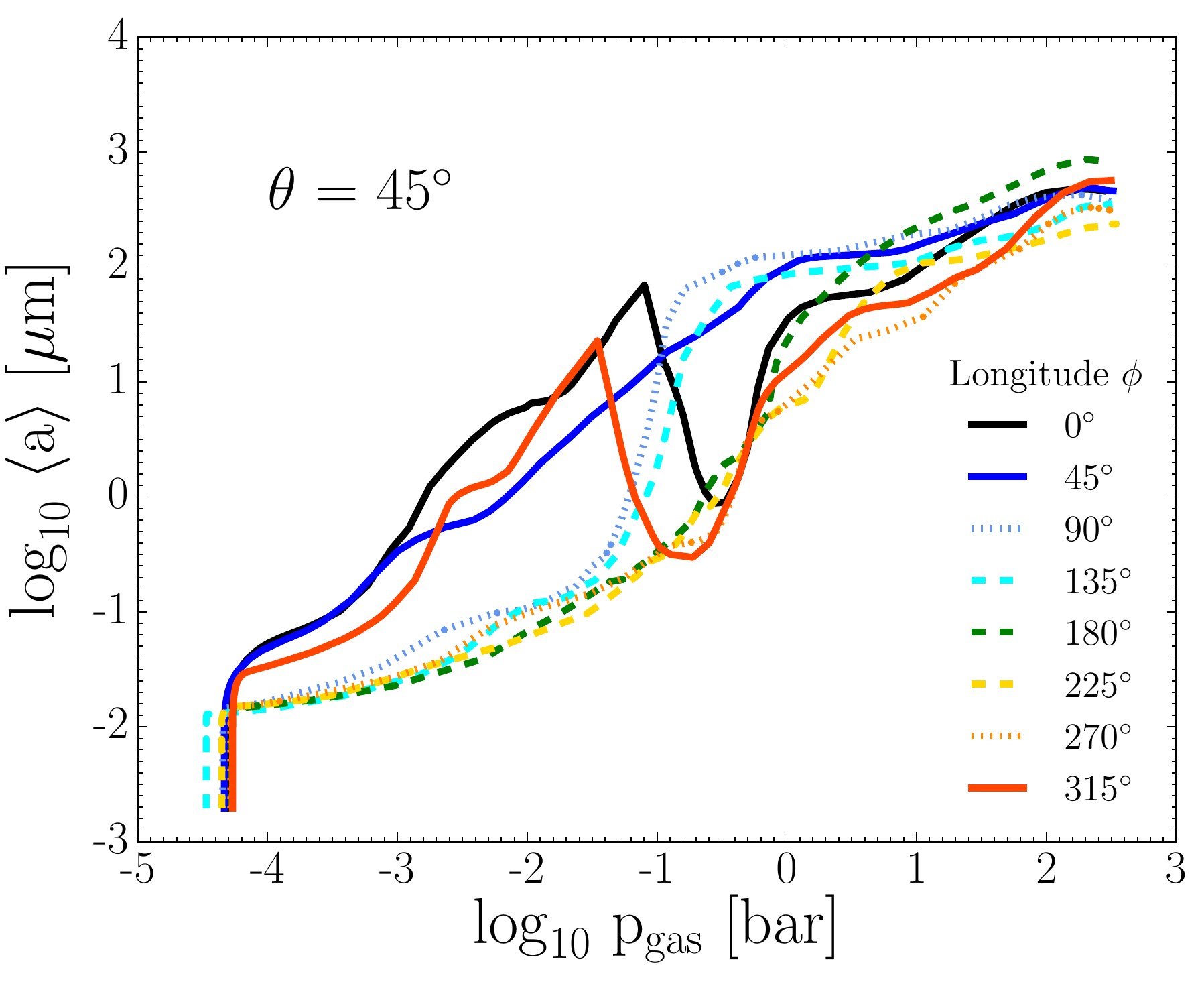}
\caption{Dust properties as a function of gas pressure for $\Delta$$\phi$ = +45\degr longitude intervals for latitudes $\theta$ =  0\degr (Left column) and 45\degr (Right column): 
\textbf{Top Row:} Nucleation rate J$_{*}$ [cm$^{-3}$ s$^{-1}$].
\textbf{Second Row:} Dust number density n$_{\rm d}$ [cm$^{-3}$].
\textbf{Third Row:} Net dust growth velocity $\chi^{\rm net}$ [cm s$^{-1}$].
\textbf{Bottom Row:} Mean grain size $\langle$a$\rangle$ [$\mu$m].
Solid, dotted and dashed lines indicate dayside, day-night terminator and nightside profiles respectively.}
\label{fig:theta0}
\label{fig:theta45}
\end{figure*}

\begin{figure*}
 \centering
\includegraphics[width=0.49\textwidth]{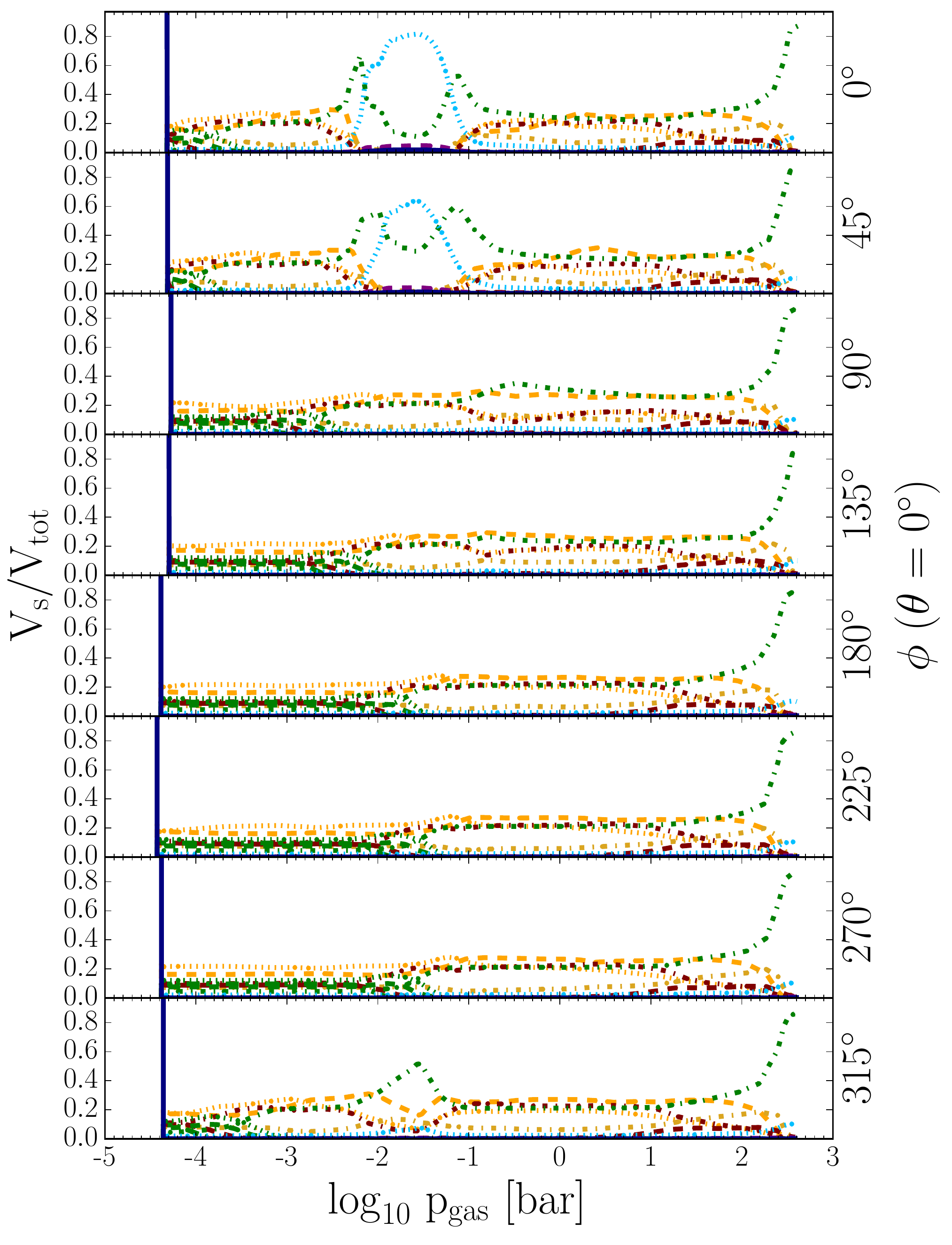}
\includegraphics[width=0.49\textwidth]{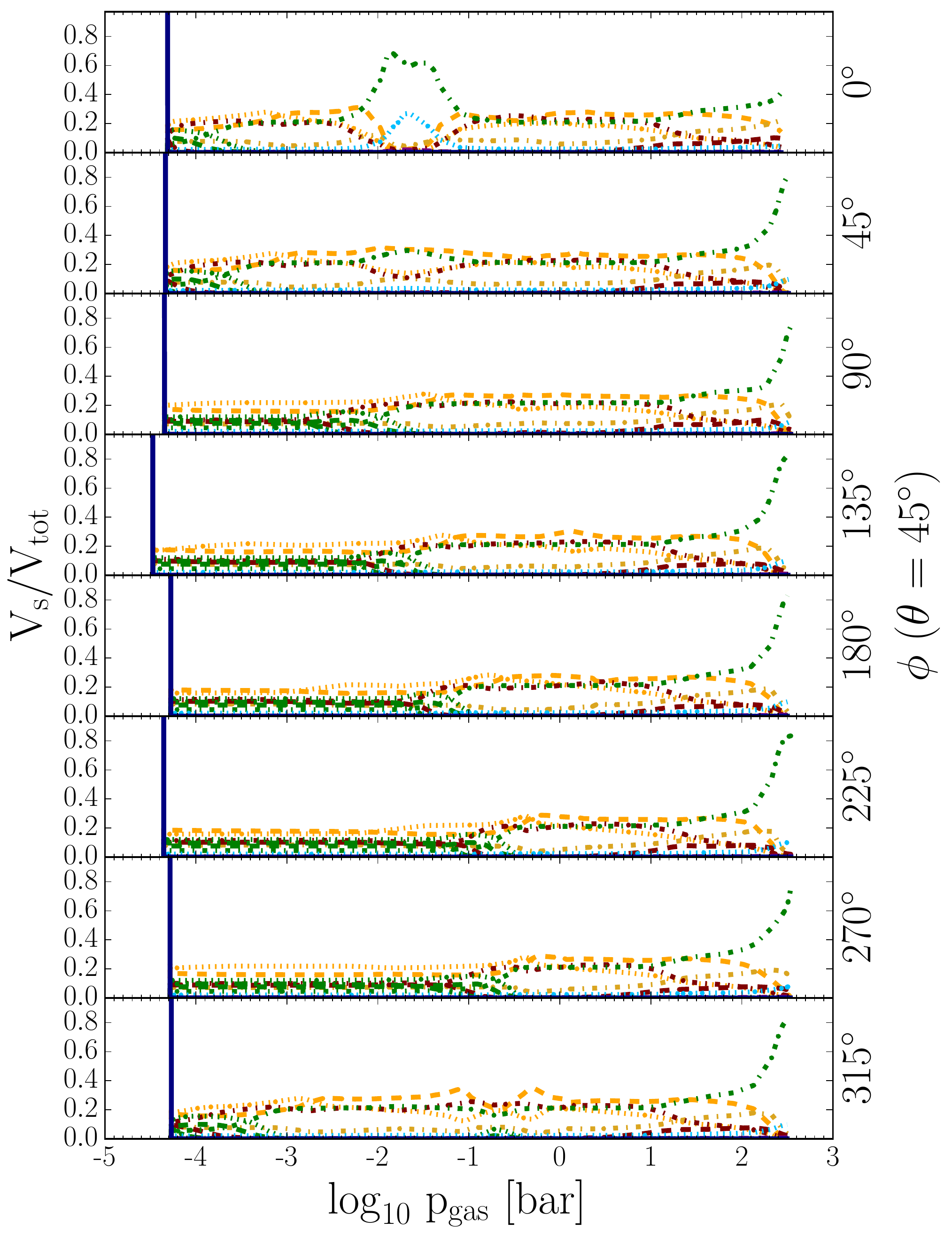}
\includegraphics[width=0.30\textwidth]{plots/HD189_Dobbs_key_paper.pdf} 
\caption{\textbf{Top:} Dust composition volume fraction V$_{\rm s}$/V$_{\rm tot}$ as a function of pressure at $\theta$ =  0\degr (Left) and $\theta$ =  45\degr (Right) in $\Delta$$\phi$ = +45\degr longitude intervals. 
Grain composition is generally dominated by silicate-oxides including MgSiO$_{3}$[s] apart from localised regions such as 10$^{-2}$$\ldots$10$^{-1}$ bar at $\phi$ = 0\degr, 45\degr, $\theta$ =  0\degr. 
Fe[s] dominates at 10$^{-2}$$\ldots$10$^{-1}$ bar at $\phi$ =   0\degr, $\theta$ =  45\degr. Fe[s] grains remain at the cloud base as the most thermally stable element.
\textbf{Bottom:} Key showing line-style and colour of our considered dust species.}
\label{fig:thetadust}
\end{figure*}

\begin{table*}
\centering
\caption{Volume fraction V$_{\rm s}$/V$_{\rm tot}$[\%] for the 12 solid species included in kinetic the cloud model.
The first row of each species corresponds to the sub-stellar trajectory ($\phi$ = 0\degr, $\theta$ = 0\degr) cloud structure.
The second row corresponds to the nightside trajectory $\phi$ = 180\degr, $\theta$ = 0\degr cloud structure. 
Note: The pressure at the cloud base is different for the two profiles.}
\label{tab:volfrac0}
 \begin{tabular}{c c c c c c c c } \hline
   \textbf{Pressure [bar]} 		& 10$^{-4}$ & 10$^{-3}$  & 10$^{-2}$  & 10$^{-1}$  & 1 	& 10  	& Cloud base	\\ \hline
 \multirow{2}{*}{$\langle$a$\rangle$ [$\mu$m]} & 0.025 &0.23 & 19.9  &307    & 146	 &275	&605 \\ 
				      & 0.018	    & 0.035 &0.15 	&12.3	    & 174 &338	&1088 \\ \hline
 \textbf{Solid species} 		& &  &  &   &  	&  	& 	\\ \hline				      
 \multirow{2}{*}{TiO$_{2}$[s]} 	& 0.03	    &	0.03	  & 1.22       &0.26        & 0.05	 &0.02	&0.22		\\ 
				      & 0.06	    &	0.04	  & 0.03	&0.03	    &0.02 	  &0.01	&0.24		\\ \hline
 \multirow{2}{*}{Al$_{2}$O$_{3}$[s]} & 2.11	    &	2.43	  &60.98	&13.99      &3.68 	 &2.84	&10.03		\\ 
				      & 2.06	    &2.06	  &2.13	       &2.42	    &2.44 	  &2.49	&10.03		\\ \hline
 \multirow{2}{*}{CaTiO$_{3}$[s]}     & 0.13	    &	0.16	  &3.56	       &0.87        &0.24 	 &0.22	&0.64		\\ 
				      &0.07 	    &0.10	  &0.14	       &0.17	    &0.17 	  &0.20	&0.77		\\ \hline
 \multirow{2}{*}{Fe$_{2}$O$_{3}$[s]} & 0.10	    &	$>$0.01	  &$>$0.01	 &$>$0.01   &$>$0.01 	 &$>$0.01&$>$0.01		\\ 
				      &9.69 	    &9.68	 &0.07	       &$>$0.01	    &$>$0.01 	 &$>$0.01&$>$0.01	\\ \hline
 \multirow{2}{*}{FeS[s]}             & 14.44	    &	0.22	  &0.03	       &0.09        &0.05 	&0.02	&0.03		\\ 
				      &12.12 	    &12.12	  &14.45	 & 0.10	    & 0.05 	 &0.02	&0.03		\\ \hline
 \multirow{2}{*}{FeO[s]}             & 8.87	    &	0.06	  &0.03		&0.07       & 0.02	 &0.02	&0.09		\\ 
				      & 7.63	    &	7.63	  &7.93	       &0.03	    &0.02 	  &0.02	&0.09		\\ \hline
 \multirow{2}{*}{Fe[s]}              & 7.87	    &	21.08	  &30.93	& 45.00     &24.29 	 &22.49	&87.15		\\ 
				      &4.52 	    &4.52	  &8.50	       &21.14	    &21.29 	 &21.75	&86.80		\\ \hline
 \multirow{2}{*}{SiO[s]}             & 2.42	    &	0.06	  &0.08	       & 0.70       &0.48 	 &3.82	&0.92		\\ 
				      &8.84 	    &9.03	  &1.12	       &0.11	    & 0.44	 &5.20	&0.99		\\ \hline
 \multirow{2}{*}{SiO$_{2}$[s]}       & 17.90	    &	20.08	  &0.75	       &7.88       &19.94 	 &20.60	&0.12		\\ 
				      &9.92 	    &10.13	  &19.35	&21.26	    &22.18 	 &20.33	&0.13		\\ \hline
 \multirow{2}{*}{MgO[s]}             & 7.73	    &	5.13	  &1.97	       & 11.10      &8.05 	 &8.87	&0.79		\\ 
				      &7.21 	    &7.28	  &7.61	       &5.40	    &6.51 	 &9.35	&0.89		\\ \hline
 \multirow{2}{*}{MgSiO$_{3}$[s]}     & 22.47	    &	24.11	  &0.14	       &8.96       &20.77 	 &15.79	&$>$0.01		\\ 
				      &21.74 	    &20.71	  &22.41	&22.02	    &21.82 	 &15.25	&$>$0.01		\\ \hline
 \multirow{2}{*}{Mg$_{2}$SiO$_{4}$[s]} & 15.93	    &	26.64	  &0.30	       &11.09       &21.41 	 &25.31	&0.02		\\ 
				      & 16.15	    &16.70	  &16.23	&27.31	    &26.59 	 &25.39	&0.04		\\ \hline
 \end{tabular}
\end{table*}



One of the main features of the radiation-hydrodynamic simulation is the equatorial jet structure which transports heat over the entire 360\degr longitude.
The presence of this jet changes the thermodynamic structure of the atmosphere which affects the resulting global cloud structure. 
We sampled the 3D RHD results in longitude steps of $\Delta$$\phi$ = +45\degr at the equatorial region $\theta$ = 0\degr to investigate if cloud properties change significantly from dayside to nightside.
Figure \ref{fig:inputq} shows that the nightside (T$_{\rm gas}$, p$_{\rm gas}$) profiles can be $\sim$200 K cooler than the dayside in the upper atmospheric regions ($\sim$10$^{-4.5}\ldots$10$^{-1}$ bar). 
Figure \ref{fig:theta0} shows the nucleation rate J$_{*}$ [cm$^{-3}$ s$^{-1}$] for dayside (solid lines), day-night terminator (dotted lines) and nightside (dashed lines) sample trajectories.
The dayside nucleation rates fall off quicker with atmospheric depth compared to the terminator and nightside profiles, particularly relevant for interpreting transit spectra.
At longitude $\phi$ = 315\degr \ a secondary nucleation region emerges, similar to the sub-stellar point structure (Fig. \ref{fig:sub-stellar}).
The terminator and nightside nucleation regions extend further into the atmosphere to $\sim$10$^{-1}$ bar.
As a consequence of the different nucleation rates between dayside and nightside, the number density of cloud particles n$_{\rm d}$ [cm$^{-3}$] is greater on the nightside down to pressures of $\sim$10$^{-1}$ bar.
At this pressure, the secondary nucleation regions on the dayside, $\phi$ = 0\degr, 315\degr profiles spike up the number density comparable to values on the nightside. 
From Fig. \ref{fig:theta0} the net growth velocity $\chi^{\rm net}$ [cm s$^{-1}$] shows that the most efficient growth regions for the grains is approximately 10$^{-3}$$\ldots$1 bar for the dayside and 10$^{-2}$$\ldots$1 bar for the nightside. 
Although the temperature of the local gas phase plays a role in increasing the growth rate, it is the increasing local density of material (as the particle falls thought the atmosphere) that is the dominating factor in determining growth rates (Eq. \eqref{eq:chinet}; Fig. \ref{fig:thetaelement}).
Consequently, the mean grain size $\langle$a$\rangle$ [$\mu$m] shows a stronger increase from $\sim$10$^{-4}\ldots$10$^{-1}$ bar on the dayside than the nightside. 
The mean grain size dips at $\sim$10$^{-1}$ bar for longitudes $\phi$ = 0\degr and 315\degr due to the increase of grain number density as now the same number of gaseous growth species have to be distributed over a larger surface area.
Figure \ref{fig:thetadust} shows the volume fraction V$_{\rm s}$/V$_{\rm tot}$ of the solid species `s'.
The dust composition is generally dominated by silicates and oxides ($\sim$60 \%) such as MgSiO$_{3}$[s], Mg$_{2}$SiO$_{4}$[s] and SiO$_{2}$[s] with ($\sim$20 \%) Fe[s] and FeO[s] content.
At the cloud base the grain is primarily composed of Fe[s] ($\sim$85\%) and Al$_{2}$O$_{3}$[s] ($\sim$10\%).
The $\phi$ = 45\degr, $\theta$ = 0\degr cloud structure contains an Al$_{2}$O$_{3}$[s] ($\sim$60 \%) and Fe[s] ($\sim$30 \%) dominant region from $\sim$10$^{-2}$$\ldots$10$^{-1}$ bar similar to the sub-stellar point structure.
Table \ref{tab:volfrac0} summarises the percentage volume fraction V$_{\rm s}$/V$_{\rm tot}$ and the mean cloud particle size $\langle$a$\rangle$ of the 12 dust species at the sub-stellar point and the $\phi$ = 180\degr, $\theta$ = 0\degr trajectory at 10$^{-4}$, 10$^{-3}$, 10$^{-2}$, 10$^{-1}$, 1 and 10 bar pressures.

\subsection{Cloud structure changes with latitude (North-South)}
\label{sec:latitude}

At higher latitudes $\theta$ $\gtrsim$40\degr the 3D RHD model produces a jet structure flowing in the opposite direction to the equatorial jet at dayside longitudes \citep{Tsai2014}.
This significantly alters the (T$_{\rm gas}$, p$_{\rm gas}$) and vertical velocity profiles (Fig. \ref{fig:inputq}).
These latitudes also contain the coldest regions of the nightside where vortexes easily form and dominate the atmosphere dynamics \citep[Fig. 1]{Dobbs-Dixon2013}.
To investigate the cloud structure at these latitudes we repeated our trajectory sampling for latitudes of $\theta$ = 45\degr in longitude steps of $\Delta$$\phi$ = +45\degr.
Figure \ref{fig:theta45} shows the nucleation rate J$_{*}$ [cm$^{-3}$ s$^{-1}$] for dayside (solid lines), day-night terminator (dotted lines) and nightside (dashed lines) sample trajectories.
The profiles are similar to the equatorial $\theta$ = 0\degr regions with double nucleation peaks at $\phi$ = 0\degr and 315\degr.
Again, there is an increase in number density n$_{\rm d}$ [cm$^{-3}$] (Fig. \ref{fig:theta45}) from 10$^{-1}$$\ldots$1 bar due to second nucleation regions at $\phi$ = 0\degr and 315\degr.
The growth velocity $\chi^{\rm net}$ is generally an order of magnitude lower on the dayside than the equatorial regions.
This results in higher latitude clouds containing smaller grain sizes compared to their equatorial counterparts.

\subsection{Element depletion, C/O ratio and dust-to-gas ratio}
\label{sec:element}

\begin{figure*}
 \centering
\includegraphics[width=0.49\textwidth]{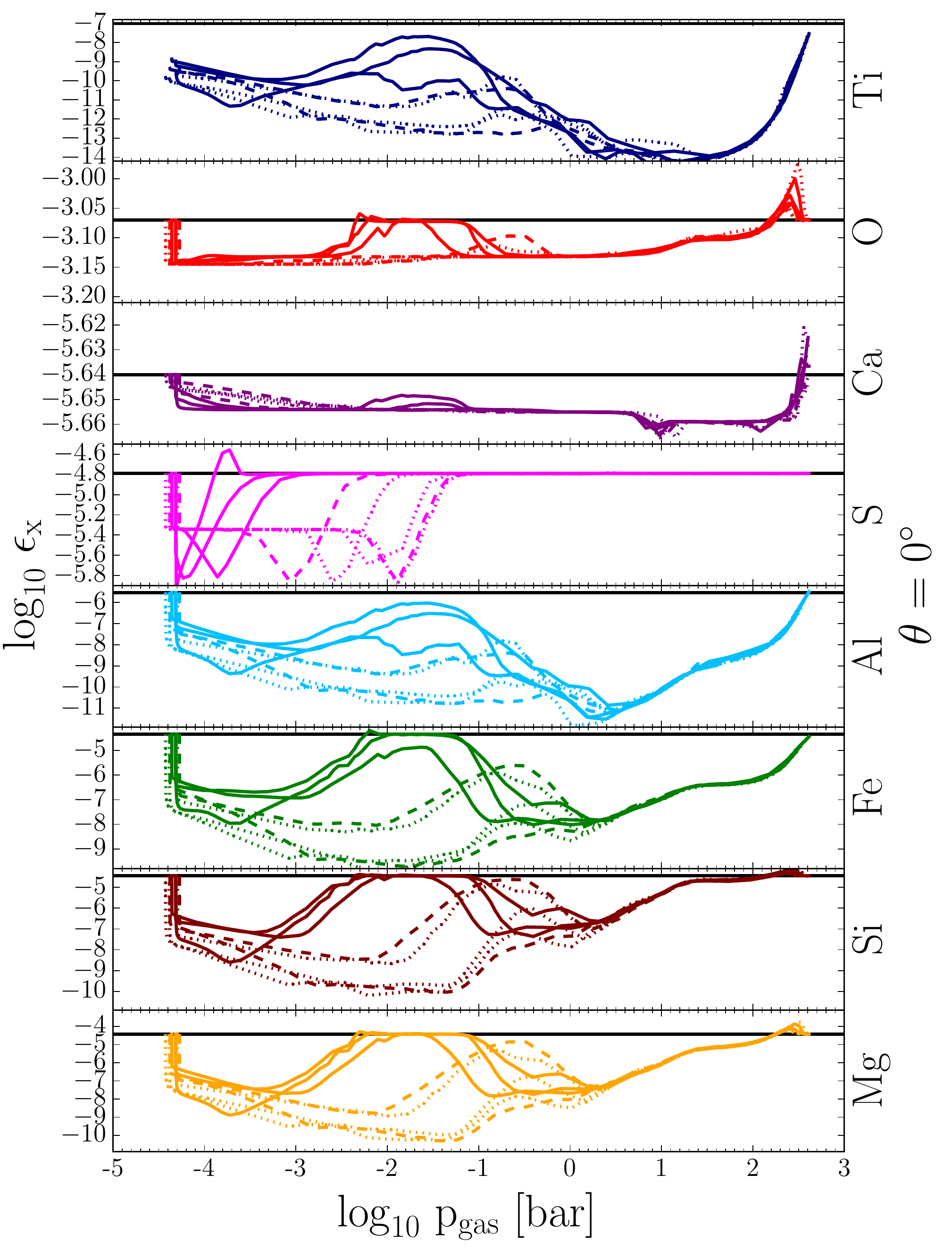}
\includegraphics[width=0.49\textwidth]{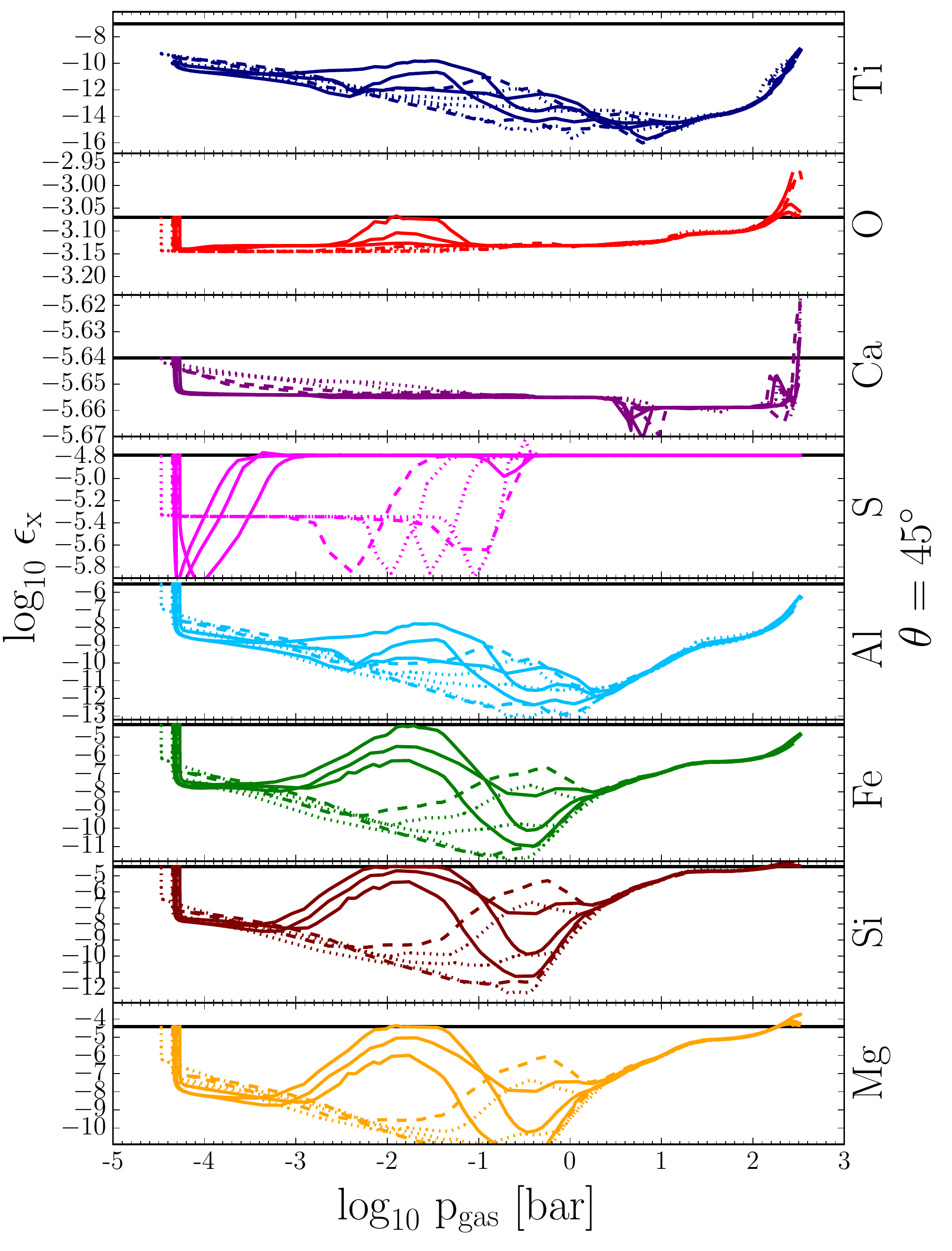}
\caption{Gas phase element abundances $\epsilon_{\rm x}$ as a function of pressure at $\theta$ = 0\degr (Left) and $\theta$ = 45\degr (Right) in $\Delta$$\phi$ = +45\degr longitude intervals.
We consider 8 elements: Mg(orange), Si(maroon), Ti(blue), O(red), Fe(green), Al(cyan), Ca(purple), S(magenta) that constitute the growth species.
Horizontal black lines indicate solar abundance $\epsilon^{0}_{\rm x}$.
A decrease in element abundance indicates condensation of growth species onto cloud particle surface. 
An increase indicates evaporation of molecules constituted of that element from the cloud particle surface.
Dayside profiles (solid) are $\phi$ = 0\degr, 45\degr, 315\degr , $\theta$ = 0\degr, 45\degr.
Day-night terminator profiles (dashed) are $\phi$ = 90\degr, 270\degr, $\theta$ = 0\degr, 45\degr.
Nightside profiles (dotted) are $\phi$ = 135\degr, 180\degr, 225\degr, $\theta$ = 0\degr, 45\degr.
}
\label{fig:thetaelement}
\end{figure*}

\begin{figure*}
\centering
\includegraphics[width=0.49\textwidth]{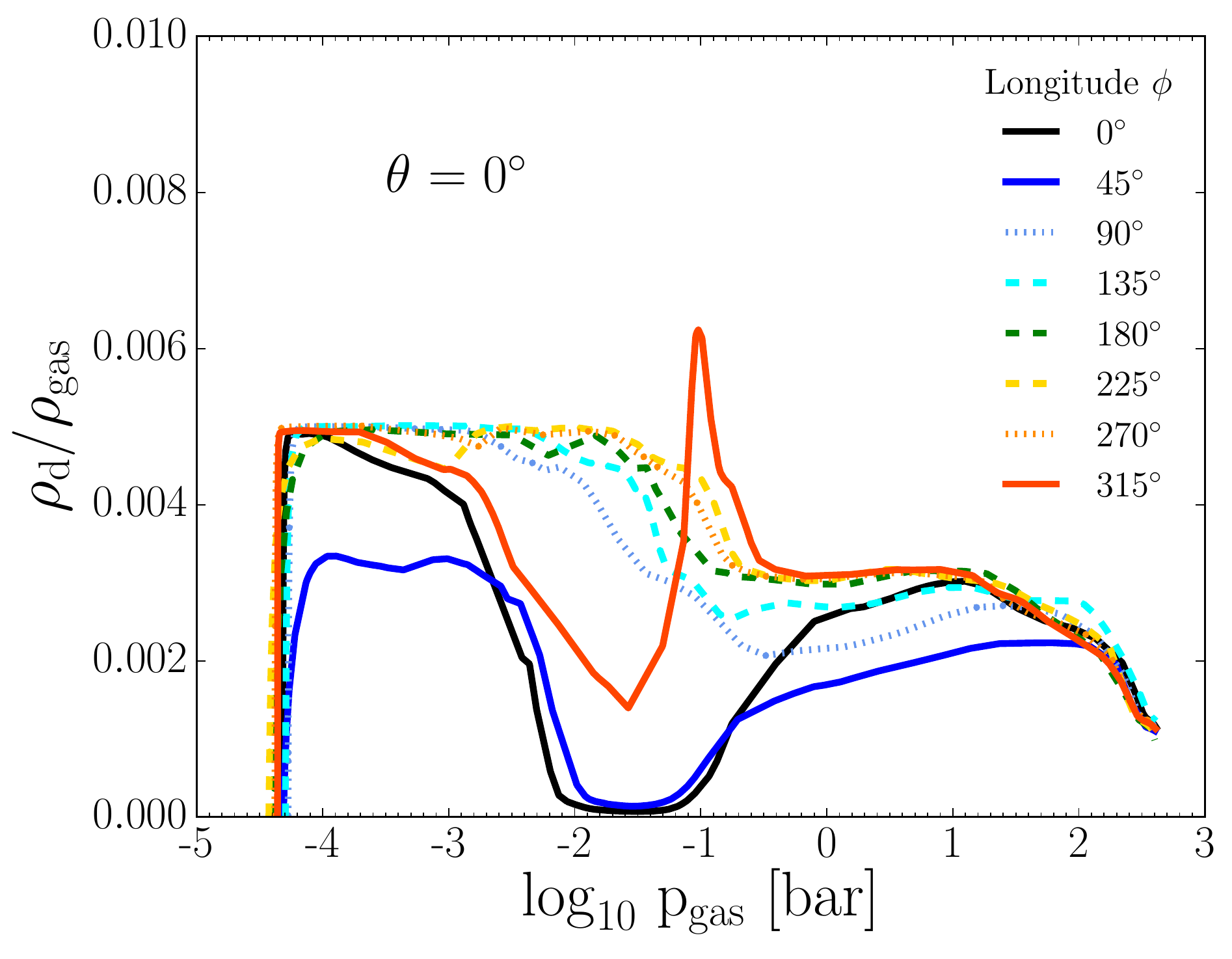}
\includegraphics[width=0.49\textwidth]{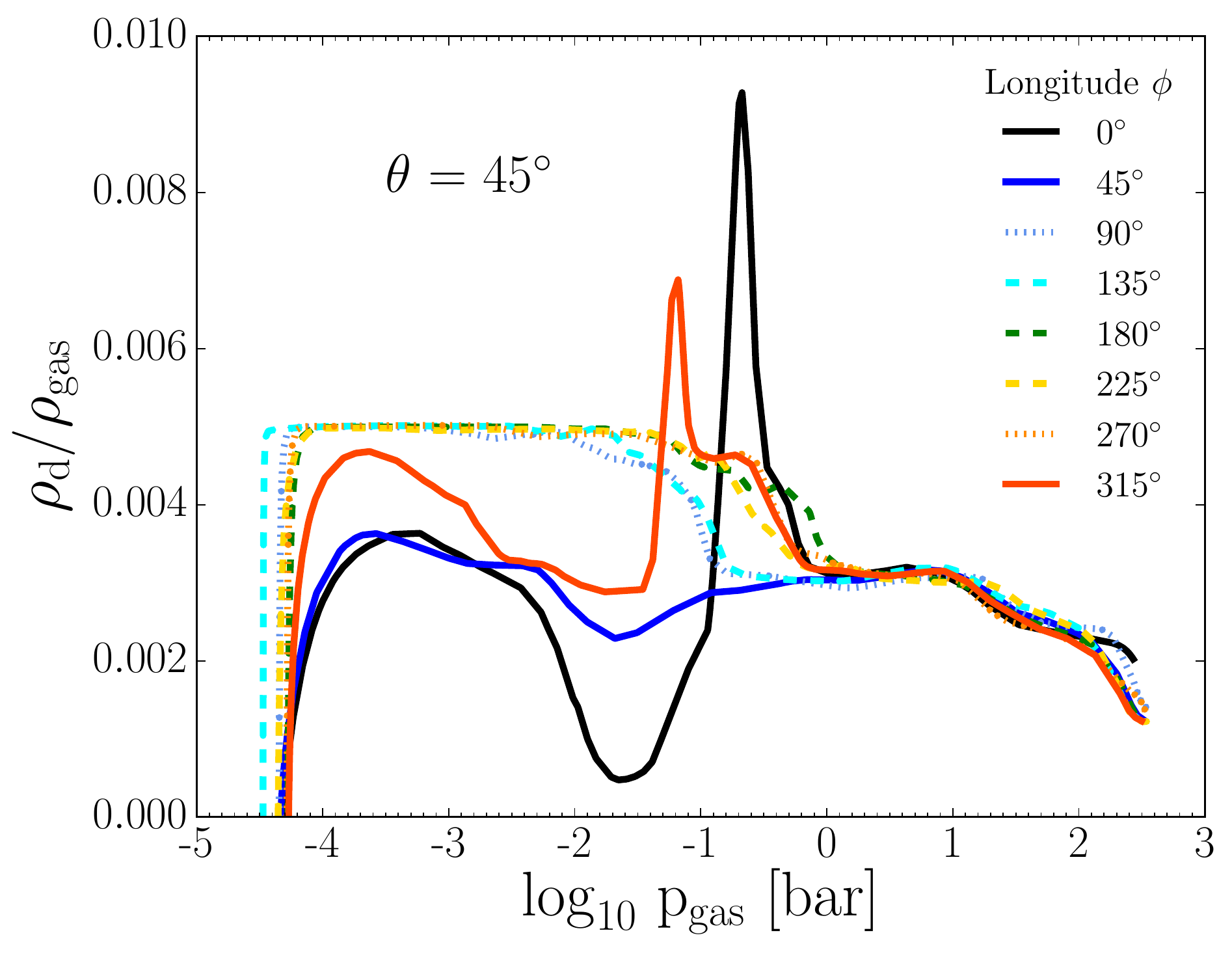}
\includegraphics[width=0.49\textwidth]{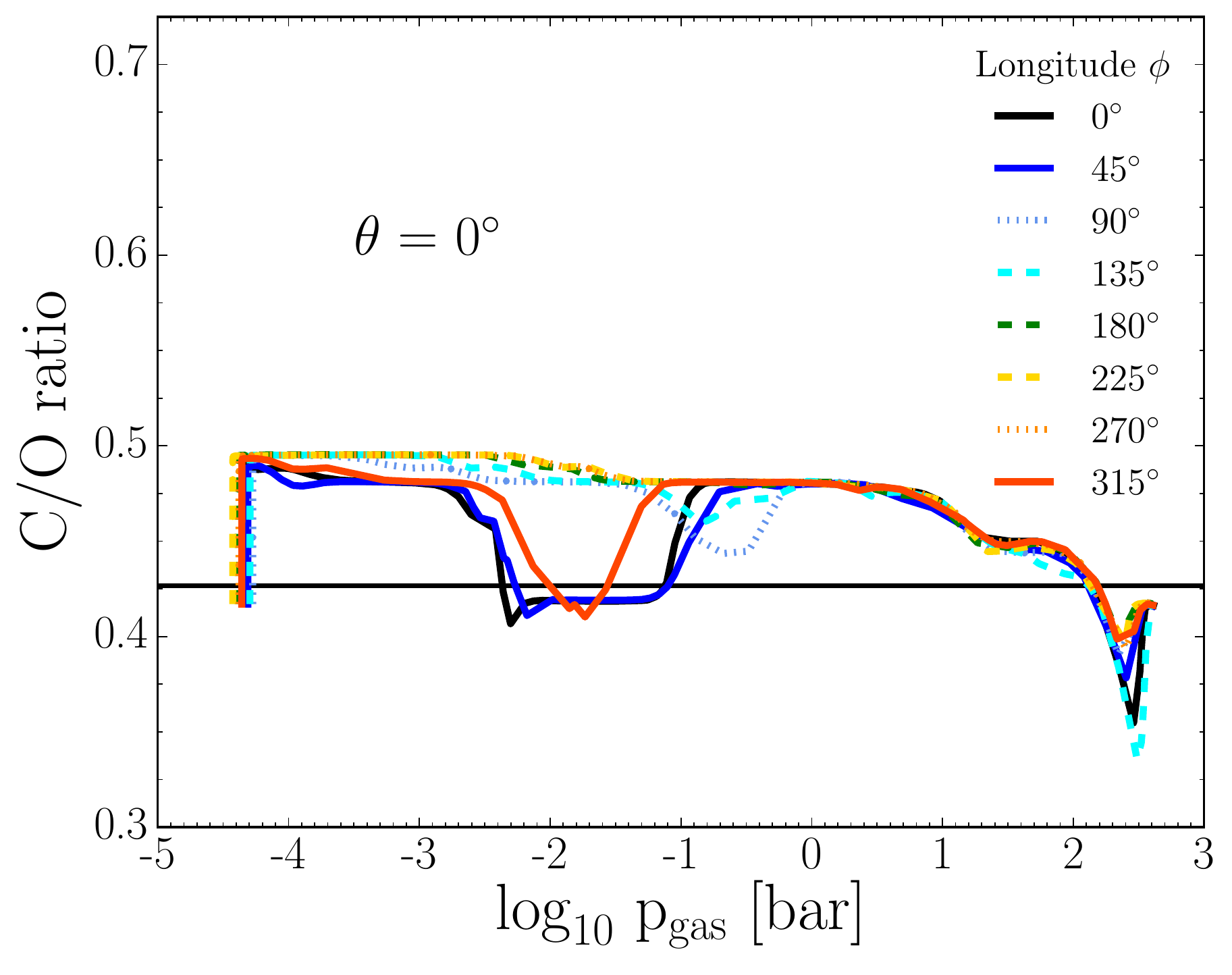}
\includegraphics[width=0.49\textwidth]{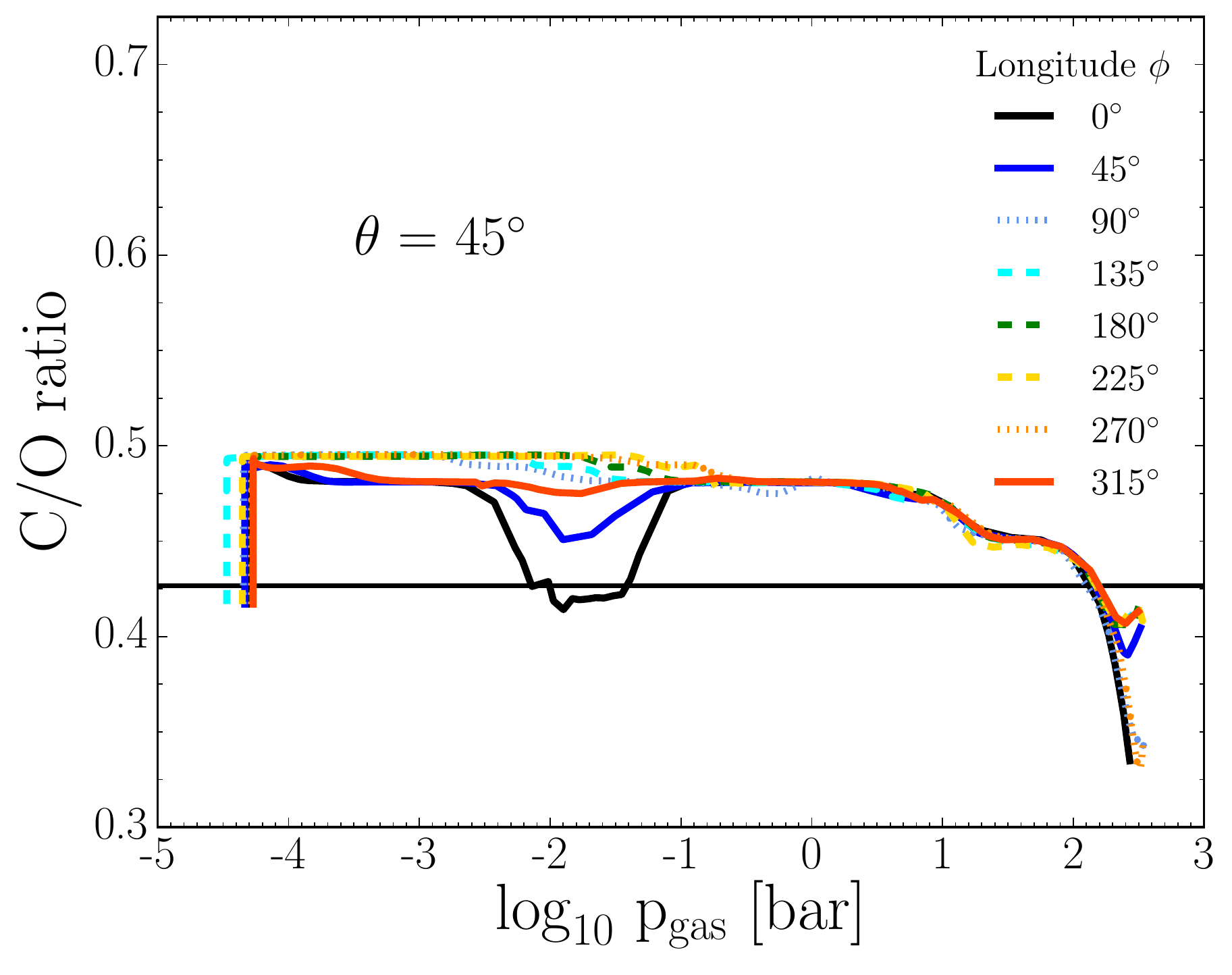}
\caption{Gas properties as a function of gas pressure for $\Delta$$\phi$ = +45\degr longitude intervals from the cloud formation process.
\textbf{Top Row:} Dust-to-gas ratio $\rho_{\rm d}$/$\rho_{\rm gas}$ at latitudes $\theta$ =  0\degr and 45\degr.
\textbf{Bottom Row:} C/O ratio at latitudes $\theta$ =  0\degr and 45\degr. 
Solid, dotted and dashed lines indicate dayside, day-night terminator and nightside profiles respectively.
The horizontal black line indicates solar C/O ratio.
Regions of decreasing $\rho_{\rm d}$/$\rho_{\rm gas}$ and C/O indicate cloud particle evaporation.}
\label{fig:thetagasprop}
\end{figure*}

The cloud formation process strongly depletes the local gas phase of elements, primarily through extremely efficient solid surface growth processes.
We consider the 8 elements that constitute the solid materials of the cloud particles, Mg, Si, Ti, O, Fe, Al, Ca and S assuming an initial solar element abundance ($\epsilon^{0}_{\rm x}$) for all layers.
Figure \ref{fig:thetaelement} shows the elemental abundance $\epsilon_{\rm x}$ (ratio to Hydrogen abundance) of each element as a function of pressure at each of the sample trajectories.
Depletion occurs due to the formation of solids made of the these element onto the cloud particle surface and by nucleation of new cloud particles.
Increase in element abundances correspond to regions of solid material evaporation.
Ti is depleted at the upper boundary due to immediate efficient nucleation.
For dayside profiles $\phi$ = 0\degr, 45\degr, 315\degr, $\theta$ = 0\degr, 45\degr, from 10$^{-4.5}$$\ldots$10$^{-3}$ bar, Mg, Ti, Si, Al and Fe are depleted by $\sim$1 order of magnitude while O, S and Ca are depleted by $\sim$ 10\%.
These profiles return to initial solar abundance values at $\sim$10$^{-2}$ bar where the solid material from the grain surface evaporates, returning elements to the gas phase.
O, Fe, Si S and Mg abundance can slightly overshoot solar abundance values at their respective maximums.
Elements are most heavily depleted in these profiles at $\sim$10$^{-1}$ bar where the most efficient surface growth occurs. 
Mg, Si, and Fe are depleted by $\sim$3 orders of magnitude, Ti by 8 orders of magnitude and Al by 5 orders of magnitude.
O, S and Ca are again depleted by $\sim$ 10\%.
Fe, Al, S and Ti return to solar abundance or slightly sub-solar abundance at the cloud base, where all materials have evaporated.
The cloud base is enriched in O, Ca, Mg and Si which are $\sim$50\% above solar abundance values.
For nightside and day-night terminator profiles $\phi$ = 90\degr, 135\degr, 180\degr, 225\degr, 270\degr $\theta$ = 0\degr, 45\degr, from 10$^{-4.5}$$\ldots$10$^{-2}$ bar, Mg, Ti, Si, Al and Fe are depleted by 4 to 8 orders of magnitude while O and S are depleted by $\sim$1 order of magnitude and Ca by $\sim$10\%.
The $\phi$ = 90\degr and 135\degr,  $\theta$ = 0\degr, 45\degr show a return to near initial abundance at $\sim$10$^{-0.5}$ bar from material evaporation.
Other nightside/terminator profiles gradually return to initial abundance from $\sim$1 bar to their respective cloud bases.
Again, O, Ca, Mg and Si are slightly above solar abundance at the cloud base.

We calculate the dust-to-gas ratio and the C/O ratio of our cloud structures.
Figure. \ref{fig:thetagasprop} shows the local dust-to-gas ratio of the sample trajectories at latitudes $\theta$ = 0\degr and 45\degr respectively.
Dayside profiles show increases and decreases in dust-to-gas ratio corresponding to regions of nucleation/growth and evaporation.
Nightside profiles show less cloud particle evaporation throughout the upper atmosphere, with only small changes in the dust-to-gas ratio which starts to drop off from $\sim$10$^{-2}$ bar.
Figure. \ref{fig:thetagasprop} shows the local gaseous C/O ratio of our sample trajectories at latitudes $\theta$ = 0\degr and 45\degr respectively.
These follow similar trends to the dust-to-gas ratio.
The C/O ratio lowers where evaporation of cloud particles releases their oxygen baring materials, replenishing the local gas phase,
The abundance of C is kept constant at $\epsilon^{0}_{\rm C}$ = 10$^{-3.45}$ (solar abundance) and is not affected by the formation of cloud particles in our model.
Dayside equatorial profiles show C/O ratio dips by $\sim$5\% below solar values at pressures of 10$^{-2.5}$$\ldots$10$^{-1}$ bar.
The $\phi$ = 0\degr, $\theta$ = 45\degr profile also shows a dip below solar values at similar pressure levels.
Apart from these localised regions of oxygen replenishment, the C/O ratio remains above solar values for the majority of the atmospheric profiles; except from the cloud base, which is enriched with oxygen by 10\%-20\% for all profiles. 

\subsection{Cloud property maps of HD 189733b}
\label{sec:interpolate}

\begin{figure*}
 \centering
\includegraphics[width=0.49\textwidth]{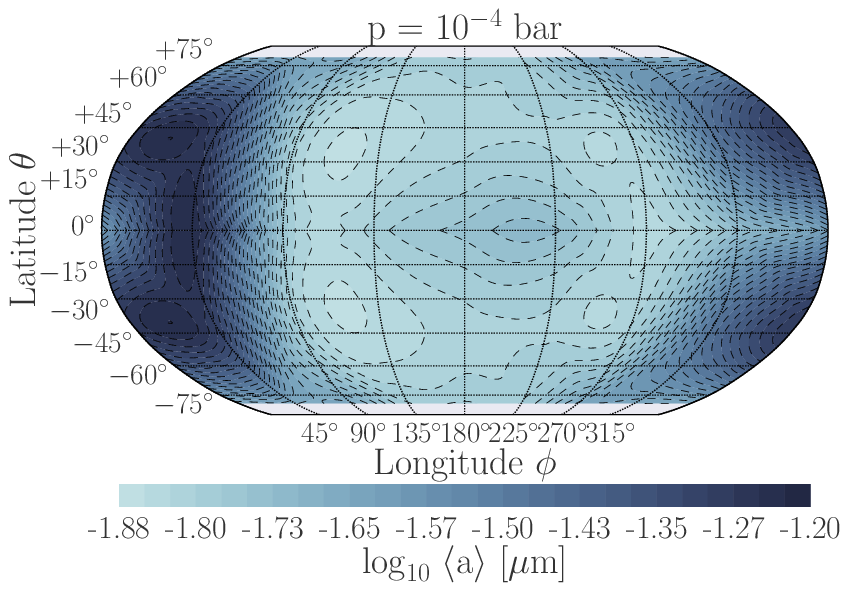}
\includegraphics[width=0.49\textwidth]{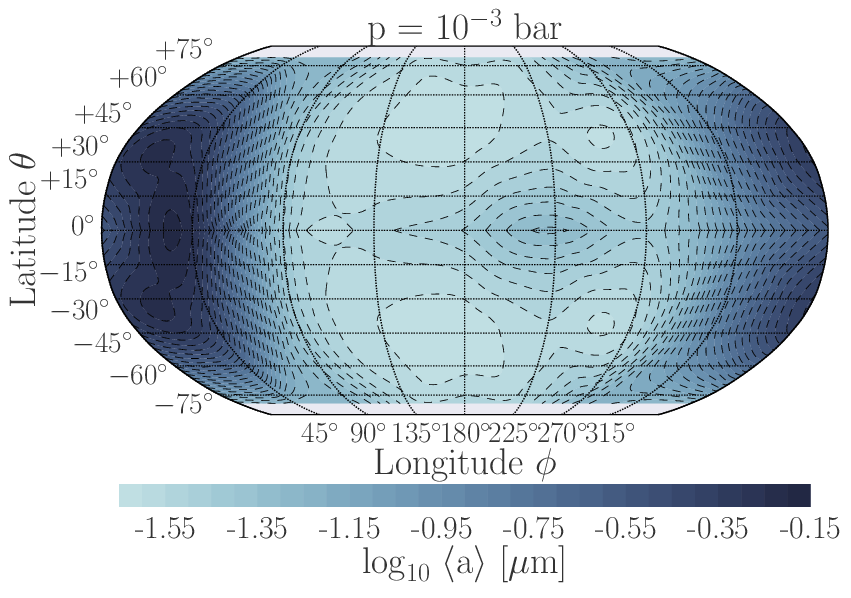}
\includegraphics[width=0.49\textwidth]{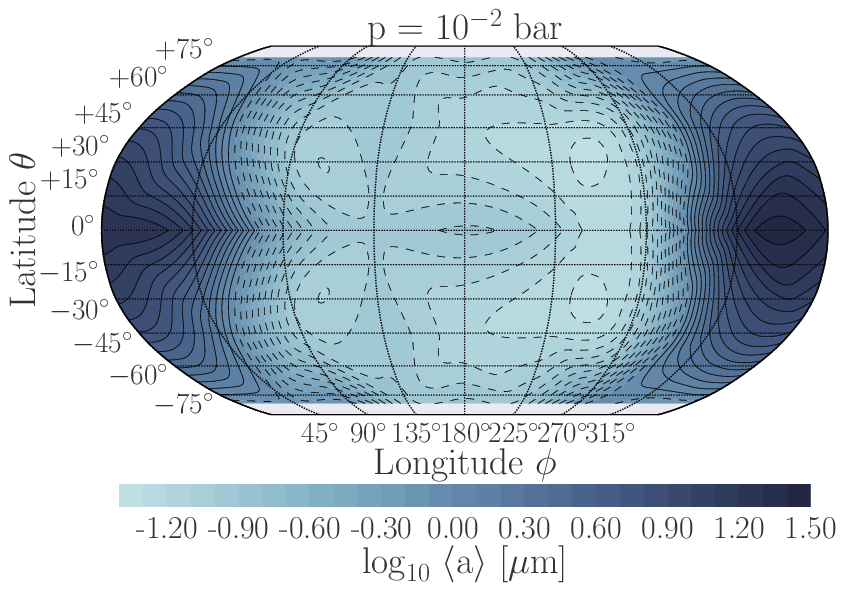}
\includegraphics[width=0.49\textwidth]{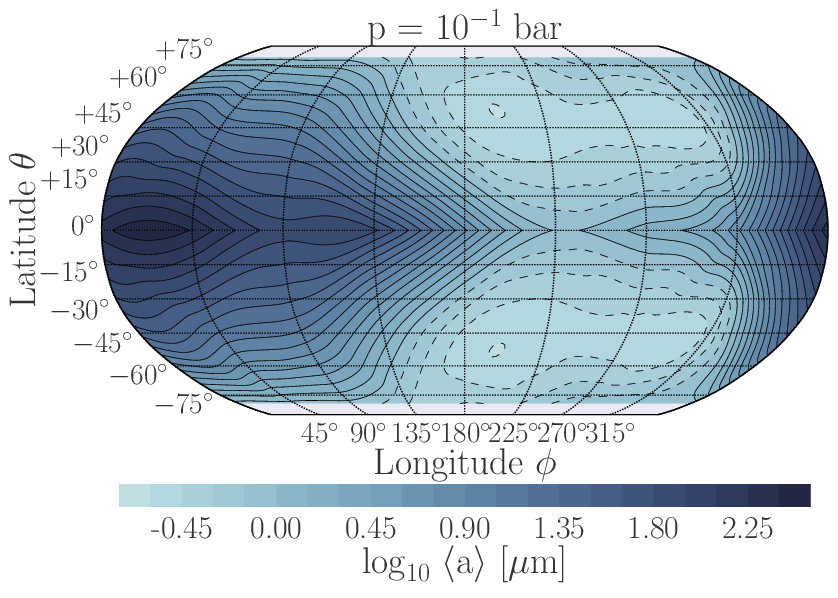}
\includegraphics[width=0.49\textwidth]{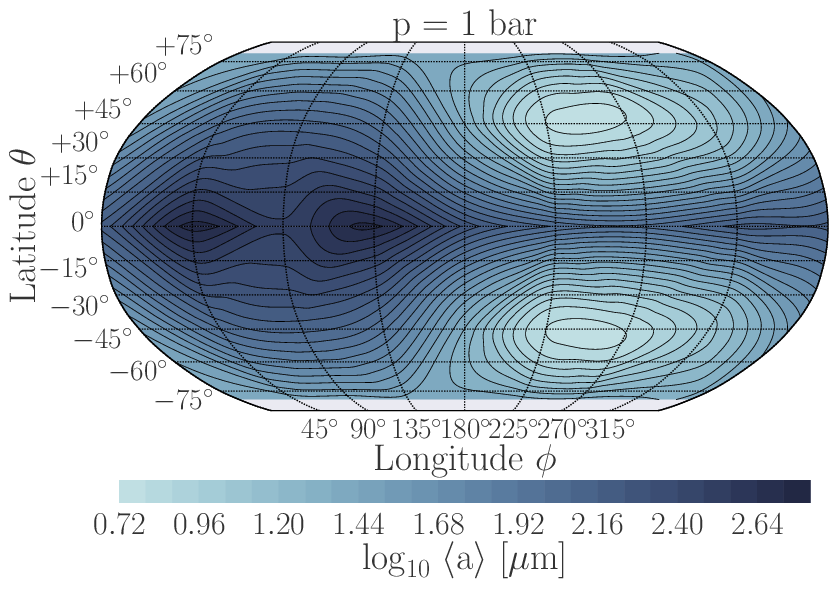}
\includegraphics[width=0.49\textwidth]{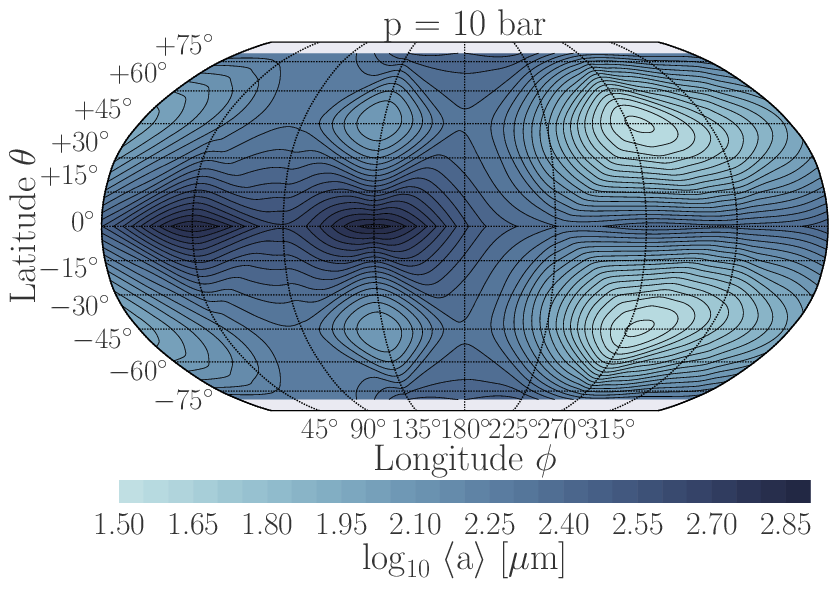}
\caption{Mean grain radius $\langle$a$\rangle$ [$\mu$m] interpolated across the globe (assuming latitudinal (north-south) symmetry) of HD 189733b.
\textbf{Top Row:} 10$^{-4}$ bar, 10$^{-3}$ bar.
\textbf{Middle Row:} 10$^{-2}$ bar, 10$^{-1}$ bar.
\textbf{Bottom Row:} 1 bar, 10 bar, respectively.
The sub-stellar point is at $\phi$ = 0\degr, $\theta$ = 0\degr. 
The largest mean grain radii of cloud particles are generally found on the dayside at all pressure levels. Consequently, cloud particles on the dayside are larger than their pressure level counterparts on the nightside.
Note: each plot contains a different colour scale.}
\label{fig:amap}
\end{figure*}

\begin{figure*}
 \centering
\includegraphics[width=0.49\textwidth]{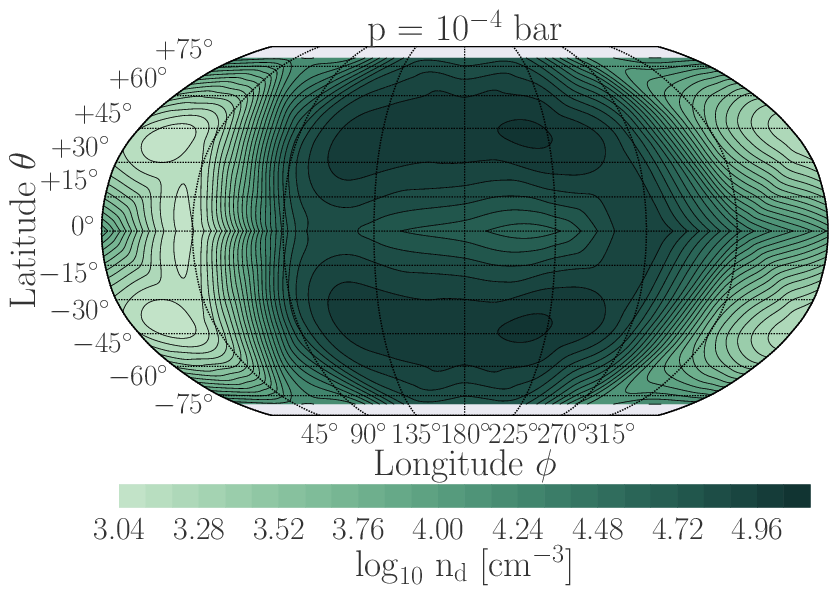}
\includegraphics[width=0.49\textwidth]{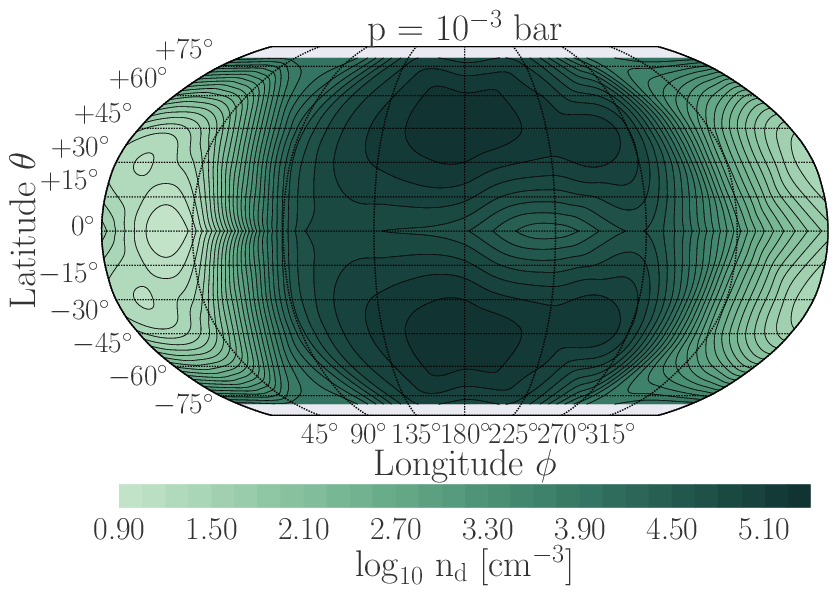}
\includegraphics[width=0.49\textwidth]{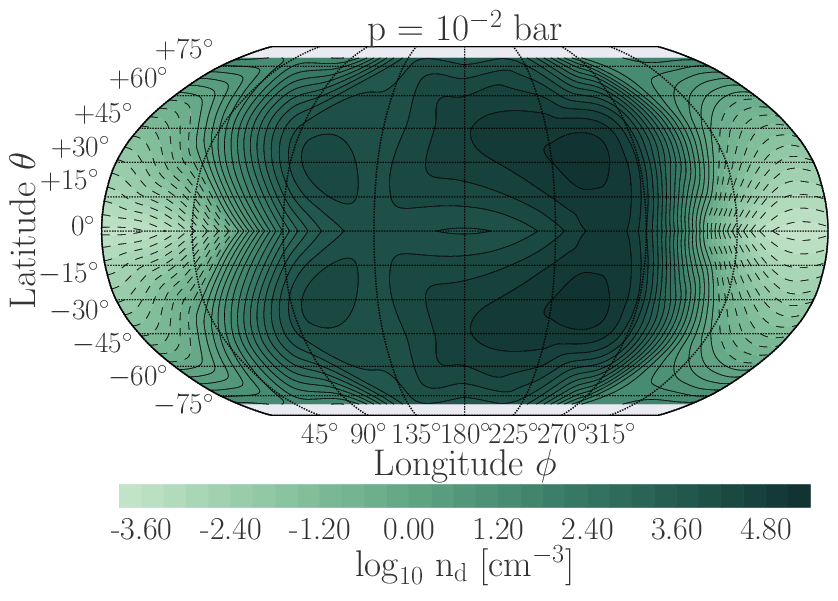}
\includegraphics[width=0.49\textwidth]{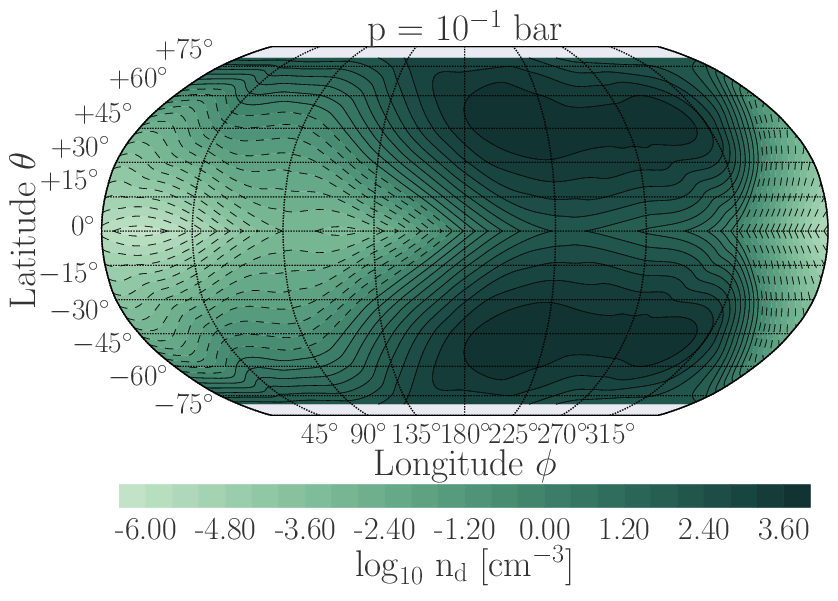}
\includegraphics[width=0.49\textwidth]{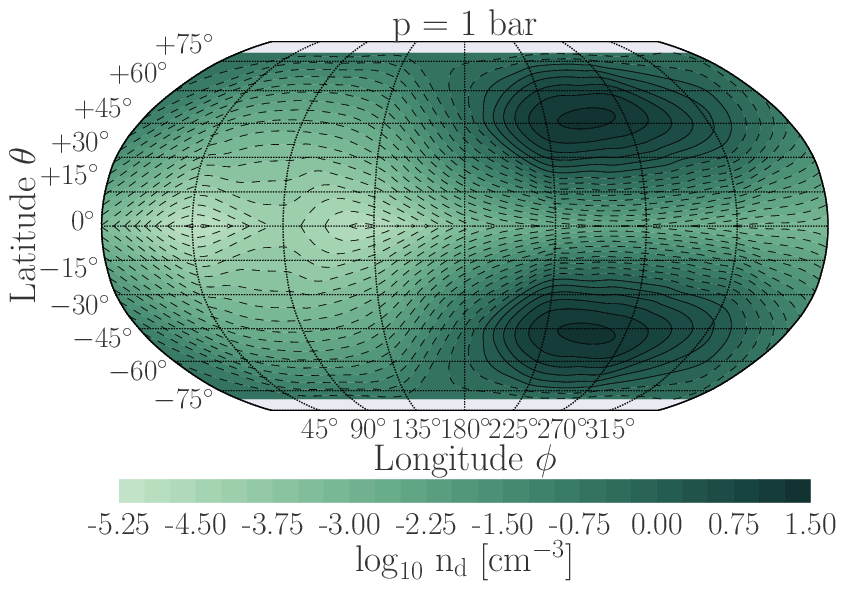}
\includegraphics[width=0.49\textwidth]{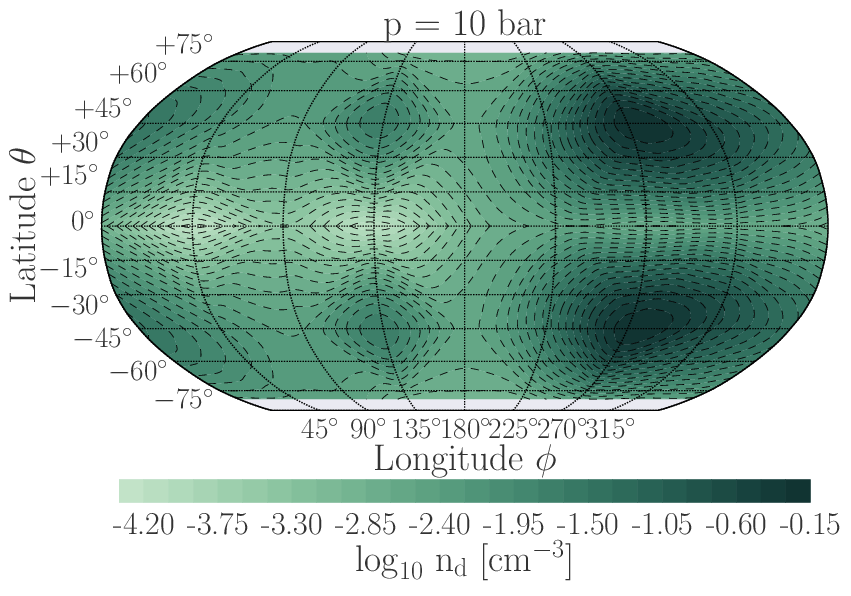}
\caption{Number density of cloud particles n$_{\rm d}$ [cm$^{-3}$] interpolated across the globe (assuming latitudinal (north-south) symmetry) of HD 189733b.
\textbf{Top Row:} 10$^{-4}$ bar, 10$^{-3}$ bar.
\textbf{Middle Row:} 10$^{-2}$ bar, 10$^{-1}$ bar.
\textbf{Bottom Row:} 1 bar, 10 bar, respectively.
The sub-stellar point is at $\phi$ = 0\degr, $\theta$ = 0\degr. 
The maximum number of cloud particles occurs on the nightside of the planet.
Note: each plot contains a different colour scale.}
\label{fig:ndmap}
\end{figure*}

Global cloud property maps of the atmosphere of HD 189733b enable the comparison between different atmospheric regions as a whole.
This has implications for interpreting cloudy observations which sample different atmospheric regions.
To produce a global cloud map of the atmosphere of HD 189733b we bi-cubically interpolate our cloud structure results across longitudes $\phi$ = 0\degr$\ldots$360\degr and latitudes $\theta$ = 0\degr$\ldots$80\degr. 
We include sample trajectories at latitude $\theta$ = 80\degr to interpolate to higher latitudes. 
We assume latitudinal (north-south) symmetry as thermodynamic conditions do not vary significantly from positive to negative latitudes \citep{Dobbs-Dixon2013}.
Figures \ref{fig:amap} and \ref{fig:ndmap} show maps of the mean grain radius $\langle$a$\rangle$ [$\mu$m] and grain number density n$_{\rm d}$ [cm$^{-3}$] at 10$^{-4}$ bar, 10$^{-3}$, 10$^{-2}$, 10$^{-1}$ bar, 1 bar and 10 bar. 
The mean grain radius and number density (along with material composition) are key values in calculating the wavelength dependent opacity of the clouds.
At all pressures there is a contrast between the dayside and nightside mean grain radii.
In the upper atmosphere ($<$ 10$^{-1}$ bar) the globally largest cloud particles occur on the dayside face.
In the deeper atmosphere ($>$ 10$^{-1}$ bar) global maximum of the cloud particle size may also occur in nightside regions.
The difference between the largest and smallest grain radii at each pressure can range from 1 to 2 orders of magnitude.
The highest difference in cloud particle sizes occurs at 10$^{-2}$ bar where grains on the dayside are 2 orders of magnitude bigger than the nightside.
This is due to dayside profiles undergoing very efficient cloud particle surface growth at $\sim$10$^{-2}$ bar, while nightside cloud particles do not grow as efficiently (Fig. \ref{fig:theta0}).  
There are also steep gradients between the maximum and minimum mean grain radius (indicating very rapid cloud formation processes) which occur at the terminator regions ($\phi$ = 90\degr, 270\degr) at pressure profiles $\lesssim$10$^{-1}$ bar.
The number density maps show similar differences between dayside and nightside profiles but with maximum values of number density generally occurring on the nightside of the planet at each pressure level.
Steep gradients in number density are also present at the terminator regions.
These results, taken as a whole, suggest that the wavelength dependent dust opacity significantly varies between the dayside and nightside.
The mixed local composition of the cloud particles will also have an effect on the dust opacity.

\section{Cloud opacities}
\label{sec:Opacity}

\begin{figure*}
 \centering
\includegraphics[width=0.42\textwidth]{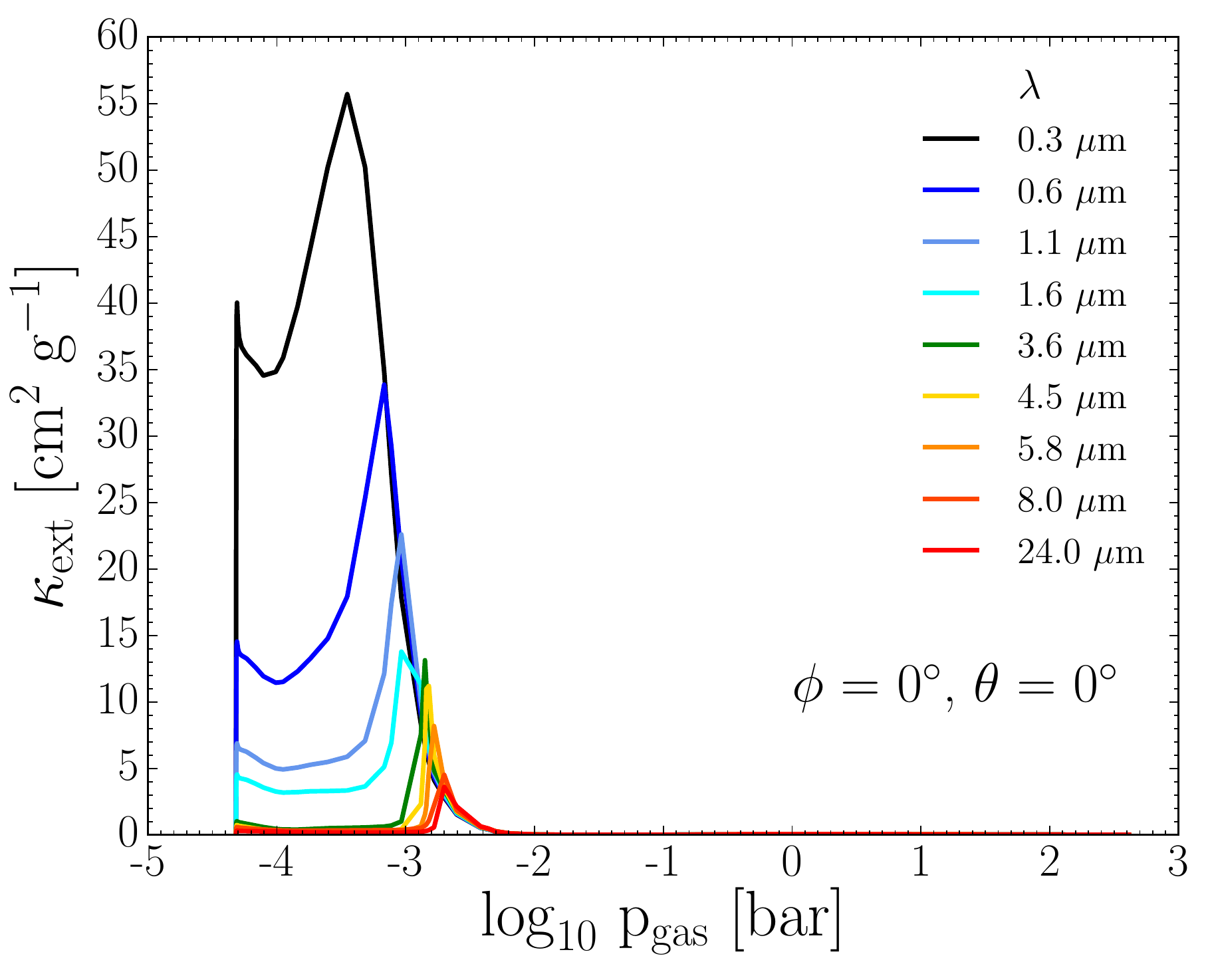}
\includegraphics[width=0.42\textwidth]{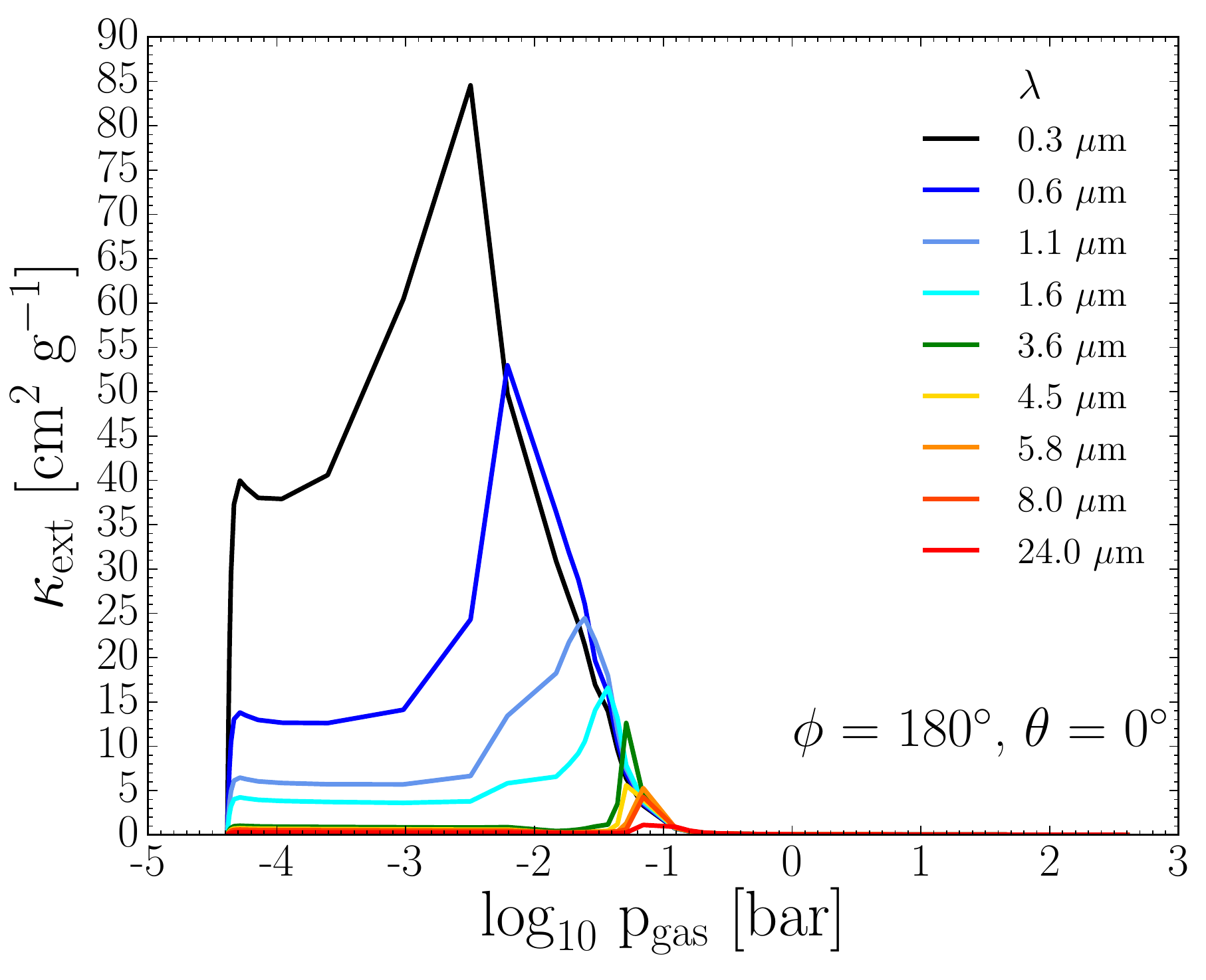}
\includegraphics[width=0.42\textwidth]{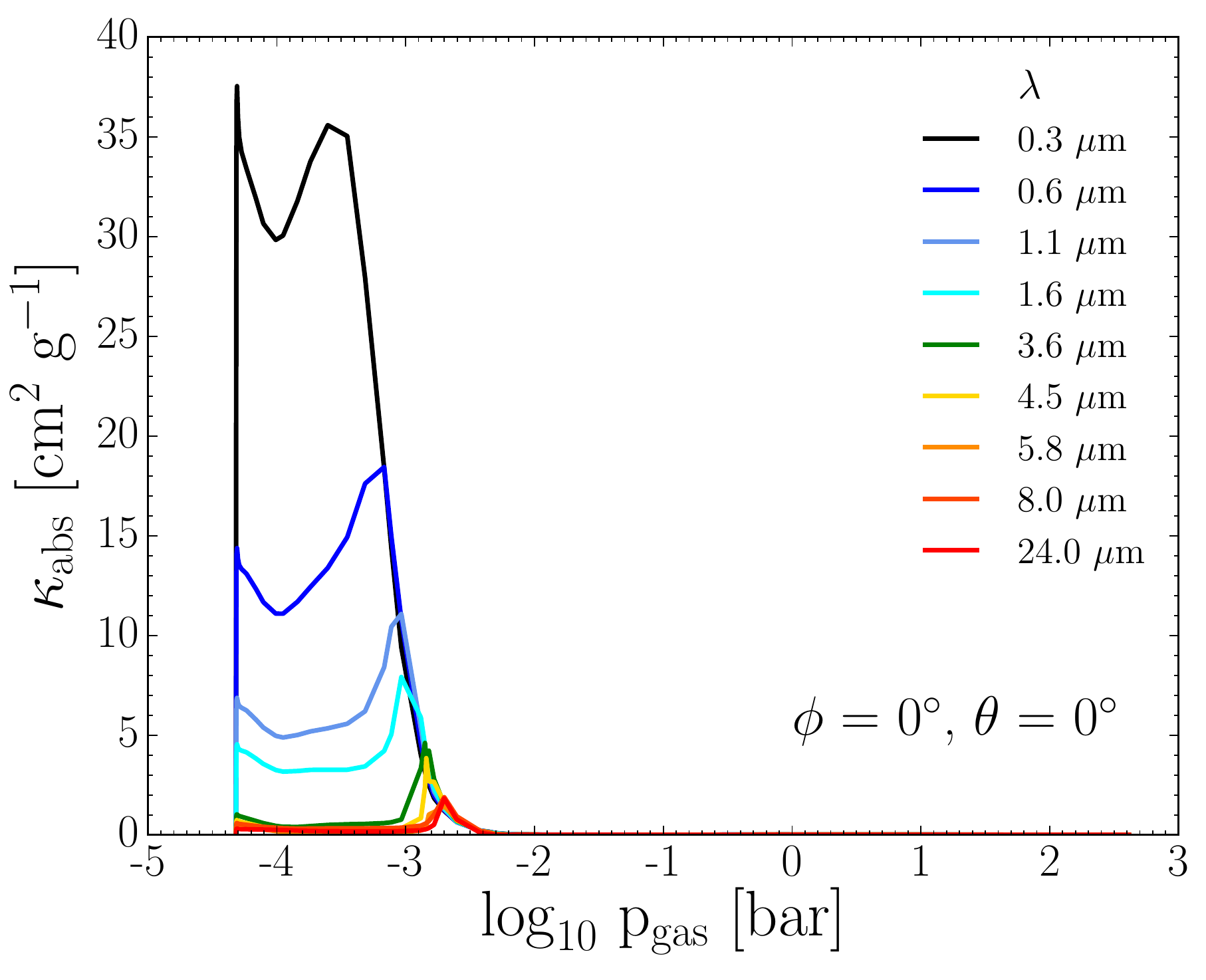}
\includegraphics[width=0.42\textwidth]{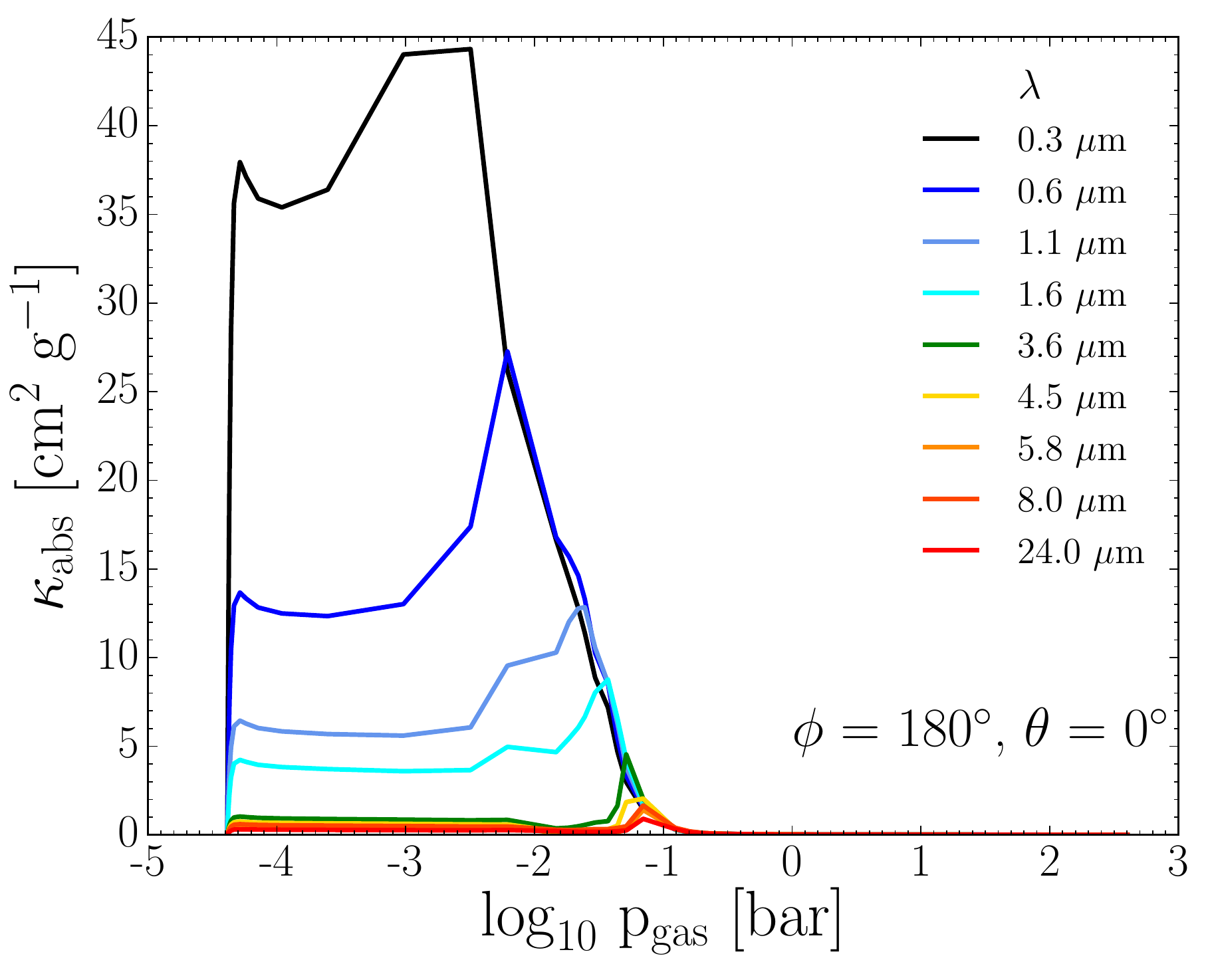}
\includegraphics[width=0.42\textwidth]{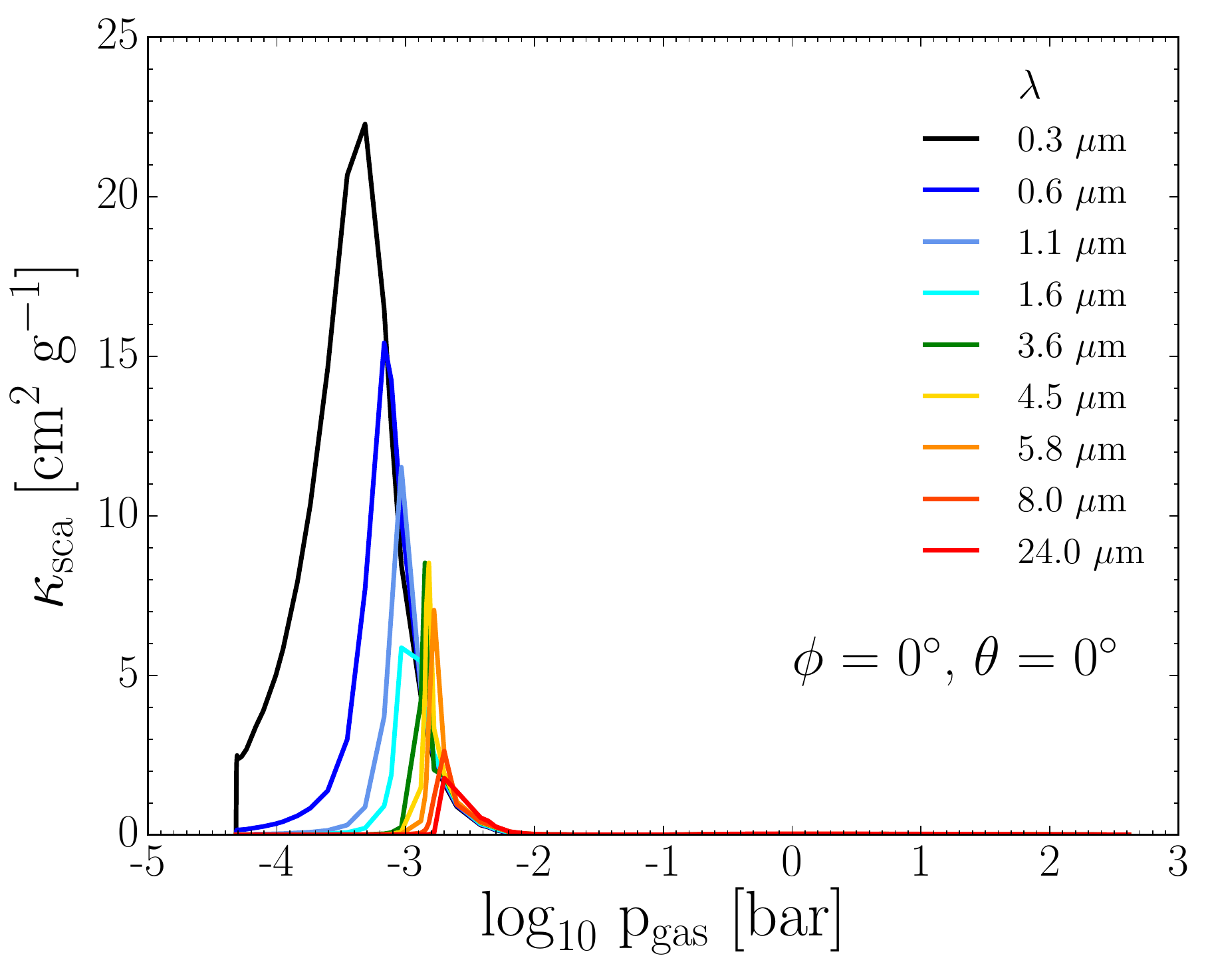}
\includegraphics[width=0.42\textwidth]{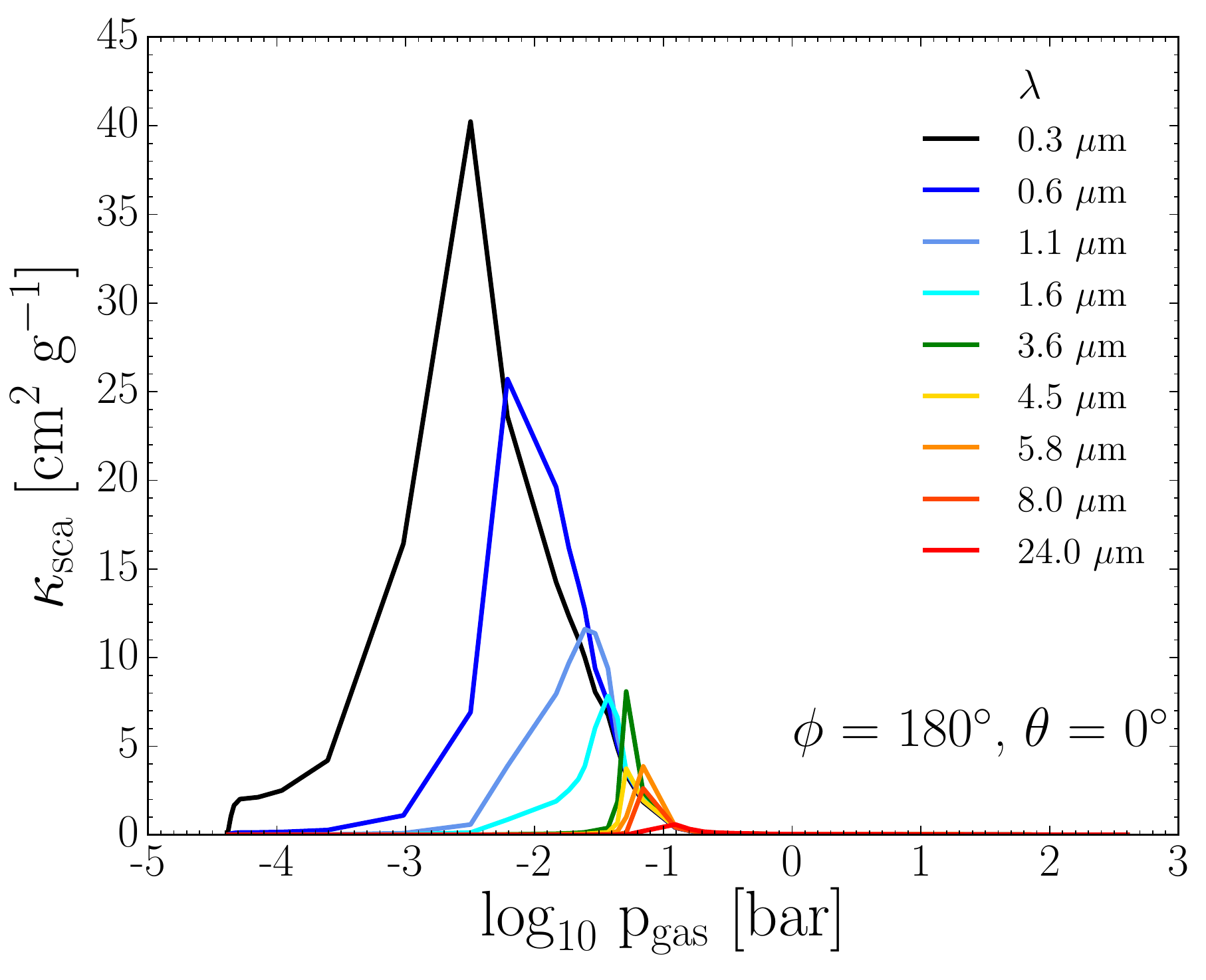}
\includegraphics[width=0.42\textwidth]{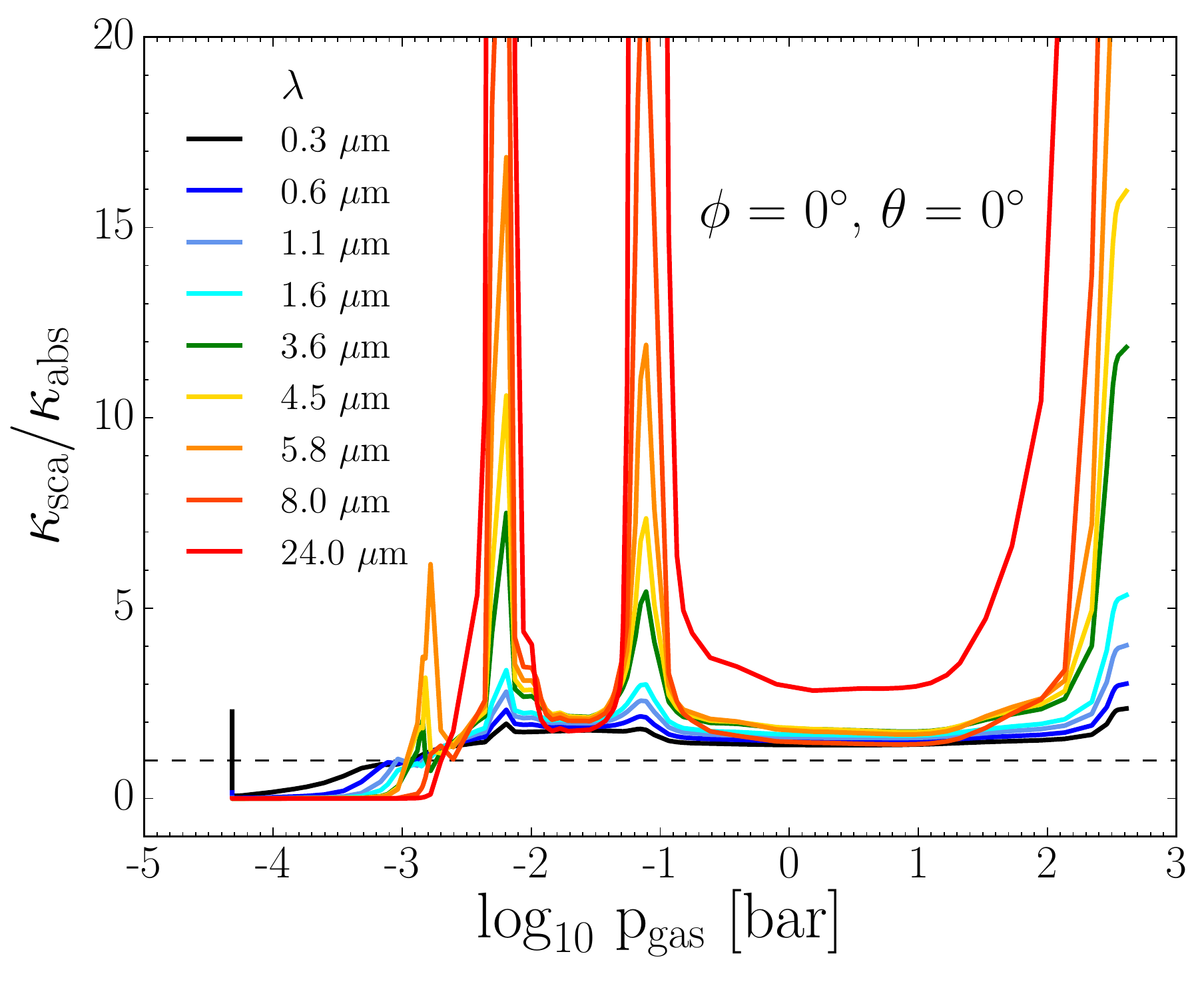}
\includegraphics[width=0.42\textwidth]{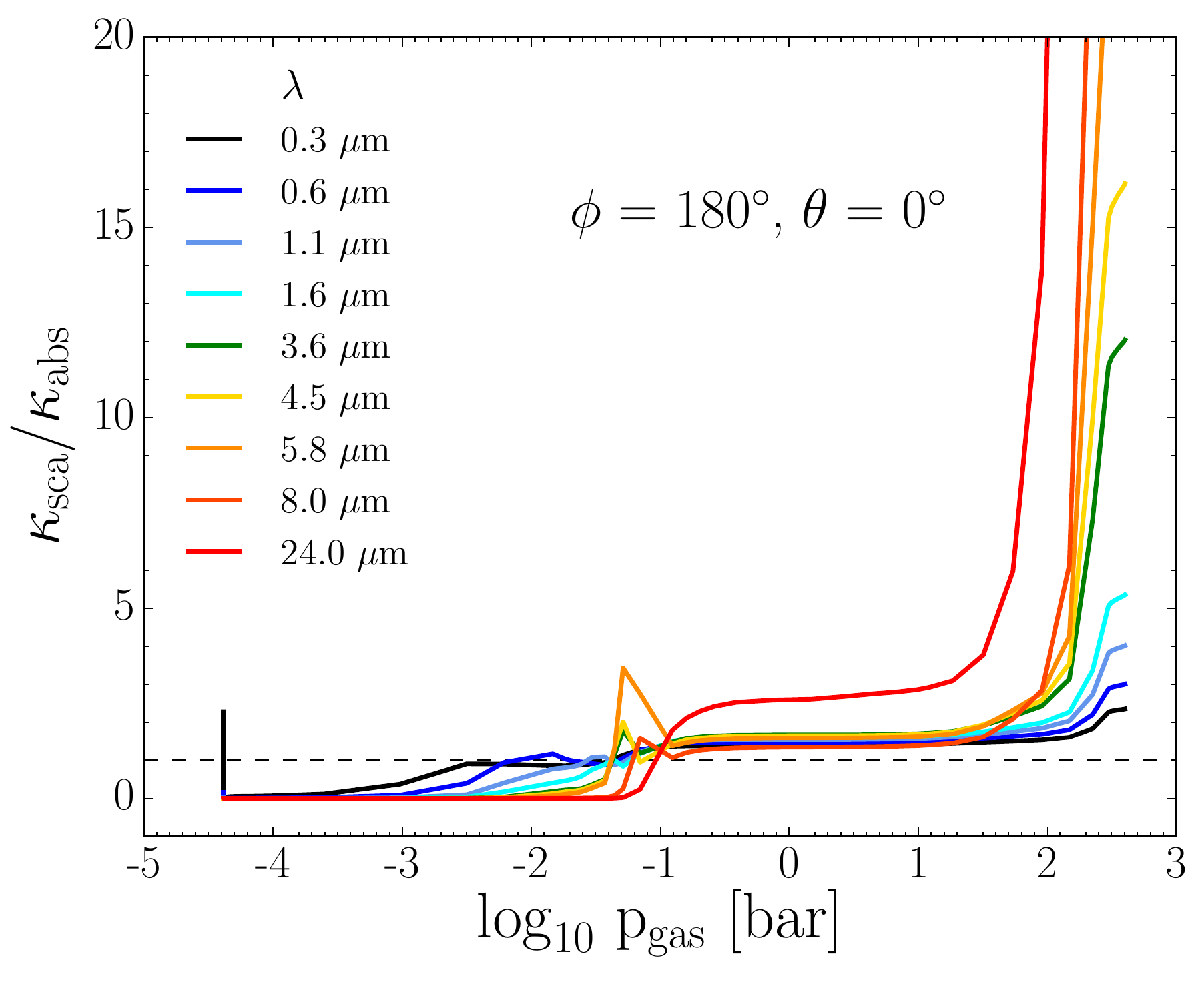}
\caption{Dust opacities $\kappa$ [cm$^{2}$ g$^{-1}$] for the cloud structure at $\phi$ = 0\degr, 180\degr, $\theta$ = 0\degr as a function of pressure. 
\textbf{Top Row:} Total extinction $\kappa_{\rm ext}$.
\textbf{Second Row:} Absorption $\kappa_{\rm abs}$.
\textbf{Third Row:} Scattering $\kappa_{\rm sca}$.
\textbf{Bottom Row:} Scattering to absorption ratio $\kappa_{\rm sca}$/$\kappa_{\rm abs}$. 
Bluer wavelengths are absorbed/scattered more efficiently than redder wavelengths in the upper atmosphere. 
The absorption dominates the total extinction in the upper regions and scattering in the deeper atmosphere. 
The spikes in scattering ratio at 10$^{-2}$ and 10$^{-1}$ bar correspond to Fe rich grain composition.}
\label{fig:opacity}
\end{figure*}

\begin{figure*}
 \centering
\includegraphics[width=0.49\textwidth]{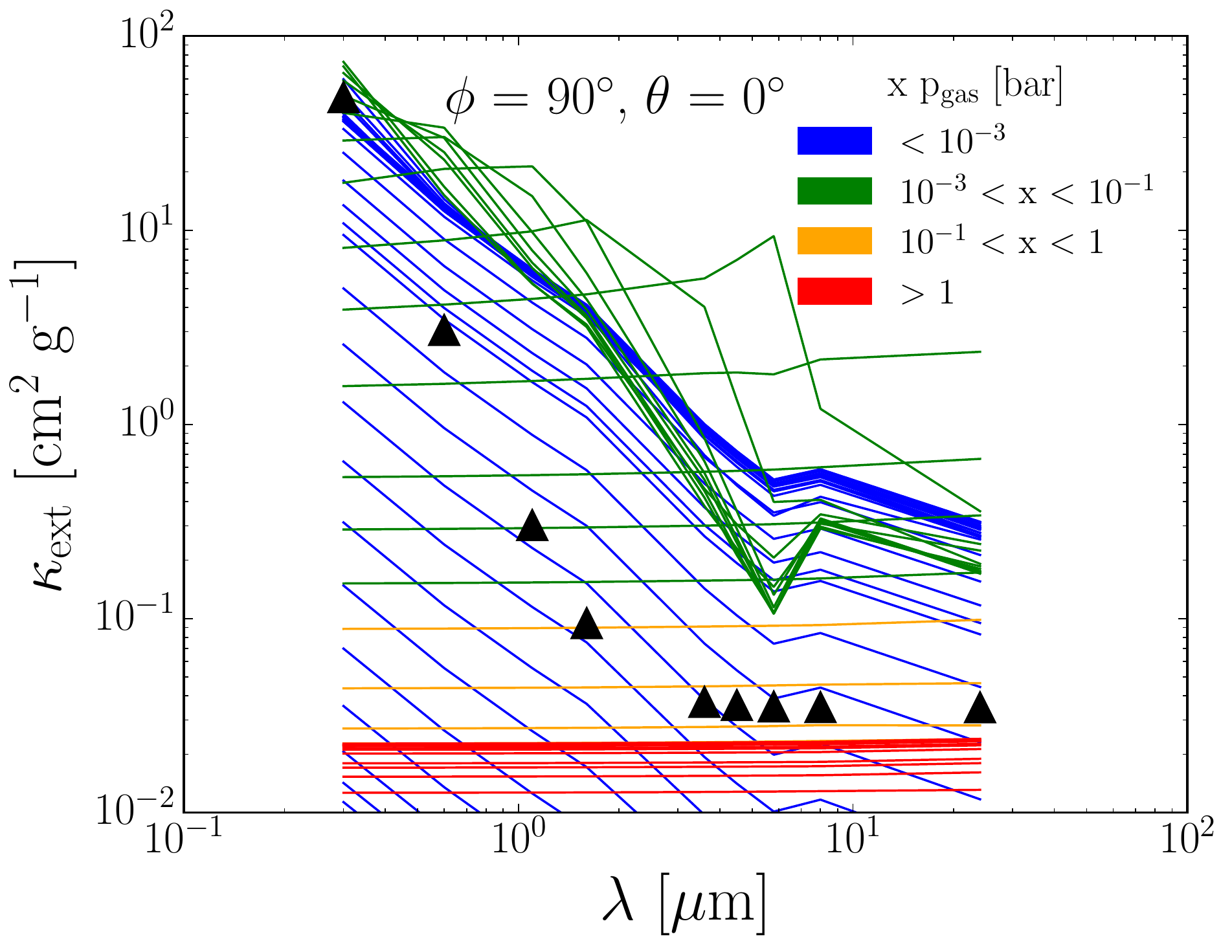}
\includegraphics[width=0.49\textwidth]{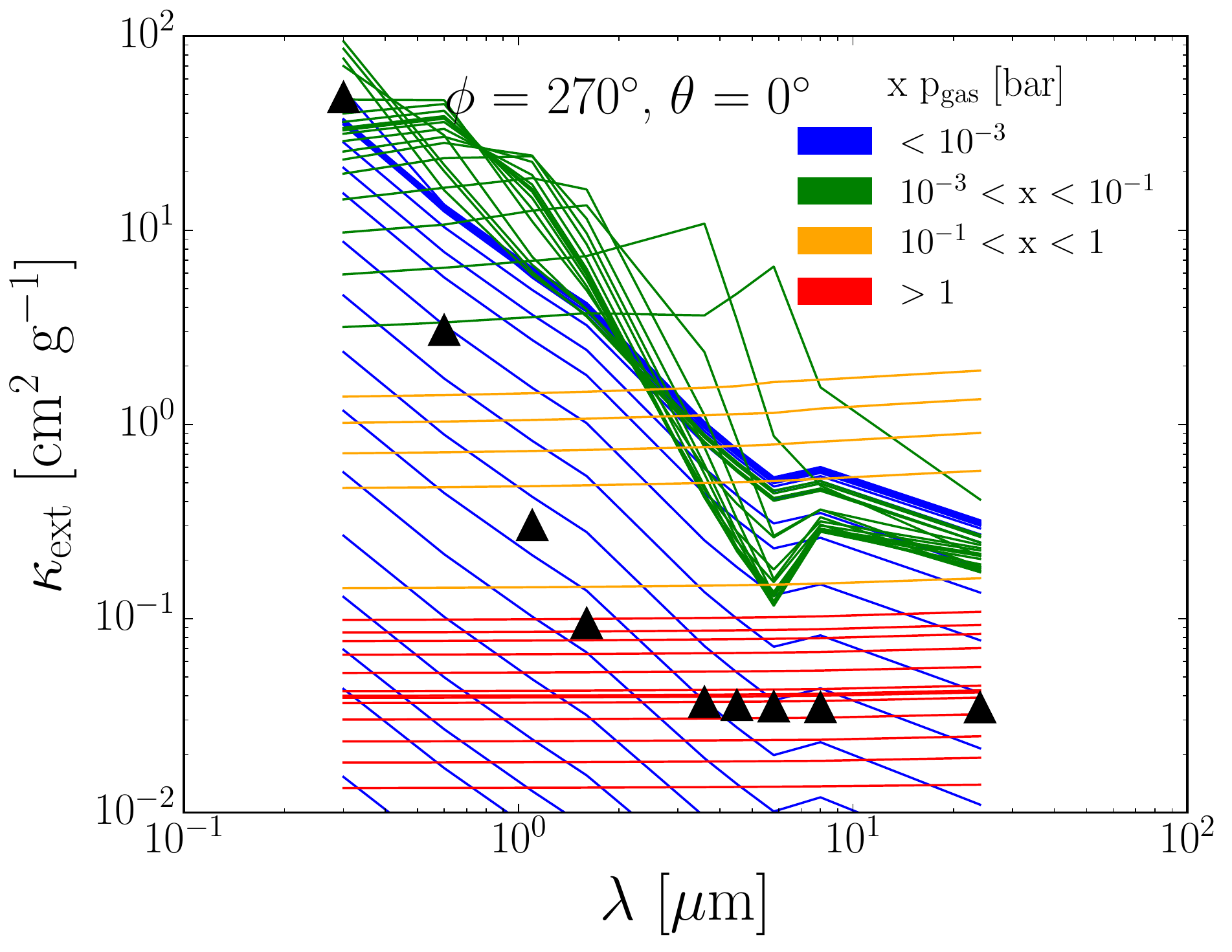}
\caption{Dust extinction $\kappa_{\rm ext}$ [cm$^{2}$ g$^{-1}$] pressure isobars at $\phi$ = 90\degr (Left), 270\degr (Right), $\theta$ = 0\degr (equatorial dayside-nightside terminator regions). 
Triangles denote the values from \citet{Dobbs-Dixon2013} assumed additional Rayleigh slope opacity $\kappa_{\rm ext}$ $\propto$ $\lambda^{-4}$. 
Isobars follow an absorption profile $\kappa_{\rm ext}$ $\propto$ $\lambda^{-2}$ at the upper atmosphere $\lesssim$10$^{-1}$ bar.
Isobars flatten deeper than $\gtrsim$10$^{-1}$ bar.
}
\label{fig:opacityisobar}
\end{figure*}

\begin{figure*}
 \centering
\includegraphics[width=0.49\textwidth]{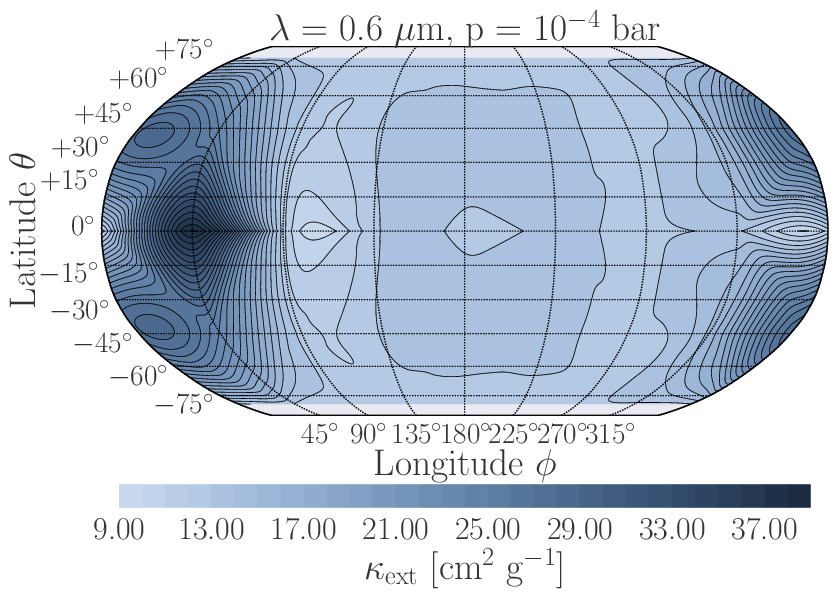}
\includegraphics[width=0.49\textwidth]{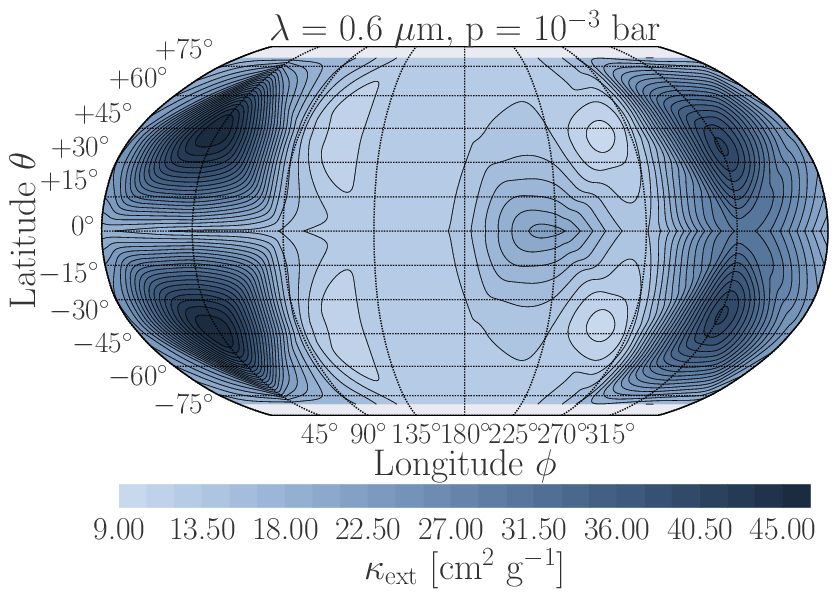}
\includegraphics[width=0.49\textwidth]{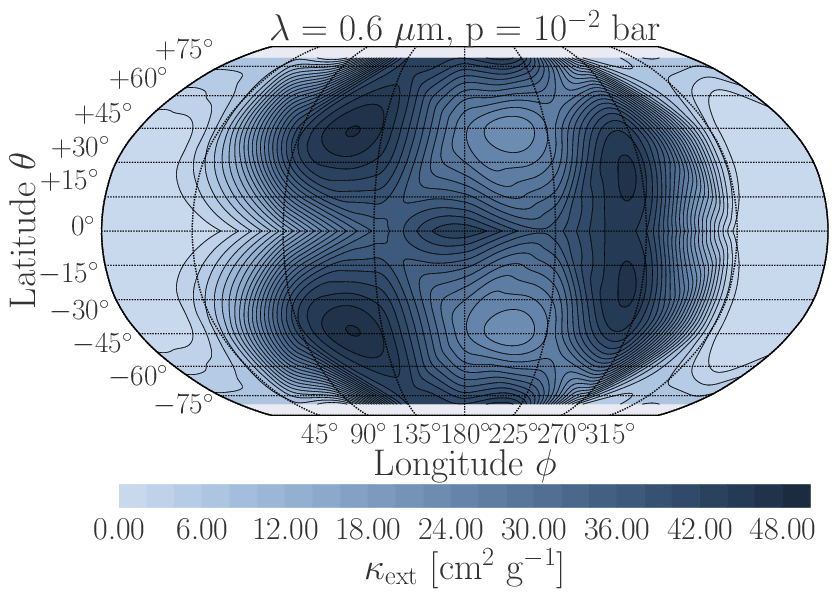}
\includegraphics[width=0.49\textwidth]{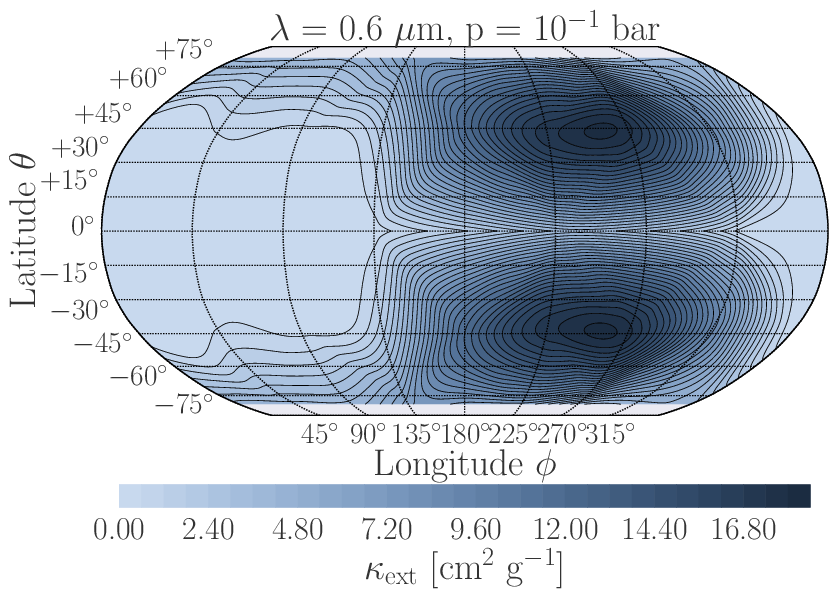}
\includegraphics[width=0.49\textwidth]{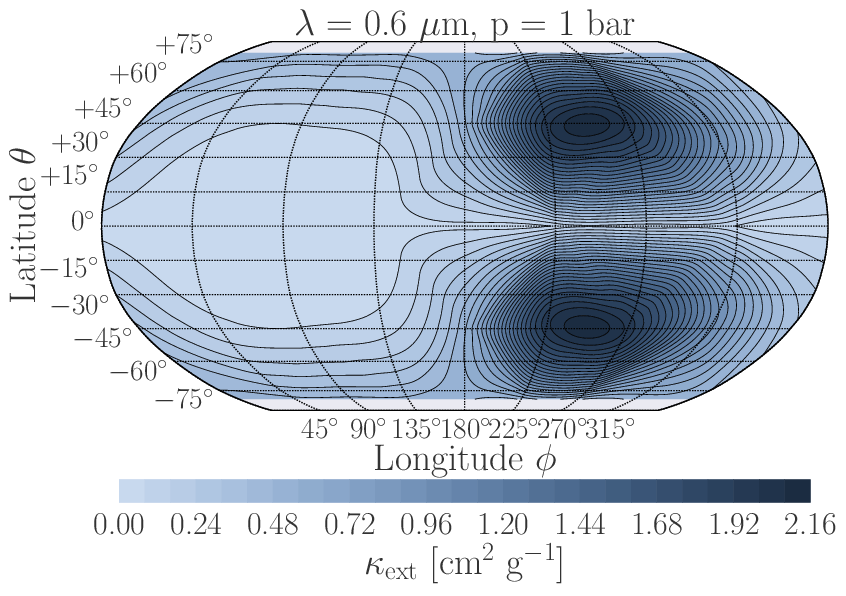}
\includegraphics[width=0.49\textwidth]{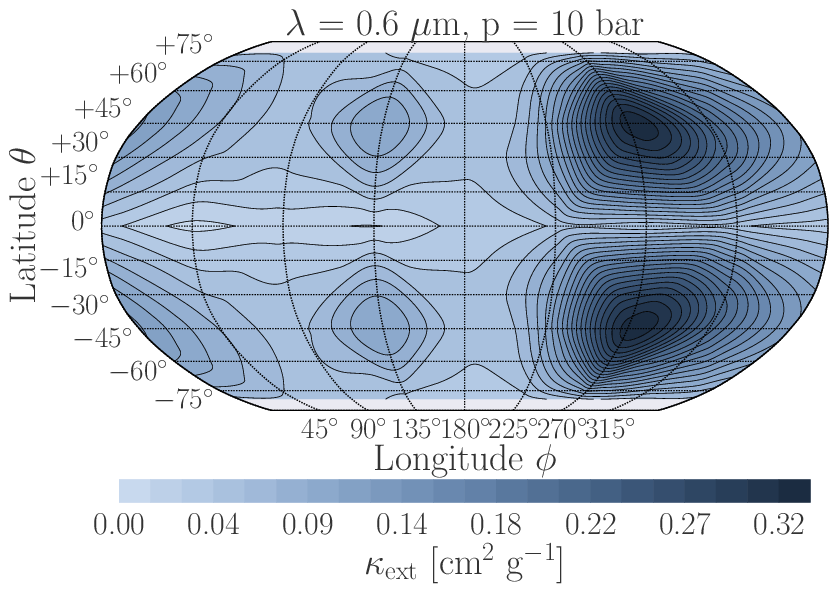} 
\caption{Dust $\kappa_{\rm ext}$ [cm$^{2}$ g$^{-1}$] opacity for 0.6 $\mu$m interpolated across the globe of HD 189733b assuming latitudinal (north-south) symmetry.
\textbf{Top Row:} 10$^{-4}$ bar, 10$^{-3}$ bar.
\textbf{Middle Row:} 10$^{-2}$ bar, 10$^{-1}$ bar.
\textbf{Bottom Row:} 1 bar, 10 bar, respectively.
We take 0.6 $\mu$m as representative of the optical wavelength extinction due to cloud.
The sub-stellar point is at $\phi$ = 0\degr, $\theta$ = 0\degr. 
The maximum extinction efficiency shifts from the dayside of the planet to the nightside with increasing depth. 
Deep in the atmosphere the most opaque region remains at $\sim$225\degr$\ldots$315\degr longitudes.
Note: The colour scale for each plot is different}
\label{fig:opacitymap}
\end{figure*}

\begin{figure*}
 \centering
\includegraphics[width=0.49\textwidth]{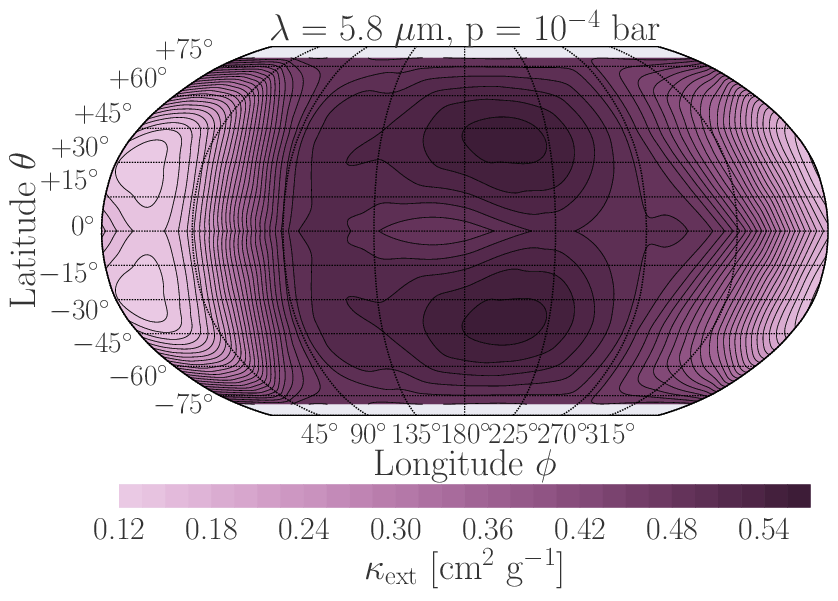}
\includegraphics[width=0.49\textwidth]{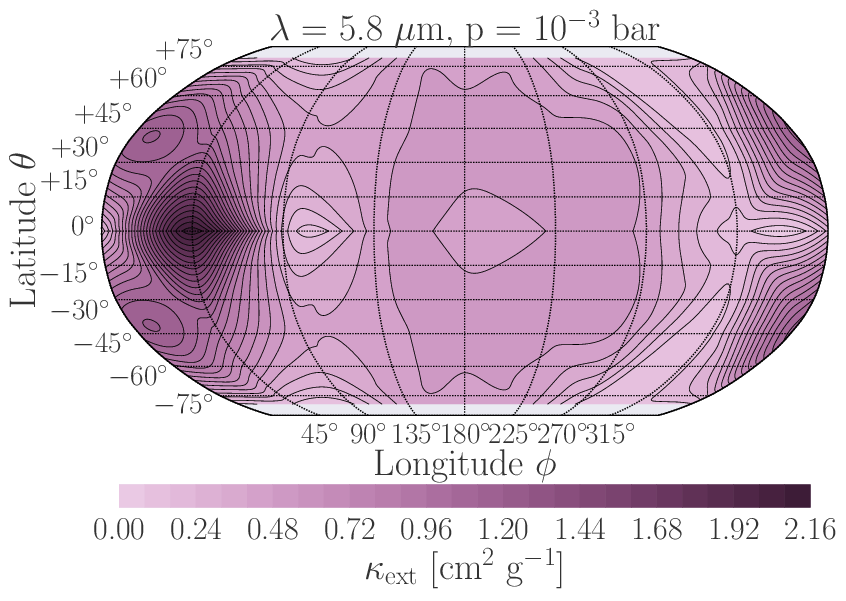}
\includegraphics[width=0.49\textwidth]{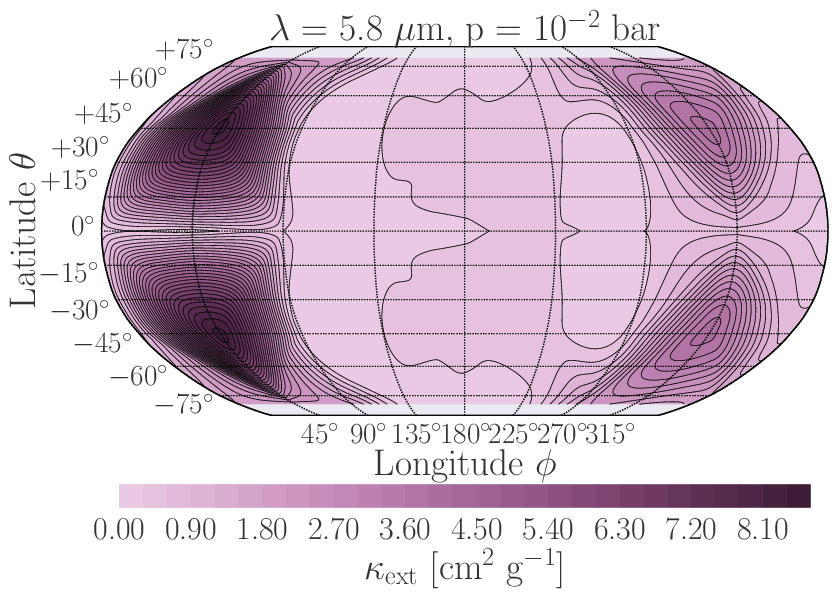}
\includegraphics[width=0.49\textwidth]{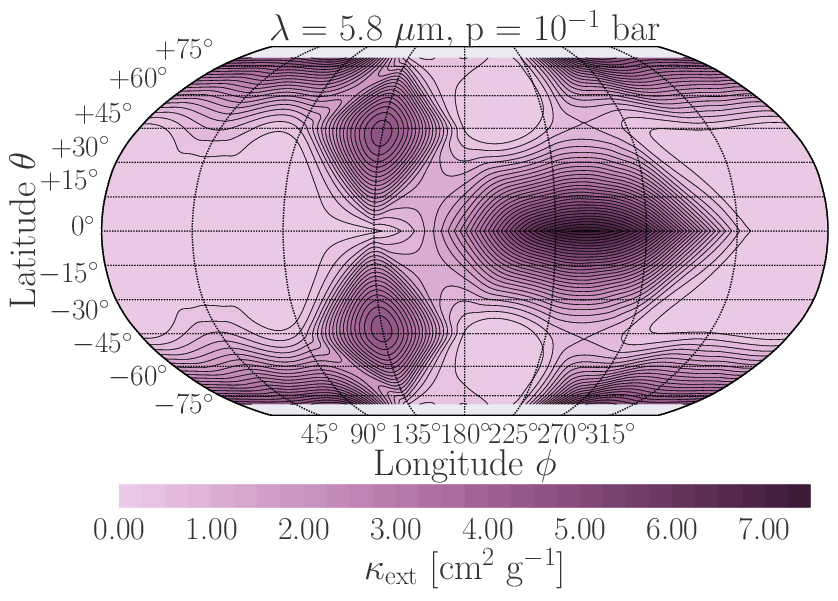}
\includegraphics[width=0.49\textwidth]{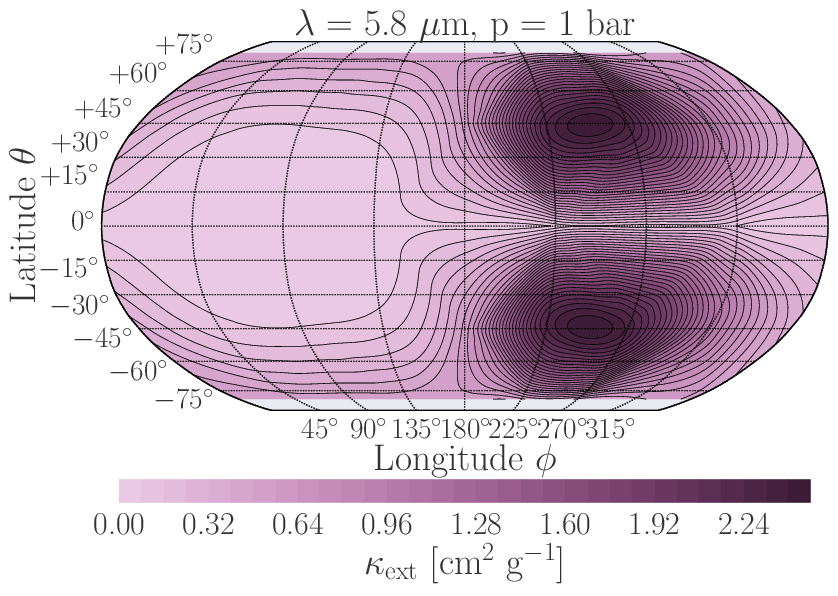}
\includegraphics[width=0.49\textwidth]{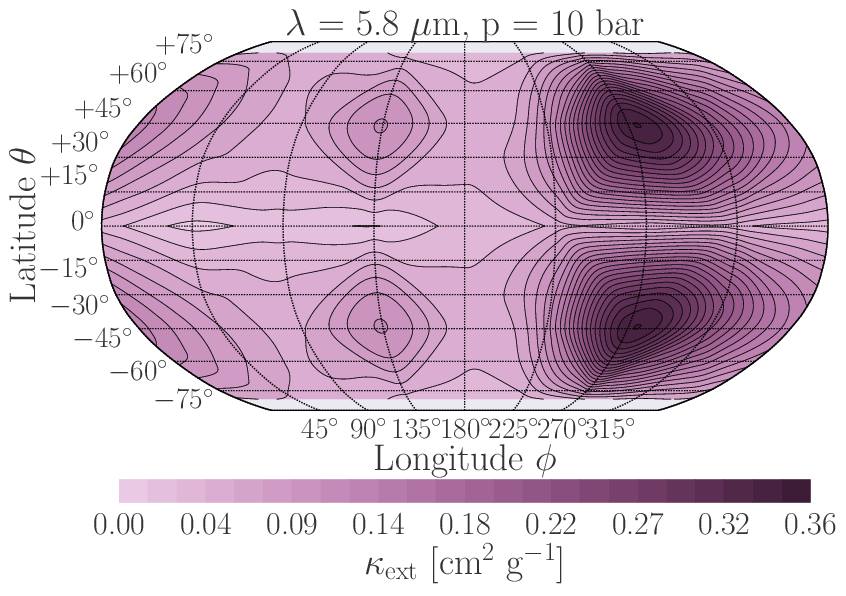}
\caption{Dust $\kappa_{\rm ext}$ [cm$^{2}$ g$^{-1}$] opacity for 5.8 $\mu$m interpolated across the globe of HD 189733b assuming latitudinal (north-south) symmetry. 
\textbf{Top Row:} 10$^{-4}$ bar, 10$^{-3}$ bar.
\textbf{Middle Row:} 10$^{-2}$ bar, 10$^{-1}$ bar.
\textbf{Bottom Row:} 1 bar, 10 bar, respectively.
We take 5.8 $\mu$m as representative of the infrared extinction due to cloud.
The sub-stellar point is at $\phi$ = 0\degr, $\theta$ = 0\degr. 
The maximum extinction efficiency shifts from the dayside of the planet to the nightside with increasing depth. 
Deep in the atmosphere the most opaque region remains at $\sim$225\degr$\ldots$315\degr longitudes.
Note: The colour scale for each plot is different.}
\label{fig:opacitymap2}
\end{figure*}

We calculated the scattering and absorption properties of our cloud particles and produce global cloud opacity maps of the HD 189733b atmosphere. 
For an illustration of the results, we present cloud opacity calculations at longitude $\phi$ = 0\degr and 180\degr at the equatorial region $\theta$ = 0\degr. 
Figure \ref{fig:opacity} demonstrates that the efficiency of light extinction depends on wavelength.
Bluer light is more heavily scattered/absorbed in the upper atmosphere compared to the infrared. 
For the $\phi$ = 0\degr trajectory the region of highest extinction extends from $\sim$10$^{-4}$$\ldots$10$^{-2}$ bar.
The $\phi$ = 180\degr trajectory extends from $\sim$10$^{-4}$$\ldots$1 bar.
The fraction of scattering to absorption is shown in Fig. \ref{fig:opacity}.
Absorption of the stellar light is the most efficient extinction mechanism in the upper atmosphere ($\sim$10$^{-3}$ bar) while scattering dominates the extinction in the deeper atmosphere $\gtrsim$10$^{-2.5}$ bar. 
There is a large increase in the scattering component corresponding to the Fe[s] rich ($\sim$50\%) grain regions at 10$^{-2}$ and 10$^{-1}$ bar in the $\phi$ = 0\degr, $\theta$ = 0\degr trajectory.
Between these Fe[s] regions a $\sim$80\% Al$_{2}$O$_{3}$[s] grain composition region occurs which is more transparent in our wavelength range than the Fe[s] surrounding regions.
For a discussion of these results, we refer to Sect. \ref{sec:Discussion}. 
From the upper atmosphere down to $\sim$10$^{-3}$ bar the extinction profile follows an absorption profile of $\kappa_{\rm ext}$ $\propto$ $\lambda^{-2}$.
A transition region from $\sim$10$^{-2}$ to 10$^{-1}$ bar occurs where profiles gradually flatten from optical to infrared wavelengths. 
This is due to the wavelength dependent extinction efficiency of cloud particles with blue light more absorbed/scattered by small, upper atmosphere grains and red light by larger, deeper atmosphere grains.
In addition, the grain composition fraction of highly opaque solid species such as Fe[s] increases from $\sim$10\% to $\sim$20\% deeper in the atmosphere ($>$10$^{-1.5}$ bar).  

A wavelength dependent `cloud opacity' global map can be produced by interpolation assuming latitudinal (north-south) symmetry of the cloud properties sample trajectories.
We choose the 0.6$\mu$m and 5.8$\mu$m wavelengths as representative of optical and infrared wavelength extinction respectively. 
Figures \ref{fig:opacitymap} and \ref{fig:opacitymap2} show the bi-cubic interpolated cloud particle extinction efficiency at 0.6$\mu$m and 5.8$\mu$m wavelengths across the globe from 10$^{-4}$$\ldots$10 bar. 
The optical 0.6$\mu$m cloud map shows the maximum extinction regions migrate from dayside to nightside regions with increasing pressure. 
The 5.8$\mu$m infrared cloud map shows the maximum extinction region also undergoes a shift, from nightside at 10$^{-4}$ bar, to dayside at 10$^{-3}$$\ldots$10$^{-2}$ bar, to nightside for $<$10$^{-1}$ bar. 
The most efficient extinction region deeper in the atmosphere (10$^{-1}$$\ldots$10 bar) for both maps occurs at longitudes $\phi$ $\sim$225\degr$\ldots$315\degr which correspond to the coldest parts of the atmosphere (Fig. \ref{fig:inputq}). 

\subsection{Reflection and sparkling of cloud particles}
\label{sec:Refspark}

The visible appearance of our cloud particles can be estimated from their scattering properties.
By estimating the relative fraction of scattered light in red, blue and green colour wavelengths, a rough RGB scale can be constructed to visualise a sparkling/reflection colour.
We use the $\kappa_{\rm sca}$ results from Sect. \ref{sec:Opacity} for cloud layers at upper regions in the atmosphere $\sim$10$^{-4.5}$$\ldots$10$^{-2}$ bar at the sub-stellar point $\phi$ = 0\degr, $\theta$ = 0\degr.
This profile was used as being best comparable to observations from secondary transit (occultation) observations in which the albedo (or colour) of the atmosphere can be determined (\citet{Evans2013}).
We linearly interpolated the $\kappa_{\rm sca}$ to proxy RGB wavelengths and calculated their relative red, blue and green scattering fractions.
This results in a deep midnight blue colour.
Deeper in the atmosphere, $>$10$^{-3}$ bar, cloud particles scatter red, blue and green light in more equal fractions which results in redder and grayer cloud particle appearance. 
\citet{Helling2009b} suggest that amorphous cloud particles can re-arrange themselves into crystalline lattice structures as they gravitationally settle.
This would allow light to refract and reflect inside the cloud particle volume producing a sparkle.
The sparkling colour is likely to be a similar colour to the reflected light.

\subsection{Reflectance of cloud particles}

\begin{figure*}
 \centering
\includegraphics[width=0.49\textwidth]{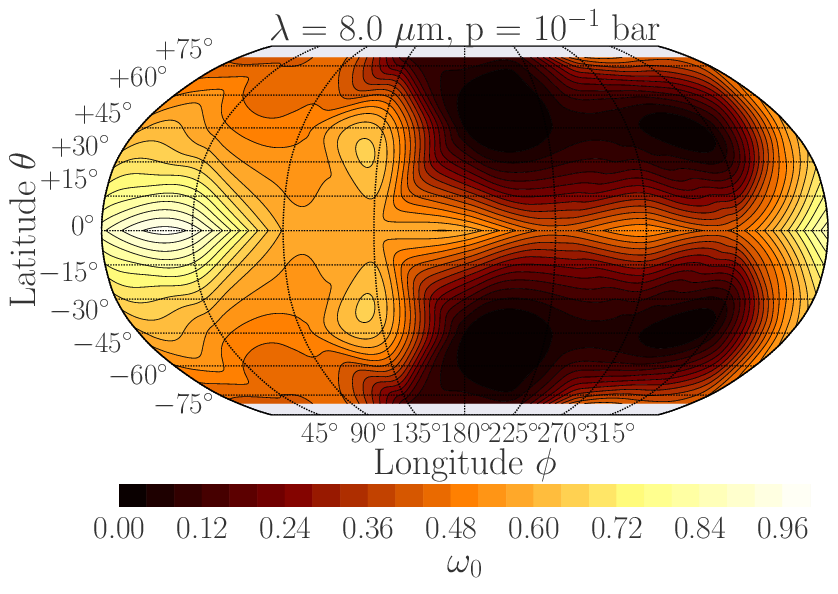}
\includegraphics[width=0.49\textwidth]{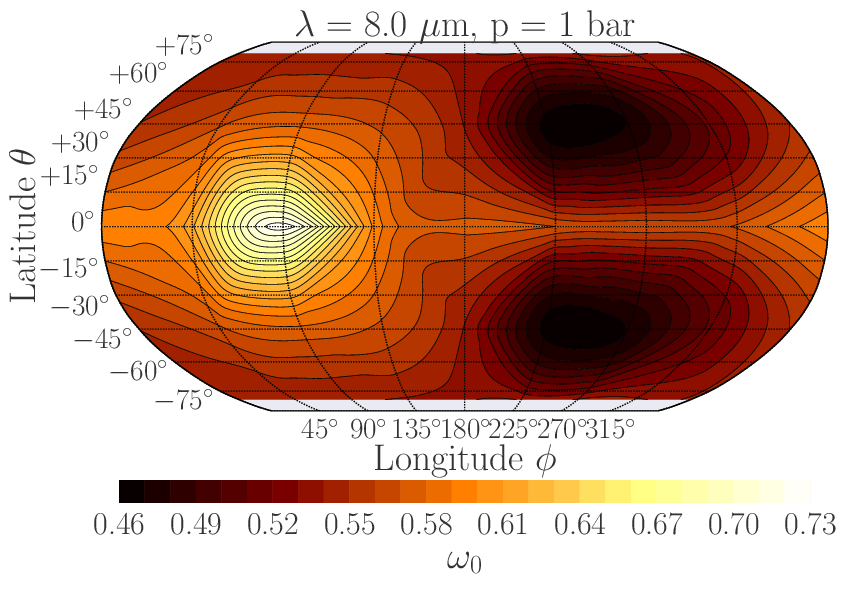}
\caption{Interpolated single scattering albedo $\omega_{0}$ (Eq. \eqref{eq:ssa}) at a wavelength of 8$\mu$m at 10$^{-1}$ bar (Left), 1 bar (Right) for our cloud properties.
The maximum of the reflectance occurs approximately from $\phi$ = 20\degr$\ldots$135\degr longitude.
These maximum reflectance regions are consistent with the 8$\mu$m \textit{Spitzer} flux maps of HD 189733b from \citet{Knutson2007}.
Note: the sub-stellar point in the diagrams is located at $\phi$ = 0\degr, $\theta$ = 0\degr, at the left side of the figures.}
\label{fig:ssa}
\end{figure*}

Cloud particles have a large effect on the observable properties by reflecting incident light on the atmosphere back into space.
We estimate the reflectance of the cloud particles by calculating the pressure dependent \textit{single scattering Albedo} $\omega_{0}$ \citep{Bohren2006} of the cloud particles defined as

\begin{equation}
\label{eq:ssa}
 \omega_{0} = \frac{\kappa_{\rm sca}}{\kappa_{\rm abs} + \kappa_{\rm sca}}
\end{equation}

This ratio indicates where the cloud particles extinction is dominated by scattering ($\sim$1) or absorption ($\sim$0).
Figure \ref{fig:ssa} shows maps of the calculated single scattering Albedo for 8$\mu$m at 10$^{-1}$ bar and 1 bar. 
The maximum of the reflectance occurs in the approximate longitude range 0\degr$\ldots$135\degr at the equatorial region.
From these maps we expect the peak of the 8$\mu$m albedo to occur from 20\degr$\ldots$40\degr East of the sub-stellar point.

\section{Discussion}
\label{sec:Discussion}

\begin{figure*}
\centering
 \includegraphics[width=0.75\textwidth]{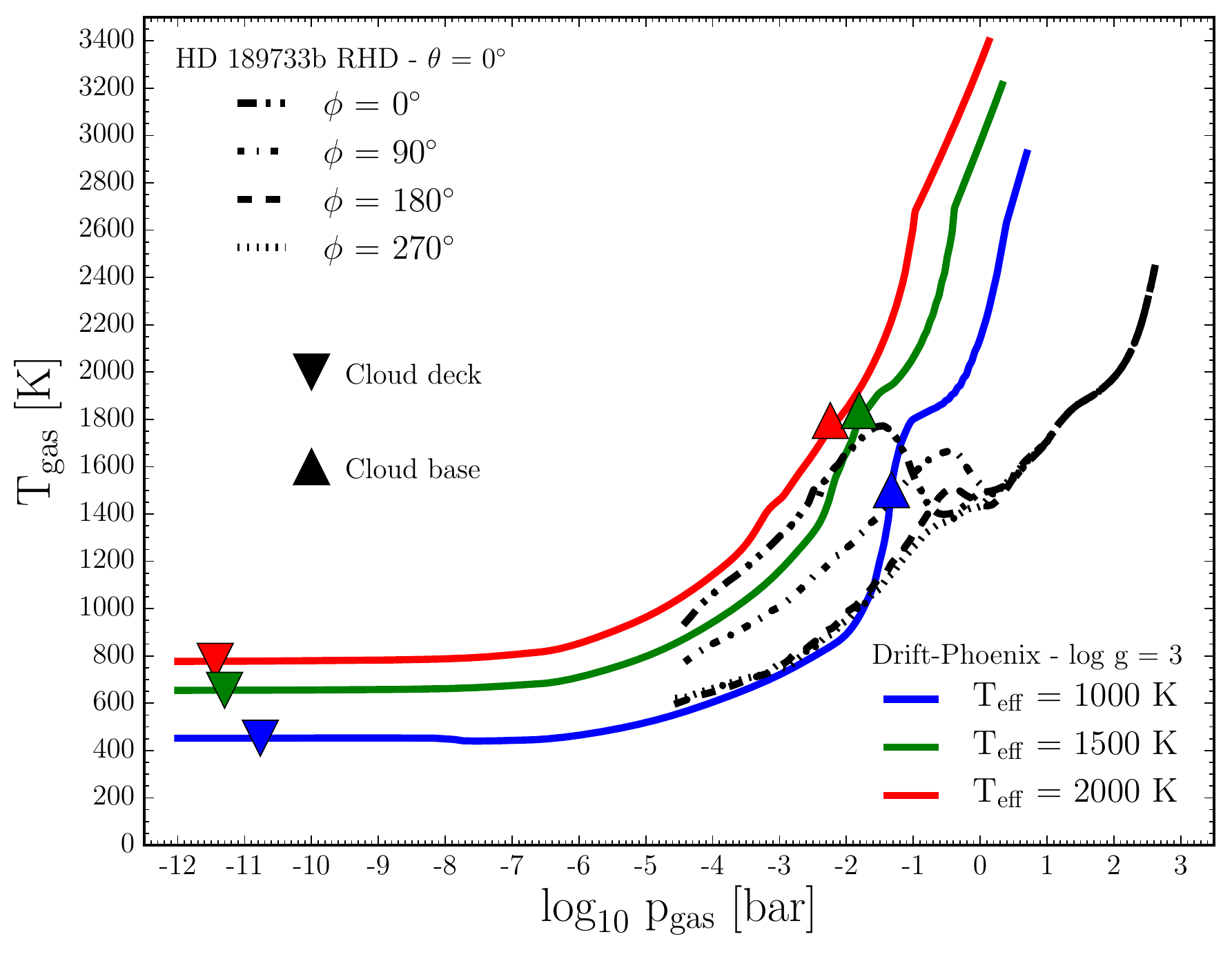}
 \caption{(T$_{\rm gas}$, p$_{\rm gas}$) profiles for Drift-Phoenix atmosphere models (coloured, solid) for log g = 3, T$_{\rm eff}$ = 1000, 1500, 2000 K and the \citet{Dobbs-Dixon2013} RHD model at $\phi$ = 0\degr, 90\degr, 180\degr, 270\degr, $\theta$ = 0\degr (black, styled). 
 Downward and upward facing triangles denote the cloud deck and base respectively. 
 The Drift-Phoenix profiles show that cloud formation begins at $\sim$10$^{-11}$ bar.}
 \label{fig:dfvsdobbs}
\end{figure*}

Our results strongly support the idea that the \textit{Hubble} and \textit{Spitzer} observations of HD 189733b described in \citet{Lecavelier2008,Sing2011,Gibson2012} and \citet{Pont2013} can be plausibly explained by the presence of cloud particles in the atmosphere. 
The cloud composition is dominated by a silicate-oxide-iron mix (Fig. \ref{fig:thetadust}) which supports the suggestion of a MgSiO$_{3}$[s] dominated cloudy atmosphere by \citet{Lecavelier2008}.
They suggest a MgSiO$_{3}$[s] dominated grain composition with sizes $\sim$10$^{-2}$$\ldots$10$^{-1}$ $\mu$m at pressures of 10$^{-6}$$\ldots$10$^{-3}$ bar.
Their grain size estimate is based on finding that Rayleigh scattering fits their observed feature-free slop in the optical spectral range.
This analysis is consistent with our detailed model results which produce grains of $\sim$20\% MgSiO$_{3}$[s] composition with sizes $\sim$10$^{-2}$$\ldots$10$^{-1}$ $\mu$m at $\sim$10$^{-4.5}$$\ldots$10$^{-3}$ bar, for all atmospheric profiles in this region.
However, Fig. \ref{fig:opacityisobar} shows that the isobars of the extinction $\kappa_{\rm ext}$ at the day-night terminator regions follow an absorption profile slope in the upper atmosphere which transitions into a flat profile deeper in the atmosphere.
The absorption dominated profile from 10$^{-4}$$\ldots$10$^{-3}$ bar does not support the Rayleigh scattering parameterisations for the observations of HD 189733b.
This does not rule out cloud particles as the source of the Rayleigh slope.
A possibility is that many smaller grains form at lower pressure regions than the boundary applied.
This would increase the scattering component of the cloud opacity.
Furthermore, in a 3D atmosphere, small grains may be more easily lofted than larger grains in regions of upward gas flow.
Our static 1D model does not include the 3D effects of a dynamic, changing atmosphere. 

Section \ref{sec:element} shows the effect that cloud particles have on local element abundances. 
Depletion of elements occur when they are consumed by cloud particle formation and replenishment occurs when these cloud particles evaporate.
In our cloud model the elements that constitute the main cloud particle composition were depleted by 3 orders of magnitude or more from $\sim$10$^{-4.5}$$\ldots$10$^{-2}$ bar.
These results suggest that the presence of cloud particles would generally flatten the spectral signatures from these elements and associated molecules in the atmosphere.
The evaporation of molecules from cloud particles transports elements from upper to lower atmospheric height.
This suggests that elements thought to contribute to upper atmosphere thermal inversion layers such as Ti (more specifically TiO; \citet{Fortney2008,Spiegel2009,Showman2009}) are transported to the deeper atmosphere by cloud particles.
The depletion and movement of TiO from the upper to lower atmosphere could reduce the intensity of a stratospheric inversion and/or move it deeper into the atmosphere or destroy it completely.
Figure \ref{fig:thetagasprop} shows that regions of the atmosphere can be oxygen poor or rich depending on the cloud processes.
This significantly alters the C/O ratio, where it is increased by the growth of oxygen baring solid materials onto cloud particle surfaces and decreased when these materials evaporate back into the gas phase.
This could have an impact on interpreting observations of over/under abundant C/O ratios in exoplanet atmospheres as well as altering the local chemistry in oxygen depleted regions.

The opacity of the clouds will strongly influence the spectral signature from the atmosphere by absorbing or scattering various wavelengths at different efficiency. 
From Fig. \ref{fig:opacity} optical wavelengths are preferentially absorbed and scattered in the upper atmosphere.
We therefore suggest that mineral clouds are responsible for the planets observed albedo and deep blue colour suggested by \citet{Berdyugina2011} and \citet{Evans2013}. 
The migration of the maximum efficiency of the cloud extinction from dayside to nightside (Fig. \ref{fig:opacitymap}, \ref{fig:opacitymap2}) at 0.6$\mu$m and 5.8$\mu$m shows a strong dayside-nightside opacity contrast. 
The terminator regions at longitude $\phi$ = 90\degr and 270\degr also show differences in dust extinction efficiency, especially deeper in the atmosphere.
Different thermodynamic conditions result in mean grain sizes, compositions and opacity that is different at each terminator region.
This has consequences for interpreting transmission spectroscopy measurements since observations of opposite limbs of the planet would have different extinction properties.
In addition, the cloud maps also show steep gradients of cloud properties at the terminator regions.
Transit spectroscopy observations that measure these regions would sample a variety of cloud particle number density, sizes and distributions dependent on wavelength. 
In Fig. \ref{fig:thetadust} the profiles at $\phi$ = 0\degr and 45\degr show a region compositionally dominated by Al$_{2}$O$_{3}$[s] cloud particles from $\sim$10$^{-2}$$\ldots$10$^{-1}$ bar with Fe[s]-rich grains on above and below this region.
The effect of this Al$_{2}$O$_{3}$[s] region is to produce a locally lower cloud opacity layer flanked by high opacity regions (Fig. \ref{fig:opacity}).
This would have a significant effect on radiation propagation through the atmosphere as the Fe[s] rich grains could shield photons reaching the Al$_{2}$O$_{3}$[s] layer from above or below.

Our qualitative calculation of the cloud contribution to the albedo provides yet another insight into the observational consequences of our cloud modelling.  
Our calculated single scattering albedo for the 8$\mu$m band (Fig. \ref{fig:ssa}) show that the maximum reflectance occurs approximately in the longitude range 0\degr$\ldots$135\degr at the equator at 10$^{-1}$ and 1 bar. 
With the peak occurring from 20\degr$\ldots$40\degr.
These maximum reflectance regions reproduce the areas of maximum 8$\mu$m flux map from \citet{Knutson2007}. 
This suggests that cloud particles influence observed phase curves of hot Jupiters depending on the position and reflectivity of cloud particles in the atmosphere.
These effects have been observed for hot Jupiter Kepler 7-b \citep{Demory2013}, where \textit{Kepler} observations showed a westward shift in optical phase curves (GCM/RHD models predict an eastward shift). 
\textit{Spitzer} phase curves showed that the shift was non-thermal in origin.
This shift was attributed to non-uniform, longitude dependent optical scattering from clouds.
Our 8$\mu$m reflectance result, albeit qualitatively, show that clouds can contribute to the infrared flux observed from the exoplanet atmosphere. 
It may not be simple to disentangle contributions by thermal emission and scattering/reflection clouds from infrared observational properties.

Figure. \ref{fig:dfvsdobbs} shows a (T$_{\rm gas}$, p$_{\rm gas}$) structure comparison between Drift-Phoenix (log g = 3, T$_{\rm eff}$ = 1000, 1500, 2000 K) and the \citet{Dobbs-Dixon2013} 3D RHD model.
Comparing trajectories from the 3D RHD model and the 1D Drift-Phoenix models suggests that clouds could be more extended into the lower-pressure regions than our present results show.
Previous studies (e.g. Drift-Phoenix; \citet{Witte2011}, \citet{Woitke2004}) involving cloud formation, the lower pressure boundary can be up to $\sim$10$^{-12}$ bar with cloud formation occurring from $\sim$10$^{-11}$ bar, these studies contain a smoother non-cloud to cloudy upper atmospheric region than our present results.
The cloud deck in the Drift-Phoenix models starts from $\sim$10$^{-11}$ bar, 7 orders of magnitude lower that the RHD upper boundary pressure.
Therefore, our results do not capture cloud formation outside the RHD model boundary conditions yet.
We also suggest that cloud formation can continue further inward than the results presented here.
Due to the increasing density and high pressure, cloud particle material remains thermally stable until considerably higher temperatures and survives deeper into the atmosphere.

We probed a 3D hot Jupiter atmospheric structure, irradiated by a host star, through selecting 16 1D atmospheres trajectories along the equator and latitude $\theta$ = 45\degr. 
We aimed to capture the main features of a dynamic atmosphere like east-west jet streams, latitudinal differences and dayside-nightside differences.
We used these thermodynamic and velocity profiles to study cloud formation in order to present consistently calculated cloud properties for HD 189733b.
Our ansatz does not yet include a self-consistent feedback onto the (T$_{\rm gas}$, p$_{\rm gas}$, $\vec{v}$$_{\rm gas}$) structure due to radiative transfer processes.
We note that due to an additional wavelength dependent dust opacity term in the radiative transfer scheme of \citet{Dobbs-Dixon2013}, this feedback is already accounted for to a significant degree in the current (T$_{\rm gas}$, p$_{\rm gas}$) profiles. 
Furthermore, we do not include non-LTE kinetic gas-phase chemistry such as photochemistry.
Departures from LTE have been detailed in non-equilibrium models of HD 189733b's atmosphere performed by \citet{Moses2011,Venot2012,Agundez2014} using thermodynamic input from the \citet{Showman2009} global circulation model.

\section{Conclusion and summary}
\label{sec:Conclusion}
We have presented the first spatially varying kinetic cloud simulation of the irradiated hot Jupiter exoplanet HD 189733b.
We applied a 2-model approach with our cloud formation model using 1D thermodynamic input from a 3D radiation-hydrodynamic simulation of HD 189733b.
Our results suggest that HD 189733b has a significant cloud component in the atmosphere, spanning a large pressure scale throughout the entire globe. 
Cloud particles remain thermally stable deep in the atmosphere up to $\sim$10 bar pressures and reach $\sim$mm sizes at the cloud base.
We suggest that cloud particles form at a lower pressure boundary than considered here, and also survive to deeper depths.
Cloud properties change significantly from dayside to nightside and from equator to mid latitudes with variations in grain size, number density and wavelength dependent opacity.
The cloud property maps show that steep gradients between dayside and nightside cloud proprieties occur at dayside-nightside transition regions.
These cloud property differences have implications on interpreting observations of HD 189733b, depending what region and depth of the planet is probed.
The single scattering albedo calculations showed that cloud particles could play a significant role in planetary phase curves. 
The reflectance of the clouds at 8$\mu$m showed that a significant fraction of infrared flux could originate from scattering/reflecting cloud particles.
Since the most reflective clouds in the infrared correspond to regions of highest temperature it may be difficult to distinguish between cloud reflections and thermal emission.
The extinction properties of the cloud particles exhibit absorption signatures in the upper atmosphere ($\gtrsim$10$^{-1}$ bar) which flatten deeper in the atmosphere ($\lesssim$10$^{-2}$ bar).
However, this does not rule out a cloud particle origin for the observations of Rayleigh scattering.
The scattering properties also suggest that the cloud particles would sparkle/reflect a midnight blue colour over the optical wavelength regime.

\begin{acknowledgements}
GL and ChH highlight the financial support of the European community under the FP7 ERC starting grant 257431.
Our respective local computer support is highly acknowledged.
We thank P. Woitke, C.R. Stark and P. Rimmer for helpful discussions and feedback.
We thank L. Neary for insightful discussion on atmospheric mixing.
\end{acknowledgements}

\bibliographystyle{aa}
\bibliography{bib2}{}

\end{document}